\documentclass{article}

\usepackage{PRIMEarxiv}

\usepackage[utf8]{inputenc} 
\usepackage[T1]{fontenc}    
\usepackage{hyperref}       
\usepackage{url}            
\usepackage{booktabs}       
\usepackage{amsfonts}  
\usepackage{amsmath}
\usepackage{nicefrac}       
\usepackage{microtype}      
\usepackage{lipsum}
\usepackage{graphicx}
\graphicspath{{media/}}  
\usepackage{xcolor}
\usepackage{soul}
\usepackage{comment}
\newcommand{\baike}[1]{\textcolor{black}{#1}}
\newcommand{\bshe}[1]{\textcolor{black}{#1}}

\title{Reverse Engineering the Reproduction Number: A Framework for Data-Driven Counterfactual Analysis, Strategy Evaluation, and Feedback Control of Epidemics
\thanks{B.S., S.S., and P.E.P. designed research; B.S. performed research, analyzed data, contributed
analytic tools, and wrote the paper; R.L.S. and I.P. contributed data and insights, and analyzed data;
B.S., S.S., and P.E.P. conceived the paper. The authors declare no competing interest. To whom correspondence should be addressed. E-mail: philpare@purdue.edu or sundara2@purdue.edu.} 
}

\author{
  Baike She \\
  Department of Mechanical and Aerospace Engineering  \\
  University of Florida \\
 Gainesville, FL\\
  \texttt{shebaike@ufl.edu} \\
   \And
  Rebecca Lee Smith \\
  Department of Pathobiology \\
  University of Illinois Urbana-Champaign \\
   Champaign, IL\\
  \texttt{rlsdvm@illinois.edu} \\
    \And
  Ian Pytlarz \\
  Institutional Data Analytics + Assessment \\
  Purdue University \\
  West Lafayette, IN\\
  \texttt{ipytlarz@purdue.edu} \\
  \And
  Shreyas Sundaram \\
  Elmore Family School of Electrical and Computer Engineering \\
  Purdue University \\
  West Lafayette, IN\\
  \texttt{sundara2@purdue.edu} \\
  \And
  Philip E. Par\'e \\
  Elmore Family School of Electrical and Computer Engineering \\
  Purdue University \\
  West Lafayette, IN\\
  \texttt{philpare@purdue.edu} \\
}

\begin{document}
\maketitle

\begin{abstract}
During the COVID-19 pandemic, different countries, regions, and communities constructed various epidemic models to evaluate spreading behaviors and assist in making mitigation policies. Model uncertainties, introduced by complex transmission behaviors, contact-tracing networks, time-varying spreading parameters, and human factors, as well as insufficient data, have posed arduous challenges for model-based approaches.
To address these challenges, we propose a novel \textit{framework for data-driven counterfactual
analysis, strategy evaluation, and feedback
control of epidemics,} which leverages statistical information from epidemic testing data instead of constructing a specific model.
Through reverse engineering the reproduction number by quantifying the impact of the intervention strategy, 
this framework tackles three primary problems:
1) How severe would an outbreak have been without the implemented intervention strategies?
2) What impact would varying the intervention strength have had on an outbreak?
3) How can we adjust the intervention intensity based on the current state of an outbreak?
Specifically, we consider the epidemic intervention policies such as the testing-for-isolation strategy as an example, which was successfully implemented by the University of Illinois Urbana-Champaign (UIUC) and Purdue University (Purdue) during the COVID-19 pandemic. By leveraging data collected by UIUC and Purdue, we validate the effectiveness of the proposed data-driven framework.
\end{abstract}

\section{Introduction}
Since 2019, the COVID-19 pandemic caused by SARS-CoV-2 has significantly affected societal work patterns~\cite{giordano2020modelling,witteveen2020economic,koelle2022changing,jones2020effective,witteveen2020economic,ioannidis2023estimates}. 
Proactive epidemic intervention policies were essential to prevent outbreaks~\cite{kucharski2020early,batteux2022effectiveness, soltesz2020effect, pradhan2020review,bubar2021model,richardson2022tracking,gostic2020estimated,runge2022modeling,kissler2020projecting,feng2020rational,accorsi2021detect,vespignani2020modelling,van2023effective}. To further assist in evaluating the effectiveness of the implemented pandemic intervention strategies and in designing feasible epidemic mitigation policies, various spreading models have been proposed and used to analyze the outbreak~\cite{cauchemez2019modelling,sharomi2017optimal,morris2021optimal,tsay2020modeling,perkins2020optimal,robert2023predicting,stoddard2023impact,dangerfield2019resource,du2022modeling}. 
However, model uncertainties, introduced by complex transmission behaviors~\cite{yesudhas2021covid,chan2022covid}, contact-tracing networks~\cite{colizza2021time}, time-varying spreading parameters~\cite{calafiore2020time}, human factors~\cite{weitz2020awareness}, and  insufficient data~\cite{zimmermann2021estimating}, have posed significant challenges for model-based approaches~\cite{mollison1994epidemics,pellis2015eight,britton2015five,riley2015five,dykes2022visualization,bertozzi2020challenges,roda2020difficult}. Meanwhile, data-driven approaches such as deep learning and reinforcement learning frameworks require extensive data and real-world trials in an actual epidemic spreading environment~\cite{sujath2020machine,bastani2021efficient,bhosale2023application,khadilkar2020optimising,hosseinloo2023data}. Due to the irreversible nature of the spread, it is impossible to restart the exact same epidemic spreading process for testing different intervention strategies and improving resource allocation strategies.
Consequently, the following questions remain unanswered: 1) How severe would an outbreak have been without the implemented intervention strategies?
2) What impact would varying the intervention strength have had on an outbreak?
3) How can we adjust the intervention intensity based on the current state of an outbreak?

To tackle these challenges, 
we propose a framework for data-driven counterfactual analysis, strategy evaluation, and feedback control of epidemics. Distinct from most existing counterfactual analyses on intervention strategy evaluation and epidemic mitigation works that assume prior knowledge of predefined epidemic models and/or require unrealistic big data, 
it is sufficient to obtain the core of our framework, i.e., the reproduction number, through limited data and statistical inference approaches.  The framework quantifies the impact of the intervention strategy on changes in the reproduction number. 
Consequently, by reverse engineering the reproduction number in a novel manner, we can reconstruct the epidemic spreading process in order to measure the impact of the chosen strategy and to test alternative intervention strategies. 
Finally, we can leverage the framework to design a feedback control algorithm for updating intervention strategies based on the assessment of past and current spreading. 


The reproduction number, inferred from spreading data, captures the average number of new infected cases generated by a single infected individual~\cite{fraser2009pandemic,vegvari2022commentary}. Leveraging the reproduction number to evaluate the epidemic spread and assist in policy-making is widely accepted~\cite{soltesz2020effect,liu2020reproductive,parag2022epidemic,dietz1993estimation,parag2021improved,inglesby2020public,thompson2019improved, linka2020reproduction,kucharski2020early, nash2022real,abbott2020epinow2,vegvari2022commentary,alimohamadi2020estimate, hasan2022new,dietz1993estimation,parag2021improved,thompson2019improved,katul2020global}.  
Our proposed framework addresses the challenge of understanding the impact of intervention strategies on the reproduction number. 
Nowadays, researchers and policy-makers mainly leverage the reproduction number to analyze and predict epidemic spreading processes.
Our mechanism opens a new door to utilize the reproduction number for assessing potential outcomes of different implemented interventions, and for adjusting implemented intervention strategies to inform future decision-making~\cite{alleyne2023control}.

We introduce and validate the framework by incorporating COVID-19 spreading data and the testing-for-isolation intervention strategy at the University of Illinois Urbana-Champaign (UIUC) and Purdue University. Using this framework, we illustrate what could have potentially happened if the strategies had not been implemented on both campuses. Meanwhile, we explore the impact of implementing a different strength of the testing-for-isolation strategy on the spread across the campuses.
Further, we investigate whether, instead of adhering to a fixed testing-for-isolation strategy, it is possible to outperform the implemented strategy by adjusting the strategy through a feedback control mechanism. Through this analysis, we show that we can adapt the intensity of the  strategy based on the severity of the epidemic spread, as indicated by the reproduction number, without the need for specific spreading models or large data sets.
\section{Background}
\subsection{A
Framework for Data-Driven Counterfactual
Analysis, Strategy Evaluation, and Feedback
Control of Epidemics}
\begin{figure}[htbp]
\centering
\includegraphics[width=1\linewidth]{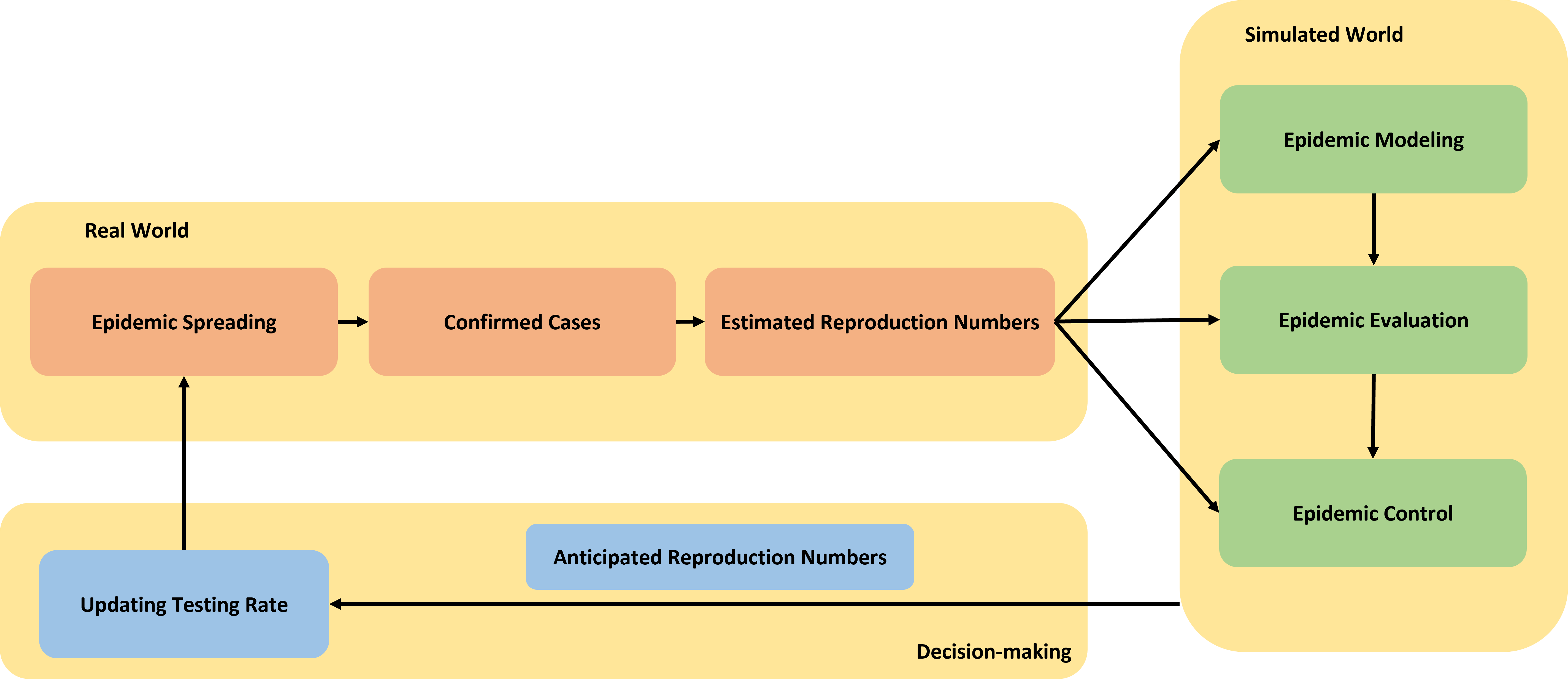}
\caption{A framework for data-driven counterfactual analysis, strategy evaluation, and feedback control of epidemics.}
\label{fig:framework}
\end{figure}

The proposed framework
is shown in Figure~\ref{fig:framework}. 
By leveraging data from a real-world epidemic spreading process, the first piece of the framework is
to \baike{estimate the reproduction number~\cite{cori2013new,huisman2022estimation, Epyestim_Python_2020, abbott2020epinow2,brockhaus2023different,gostic2020practical}}. 
It is necessary to leverage the statistical information from the data to estimate the reproduction number, including the 1) initial infection profile of the spread~\cite{cori2013new, gostic2020practical, lehtinen2021relationship} and 
2) quantification of intervention strategy on the spreading. 
We further explain these concepts later in the paper.

As shown in Figure~\ref{fig:framework}, the epidemic reconstruction, intervention strategy evaluation, and feedback control methodologies rely on the estimation of the reproduction number from real-world data. \baike{By leveraging the estimated reproduction number,} the framework first reconstructs the spreading processes under the exact same 
\bshe{intervention} strategies.
Then, we reverse the impact of the interventions strategies to reconstruct the worst-case conditions for the outbreak, if no actions had been taken.  
Thus, the framework allows us to implement intervention strategies of different strengths by leveraging the reconstructed spread. We achieve the procedure by proposing a mechanism to scale the reproduction number through the corresponding intervention strategy. This reconstruction not only allows for an evaluation of the effectiveness of the employed intervention strategies but also provides a testing environment to compare intervention strategies of different intensities. 
Further, we use the reproduction number as  feedback to indicate the severity of the pandemic to adjust the intensity of the intervention strategy via the reproduction number. 


In particular, we consider the intervention strategies in Figure~\ref{fig:framework} as the 
testing-for-isolation strategy that removes
the infected  individuals from the whole population~\cite{casella2020can,maya2023covid}. Similar to vaccination strategies, which remove the susceptible population from the mixed group \cite{grundel2021coordinate,du2022modeling,bubar2022sars,hellewell2020feasibility}, the testing-for-isolation strategy is another widely adopted method for epidemic mitigation~\cite{casella2020can,acemoglu2021optimal,olejarz2023optimal,she2022optimal}. Inspired by the successful implementation of testing-for-isolation strategies during the COVID-19 pandemic, in particular at the University of Illinois Urbana-Champaign and Purdue University, we leverage the data collected by both universities to validate our framework. 

~\baike{Although we introduce the methodologies and framework by considering the testing-for-isolation strategy as the intervention strategy, other intervention strategies will also fit within this framework.}
\bshe{Since, if another intervention strategy other than the testing-for-isolation strategy is considered, the strength of the testing-for-isolation strategy can be replaced with other factors that affect the reproduction number~\cite{zhang2020changes}. For instance, if we study the impact of a lockdown intervention, we substitute the intervention strategies in Figure~\ref{fig:framework} with the strength of lockdown intervention.} 
\subsection{Testing-For-Isolation Strategies Implemented by UIUC and Purdue}
We leverage the aggregated data collected by
Purdue and UIUC to
validate the framework. 
Plenty of universities, including Cornell University, Emory University, Purdue University, and the University of Illinois Urbana-Champaign, implemented various nonpharmaceutical interventions (NPIs) such as social distancing, face-masking, and remote working, among others~\cite{frazier2022modeling,lopman2020model, Purdue_testing, ranoa2022mitigation, kharkwal2021university,lopman2020model,wrighton2020reopening,ghaffarzadegan2021diverse,gressman2020simulating, brown2021simple,bahl2021modeling,borowiak2020controlling}. A critical mitigation policy adopted by these universities was testing-for-isolation (or testing-for-quarantine). This strategy actively tests a proportion of the university’s population and quarantines those who test positive to prevent the infected population from spreading the virus. More discussion on unique methodologies implemented by different universities is in SI-1-A.

Compared to the implemented intervention strategies from other universities,
Purdue University and UIUC implemented pre-arrival testing, regular screening, and voluntary testing, as well as case isolation, contact tracing, and quarantine to mitigate the spread of infection. We collected and studied testing data from both universities to investigate their testing strategies during the COVID-19 pandemic through close collaboration with the 
Institutional Data Analytics + Assessment team (IDA+A) at Purdue University and the SHIELD: Target, Test, Tell team (SHIELD) at the University of Illinois Urbana-Champaign. 
Both teams adopted testing-for-isolation strategies  to assess the severity of the pandemic and made necessary adjustments to their plans.
Consequently,
Purdue and UIUC effectively maintained the infected population  at levels that allowed for safe operation throughout the semesters, despite experiencing periodic spikes. 
This success highlights the potential of testing-for-isolation strategies in mitigating current and future pandemics.



\begin{figure}[htbp]
\centering
\includegraphics[width=1\linewidth]{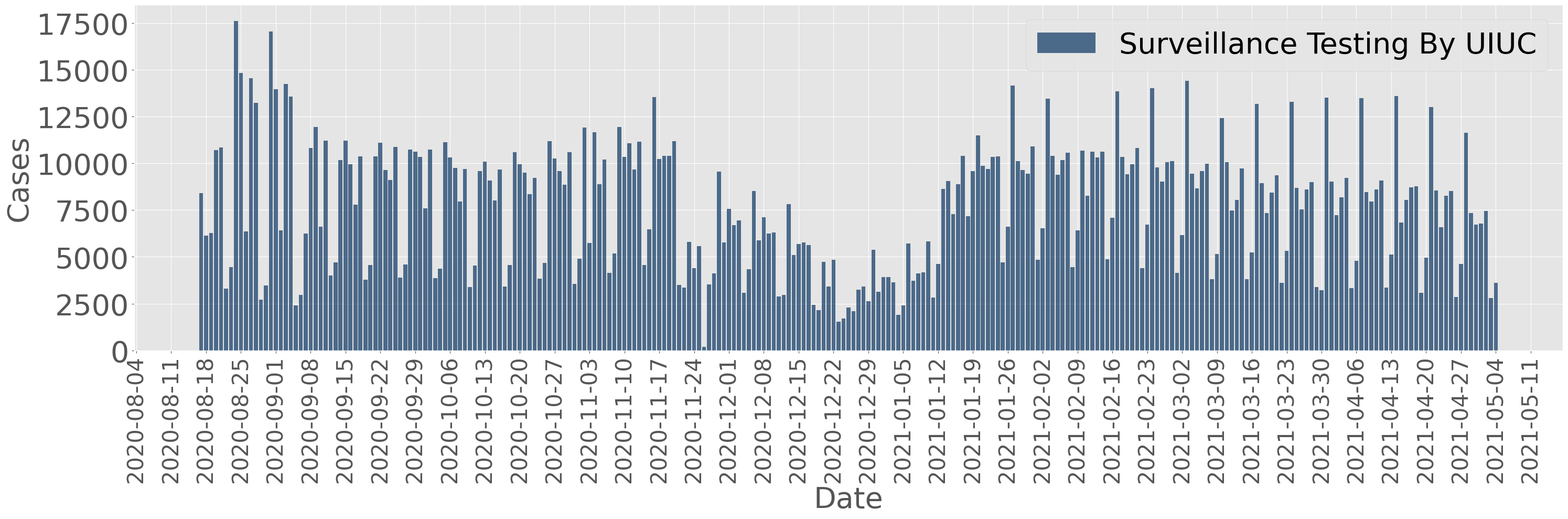}
\caption{Daily surveillance tests by UIUC during Fall 2020 and Spring 2021.}
\label{fig:SVL_UIUC_T}
\end{figure}
\begin{figure}[htbp]
\centering
\includegraphics[width=1\linewidth]{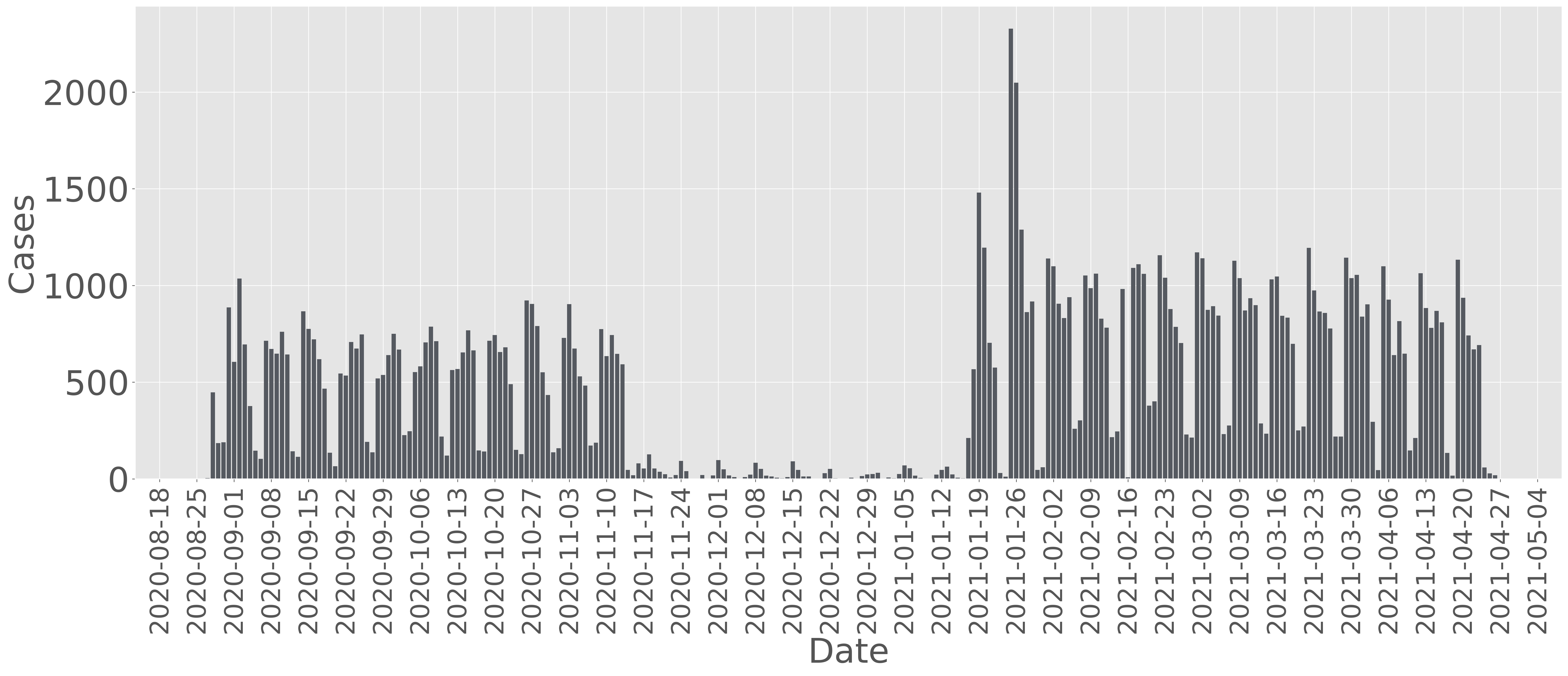}
\caption{Daily surveillance tests implemented by Purdue for sampling asymptomatic cases during Fall 2020 and Spring 2021.}
\label{fig:SVL_T_P}
\end{figure}

To validate our framework, we specifically focus on the early stage of the pandemic, between Fall 2020 and  Spring 2021, when pharmaceutical interventions were not yet available. To prevent the spread of the epidemic and ensure the safety of students, faculty, and staff, UIUC implemented a policy of testing the entire campus 2-3 times a week during Fall 2020 \baike{and Spring 2021}.
\baike{The surveillance testing of UIUC is shown in Figure~\ref{fig:SVL_UIUC_T}}.
This approach helped \baike{identify and isolate infected individuals, and therefore, prevented} further transmission. In summary,
Purdue encouraged individuals with COVID-19 symptoms to conduct voluntary testing. Additionally, to identify and isolate positive cases among asymptomatic individuals, Purdue also implemented \baike{a surveillance testing policy} by sampling between 8\% and 12\% of the total population each week during Fall 2020, as illustrated by Figure~\ref{fig:SVL_T_P}. 

\bshe{
Moreover, both UIUC and Purdue distinguish between isolation and quarantine.  In this work, we refer to isolation as a measure to prevent those who tested positive from further spreading the virus. We do not consider quarantine.
For simplicity, we consider uniformly random sampling at both universities. Additionally, since we use daily aggregated data, we do not consider contact-tracing-based analysis using spatial data or networks.} 
More detailed information about \baike{the testing methods and resources implemented by UIUC and Purdue} can be found in SI-1-B and SI-1-C. 
\section{Data-Driven Counterfactual Analysis: A Case Study}
\subsection{Quantification of Testing Strategy on Epidemic Spread}
\label{SEC-A}
One standard way of characterizing epidemic spread is to use contact tracing data to create an \textit{infection profile}, also known as the generation time interval~\cite{fraser2007estimating,lehtinen2021relationship, griffin2020rapid,knight2020estimating}. 
The infection profile represents the average time between the onset of infections of a primary case and its secondary cases~\cite{cori2013new,knight2020estimating,gostic2020practical}. It is also used to estimate critical epidemiological parameters such as the reproduction number, generation time, and attack rate~\cite{vink2014serial,lehtinen2021relationship, griffin2020rapid,prem2020effect, he2020temporal}. 
The research~\cite{ali2020serial} discovered that changes in contact patterns and the implementation of public health interventions can alter the infection profile.
Hence, based on the widely adopted testing-for-isolation strategies, we first propose a methodology to quantify the influence of these strategies on infection profiles~\cite{ranoa2022mitigation}, 
and further on the reproduction number,
in order to \baike{lay a foundation for}
reverse engineering the reproduction number.




Epidemics such as COVID-19 can result in both symptomatic and asymptomatic infections. 
We define infection profiles for
symptomatic and asymptomatic infections separately. 
Due to the existence of incubation period where an infected individual is not infectious, consider the day when a symptomatic case becomes infectious as day $1$. We define $\underline{v}_i\in \mathbb{R}_{>0}$ as the average infected cases caused by a single symptomatic case on day $i$ since day $1$,  $i\in\{1,2,3, \dots, n\}$,  where  $n\in \mathbb{N}_{>0}$ is the number of days during which an symptomatic case is infectious. Hence, the infection profile of symptomatic cases is defined as a vector 
\begin{equation}
\label{eq:inf_prof_sym}
  \underline{v}=[\underline{v}_1, \underline{v}_2, \cdots, \underline{v}_n]. 
\end{equation}
Similarly, we define $\overline{v}_i\in \mathbb{R}_{>0}$ as the average infected cases caused by a single asymptomatic case on day $i$ since day $1$, $i\in\{1,2,3, \dots, m\}$, where  $m\in \mathbb{N}_{>0}$ is the number of days during which an asymptomatic case is infectious.
The infection profile of asymptomatic cases is defined as a vector 
\begin{equation}
\label{eq:inf_prof_asym}
    \overline{v}=[\overline{v}_1, \overline{v}_2, \cdots, \overline{v}_m].
\end{equation}
Then, the reproduction number of symptomatic and asymptomatic cases can be obtained by
\begin{equation} 
    \underline{\mathcal{R}} = \sum_{i=1}^n \underline{v}_i \ \mbox{and} \ 
    \overline{\mathcal{R}}= \sum_{i=1}^m \overline{v}_i,
    \label{Eq: R_total_1}
\end{equation}
respectively \cite{cori2013new}.
For an epidemic with both symptomatic and asymptomatic infected cases, if \textit{the proportion of the symptomatic infection} is $\theta\in [0,1]$, \baike{then,} the proportion of  the asymptomatic infection will be $(1-\theta)\in [0,1]$. \baike{Consequently,} the reproduction number of the spreading process is given by 
\begin{equation}
\label{eq:R_total}
    \mathcal{R} = \theta \sum_{i=1}^n \underline{v}_i+ (1-\theta)\sum_{i=1}^m \overline{v}_i=\theta \underline{\mathcal{R}}+ (1-\theta)\overline{\mathcal{R}}.
\end{equation}
Note that $\mathcal{R}$ denotes the basic reproduction number of an infection if the infection profiles $\underline{v}$ and $\overline{v}$ are estimated in a 
nearly fully susceptible population. \baike{In addition, the reproduction number $\mathcal{R}$ is heavily determined  by the ratio of the symptomatic infection.}
In this study, we use the same infection profiles for both symptomatic and asymptomatic cases at UIUC and Purdue, as previous research on epidemic spread over the UIUC campus did not differentiate between the infection profiles for these two types of infections~\cite{ranoa2022mitigation}.
More discussion on this topic can be found in SI-2-A-1. 

When mentioning testing everyone twice a week during Fall 2020 at the UIUC, the testing was not implemented on the same day, since the daily laboratory capacity to perform the test on the UIUC campus was 10,000 tests. As illustrated by Figure~\ref{fig:SVL_UIUC_T} and~\ref{fig:SVL_T_P}, 
testing-for-isolation strategies can be considered as a policy implemented on a weekly basis and distributed evenly throughout the week.
Furthermore, we consider that testing only records the number of confirmed infectious cases. The intervention that can mitigate the potential outbreak is the isolation intervention after testing positive. Ideally, the \textit{daily isolated population} should equal the daily confirmed cases. However, because isolation is not mandatory, we use the \textit{isolation rate} instead of the testing rate to describe the effectiveness of the testing-for-isolation strategy when they are different.

If the weekly isolation rate is denoted as $7\times \alpha$, the average daily isolation rate during the week \baike{is given by} $\alpha$, $\alpha\in\mathbb{R}_{\geq 0}$. 
We propose a mechanism to quantify the impact of the isolation rate on the infection profile, and thus, on the reproduction number. Consider testing-for-isolation strategies for asymptomatic infection, where the infection profile is given by~\eqref{eq:inf_prof_asym}.
If we have $k\in\mathbb{N}_{>0}$ asymptomatic
infectious cases on day 1,
without testing-for-isolation strategies, 
these $k$ infectious cases will generate an average number of $k\overline{v}_i$ cases on day $i$, $i\in \{1,2,\dots, m\}$. However, consider the same number of $k$ asymptomatic cases under a testing-for-isolation strategy. 
If we test and then isolate $k\overline{\alpha}$ asymptomatic cases on day $1$ from the $k$ asymptomatic cases, 
there will be $k\overline{v}_1(1-\overline{\alpha})$ new infected cases that are generated by the $k(1-\overline{\alpha})$ cases. On day $2$, there will be $k\overline{v}_2(1-\overline{\alpha})^2$ cases generated by $k(1-\overline{\alpha})^2$ infectious asymptomatic cases. Consequently, the new infected cases caused by the original $k$ asymptomatic cases are $k\overline{v}_i(1-\overline{\alpha})^i$ on day $i$, $i\in\{1,2,\dots, m\}$. Thus, the average number of infected cases generated by a single asymptomatic individual on day $i$, $i\in\{1,2,\dots, m\}$, is given by $k\overline{v}_i(1-\overline{\alpha})^i$. 
We obtain the asymptomatic infection profile under the impact of the isolation rate $\overline{\alpha}$, which is given by
\begin{equation}
\label{Eq:Infection_Prof}
    \overline{v}(\overline{\alpha})=[\overline{v}_1(1-\overline{\alpha}), \overline{v}_2(1-\overline{\alpha})^2, \cdots, \overline{v}_m(1-\overline{\alpha})^m].
\end{equation}
\eqref{Eq:Infection_Prof} quantifies the connection between the isolation rate and the infection profile, and 
consequently, the reproduction number. Note that the same mechanism is applicable to modify the infection profile of symptomatic infections, and we use $\underline{\alpha}$ and $\underline{v}(\underline{\alpha})$ to represent the isolation rate of symptomatic cases and the infection profile of symptomatic infections under the isolation rate, respectively.
Similar to the computation of the reproduction number of the mixed population in~\eqref{eq:R_total}, the modified infection profile with both symptomatic and asymptomatic infection of the population is given by 
\begin{equation}
\label{Eq:infection_t}
    v(\alpha) = \theta \underline{v}(\underline{\alpha}) +(1-\theta) \overline{v}(\overline{\alpha}),
\end{equation}
\baike{quantifying the impact of the overall isolation rate $\alpha = \theta\underline{\alpha}+(1-\theta)\overline{\alpha}$
on the infection profile.}

\begin{figure}[htbp]
\centering
\includegraphics[width=1\linewidth]{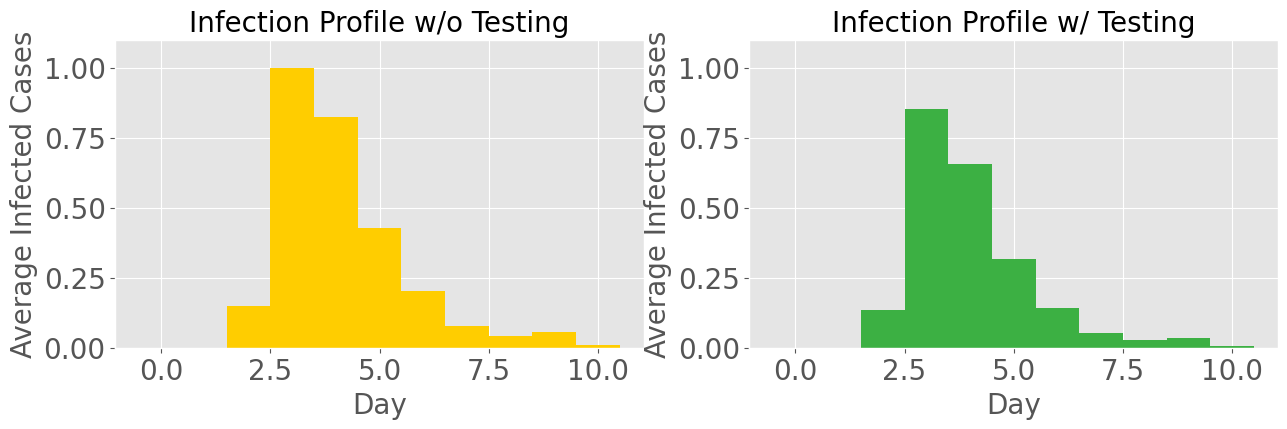}
\caption{The infection profile of COVID-19 w/ (left) and w/o (right) testing strategies.}
\label{fig:Inf_Pro}
\end{figure}

After quantifying the impact of testing-for-isolation strategies on infection profiles, we consider implementing the mechanism in~\eqref{Eq:infection_t} on the following infection profile. 
As demonstrated by the SHIELD team at the UIUC~\cite{ranoa2022mitigation}, 
in this work, we use  
the following
infection profile to capture epidemic spread at Purdue and UIUC~\cite{ranoa2022mitigation,goyal2021viral}: 
\begin{align}
    \underline{v}=\overline{v} = [0.148, 1.0, 0.823, 0.426, 0.202,\nonumber \\ 
    0.078, 0.042, 0.057, 0.009].
    \label{eq:inf_pro_ref}
\end{align}
The infection profile in~\eqref{eq:inf_pro_ref} is shown in Figure~\ref{fig:Inf_Pro}~(left). 
Then, we study the impact of Purdue's testing-for-isolation strategy on the infection profile. \baike{Unlike UIUC's testing strategy, Purdue performed voluntary testing for symptomatic infection while surveillance testing for asymptomatic infection.}
Based on the Purdue testing data from Fall 2020 \baike{and Spring 2021}, there were 55\% symptomatic infections ($\theta = 0.55$)~\cite{CDC_sym_2021}. We refer the readers to SI-2-E-3 for more information on this ratio. 
To match Purdue's record, we also consider the positive cases that caught by the voluntary testing at Purdue as the symptomatic cases. We further consider cases caught through voluntary testing are isolated once they test positive within a week (since they are cautious and willing to be tested), spread evenly over the week. Consequently, we treat the daily isolation rate for symptomatic cases, i.e., positive cases that were caught by voluntary testing, as $\underline{\alpha}_P = 1/7$. Based on~\eqref{Eq:Infection_Prof}, the infection profile of symptomatic cases under the voluntary testing-for-isolation strategy is 
\begin{align*}
    \underline{v}(\underline{\alpha}_P) = [  0.127, 0.735,
 0.518, 0.229, \\ 
 0.093, 0.031,
 0.014, 0.017, 0.002].
\end{align*}
In addition to voluntary testing,
Purdue tested 10\% of the whole population  on campus to catch the asymptomatic cases (cases that were not tested voluntarily), shown in Figure~\ref{fig:SVL_T_P}. 
However, due to the implemented strategies by Purdue IDA+A to select sampling targets, we consider that the isolation rate for asymptomatic cases $\overline{\alpha}_P$ is higher than $10\%$. 
We consider $\overline{\alpha}_P = 0.3/7$\footnote{We also study the situation under different isolation rates in the supplemental material.}. This 
prerequisite implies that we can detect and isolate 30\% of the asymptomatic infected population. 
More information on the isolation rate can be found in SI-2-C.

Based on \eqref{Eq:Infection_Prof}, the infection profile of asymptomatic cases at Purdue is 
\begin{align*}
    \overline{v}(\overline{\alpha}_P) = [ 0.147, 0.916,
 0.722, 0.356, \\0.162, 0.06,
 0.03, 0.04, 0.006].
\end{align*}
Then we generate the combined reproduction number of the infection profile of the whole population through \eqref{Eq:infection_t}, shown in Figure \ref{fig:Inf_Pro}~(right). Based on~\eqref{Eq: R_total_1} and~\eqref{eq:R_total}, the reproduction number is given by $\mathcal{R}(\alpha_{P}) = 2.07$, where $\alpha_P$ is the overall isolation rate. Based on~\eqref{eq:inf_pro_ref}, the original reproduction number of the infection profile $v$ is $\mathcal{R} = 2.785$. Therefore, the testing-for-isolation strategy that is implemented by Purdue scales the reproduction number by
\begin{equation}
\label{eq:S_F}
    \mathcal{F} = \mathcal{R}(\alpha_P)/\mathcal{R} =0.742,
\end{equation}
where we define $\mathcal{F}$ as the \textit{scaling factor} of the reproduction number under the overall isolation rate $\alpha_P$ (see SI-2-C).  
 
 To analyze the impact of the testing-for-isolation on the UIUC campus,  we treat everyone as asymptomatic or symptomatic cases since we leverage the the same infection profile. 
Since UIUC tested the whole campus twice a week during Fall 2020, we consider all detected symptomatic and asymptomatic cases could be isolated directly. Therefore, we consider the isolation rate to be the same as the testing rate.
Using our proposed mechanism in~\eqref{Eq:Infection_Prof}, we find that during Fall 2020, the strategy of testing everyone twice a week and then isolating all positive cases evenly spread throughout the week, scales the reproduction number down by $38.85\%$, i.e.,~$\mathcal{F}=0.389$. 

Compared to Purdue's strategy, UIUC's testing-for-isolation strategy can reduce the reproduction number almost twice as much as Purdue's ($\mathcal{F}=0.742$), at the cost of more testing resources, as illustrated in~Figures~\ref{fig:SVL_UIUC_T} and \ref{fig:SVL_T_P}. Based on the confirmed cases at UIUC (Figure~\ref{fig:UIUC_P}) and Purdue (Figure~\ref{fig:Purdue_P}), Purdue had more confirmed cases in general, with higher spikes, compared to UIUC. This observation is under the fact that UIUC caught and isolated more infected cases through higher testing rates, while Purdue only caught a proportion of the infected cases daily. We validate the
proposed mechanism in~\eqref{Eq:Infection_Prof} and~\eqref{Eq:infection_t}, which establishes a way of measuring the strength of intervention strategies (i.e., the testing-for-isolation strategy) on epidemic spreading processes in terms of modifying the infection profile and the reproduction number. 
More discussion on how to quantify the impact of testing-for-isolation strategy on the infection profile and the reproduction number can be found in SI-2-C. 
\subsection{Analysis of Epidemic Spread at UIUC and Purdue}
\label{Sec-B} 
In this subsection, 
we analyze the epidemic spread at Purdue and UIUC. We leverage the confirmed cases from both universities, and their corresponding modified infection profiles to estimate the reproduction number, as illustrated by the framework in the dashed lines in Figure~\ref{fig:framework}.
In order to estimate the reproduction number, it is important to distinguish between infected cases and confirmed cases~\cite{cori2013new,gostic2020practical}. 
An infected case means that an individual is already infected by the virus, but they may not be contagious yet due to the existence of an incubation period. Therefore, an infected case is not equivalent to an infectious case. Meanwhile, a confirmed case refers to a case that has been reported as infected, but there may be delays between the time of infection and confirmation due to testing and reporting delays. All the data we obtained from UIUC and Purdue are considered confirmed cases rather than infected cases.
We use the Bayesian inference techniques proposed in EpiEstim~\cite{Epyestim_Python_2020,huisman2022estimation,bhatia2023extending} to estimate the reproduction number, where we leverage the deconvolution techniques in~\cite{huisman2022estimation,Epyestim_Python_2020}
to obtain infected cases through confirmed cases\footnote{We leverage the package Epyestim developed by~\cite{Epyestim_Python_2020}, which is a realization of the R package EpiEstim CRAN~\cite{cori2013new}~in Python.}. 
We implement a delay distribution of 10-day mean to capture the delay from infection-to-confirmation~\cite{huisman2022estimation,brauner2021inferring,Epyestim_Python_2020}. 
The infection-to-confirmation delay distribution is a convolution from
the incubation period distribution~\cite{brauner2021inferring, brauner2020effectiveness} and the testing-to-confirmation distribution~\cite{tariq2020real}.
Detailed methodologies on obtaining infected cases through confirmed cases can be found in~\cite{huisman2022estimation,cori2013new, gostic2020practical, nash2022estimating,bhatia2023extending, abbott2020epinow2,Epyestim_Python_2020} and in SI-2-B and SI-2-D.

One critical step to leverage Bayesian inference to estimate the reproduction number is the normalized infection profile ~\cite{cori2013new,huisman2022estimation, bhatia2023extending,abbott2020epinow2,Epyestim_Python_2020}, which is usually referred to as the serial interval distribution~\cite{flaxman2020estimating}. 
Following the modified infection profile in~\eqref{Eq:Infection_Prof}, we define serial interval distributions of symptomatic and asymptomatic infections as 
\begin{align}
\underline{w}=\underline{v}/\underline{\mathcal{R}}=[\underline{v}_1/\underline{\mathcal{R}}, \underline{v}_2/\underline{\mathcal{R}}, \cdots, \underline{v}_n/\underline{\mathcal{R}}],\\
    \overline{w}= \overline{v}/\overline{\mathcal{R}}=[\overline{v}_1/\overline{\mathcal{R}}, \overline{v}_2/\overline{\mathcal{R}}, \cdots, \overline{v}_m/\overline{\mathcal{R}}],
\end{align}
respectively. 
The modified serial interval distribution of a spreading process with both symptomatic and asymptomatic infections under isolation rates $\underline{\alpha}$ and $\overline{\alpha}$, with the ratio of the symptomatic infection equals $\theta$ is defined as $w(\alpha)$, where
\begin{equation}
\label{eq:serial_int}
     w_i(\alpha)=(\theta\underline{v}_i(1-\underline{\alpha})^i+(1-\theta)\overline{v}_i(1-\overline{\alpha})^i)/\mathcal{R},
\end{equation}
$i\in \{1,2,\dots, n\}$ (considering $m=n$). Based on the implemented isolation rates and the predefined infection profile in~\eqref{Eq:infection_t}, we can obtain the modified infection profile, and then obtain the modified serial interval in~\eqref{eq:serial_int}. Since the serial interval distribution is the normalized infection profile, we refer readers to SI-2-C to check how to compute the scaling factor $\mathcal{F}$ directly through the serial interval distribution. 

\begin{figure}[htbp]
\centering
\includegraphics[width=1\linewidth]{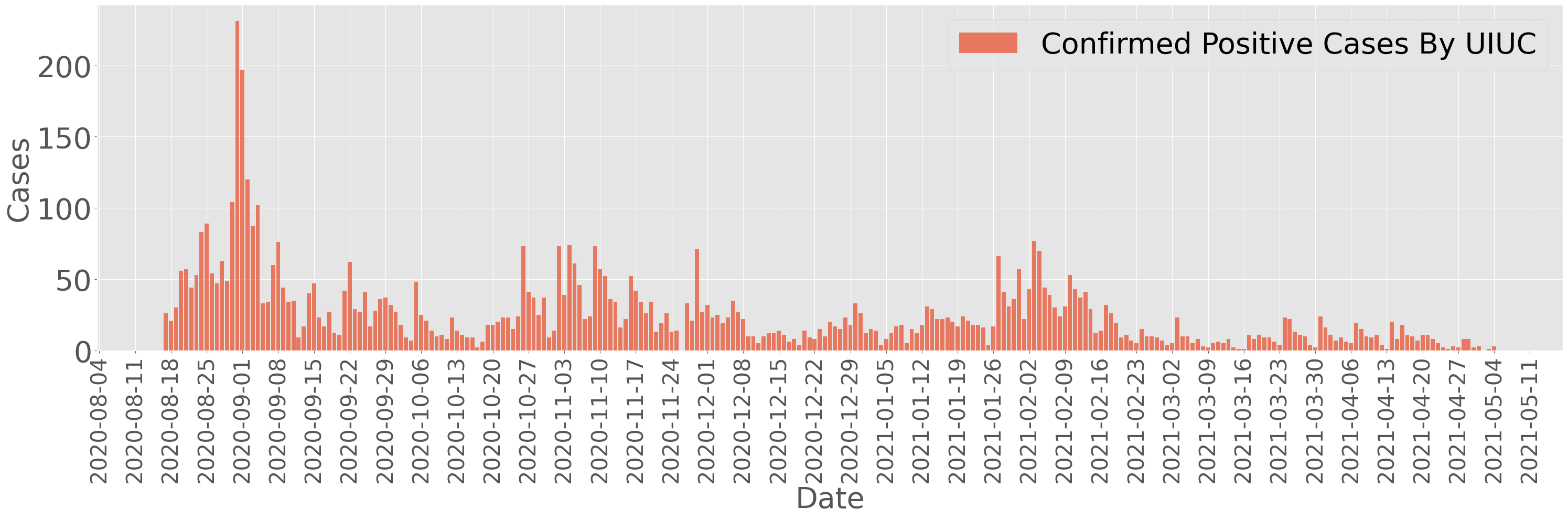}
\caption{Daily confirmed positive cases at the UIUC from Fall 2020 to Spring 2021 captured by the SHIELD team at the UIUC.}
\label{fig:UIUC_P}
\end{figure}

\begin{figure}[htbp]
\centering
\includegraphics[width=1\linewidth]{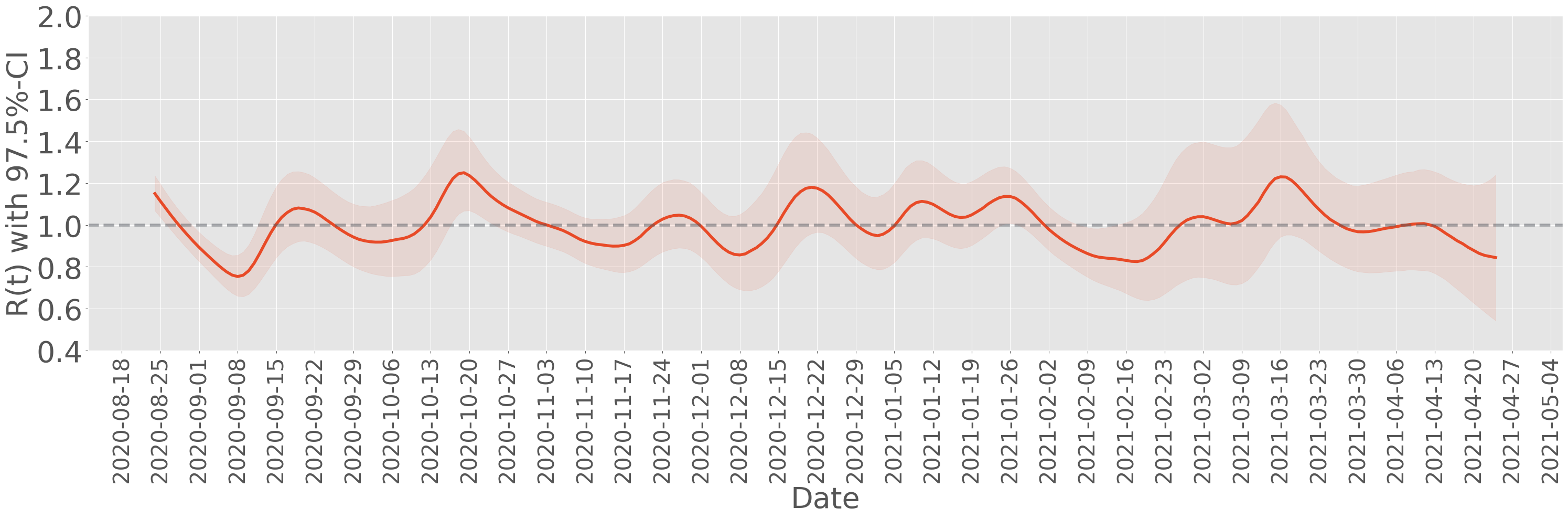}
\caption{Estimated reproduction number at the UIUC from Fall 2020 to Spring 2021.}
\label{fig:UIUC_R}
\end{figure}

We estimate the reproduction number for both universities based on the modified serial interval distribution~\cite{ranoa2022mitigation}, 
accounting for the testing-for-isolation strategies employed by UIUC and Purdue, respectively. We leverage the daily confirmed cases from UIUC during Fall 2020 and Spring 2021 (Figure~\ref{fig:UIUC_P}) 
to estimate the reproduction number, 
illustrated in Figure~\ref{fig:UIUC_R}, 
along with the $97.5\%$ confidence interval. 
One key observation during Fall 2020 is that there were four periods where the reproduction number was above 1, matching the four main spikes during that semester in the testing data in Figure~\ref{fig:UIUC_P}. 
The most notable spike during Fall 2020 in Figure~\ref{fig:UIUC_P} was during the period from August 11th, 2020 to August 31st, 2020. At the beginning of the Fall 2020 semester, when students returned to campus, a significant number of infected cases were identified by the SHIELD team. Additionally, the reproduction number experienced a significant spike during mid-October to early November. Research by the SHIELD team at UIUC attributed this increase to the return of the BIG TEN football season, when 
students violated the social distancing policy and began to gather at parties. Notably, there is a delay between the spikes in confirmed cases and the estimated reproduction number due to the existence of the incubation period and testing-to-confirmation delay. Further discussion about the impact of different delays on the estimation of the reproduction number can be found in SI-2-D-3. 

\begin{figure}[htbp]
\centering
\includegraphics[width=1\linewidth]{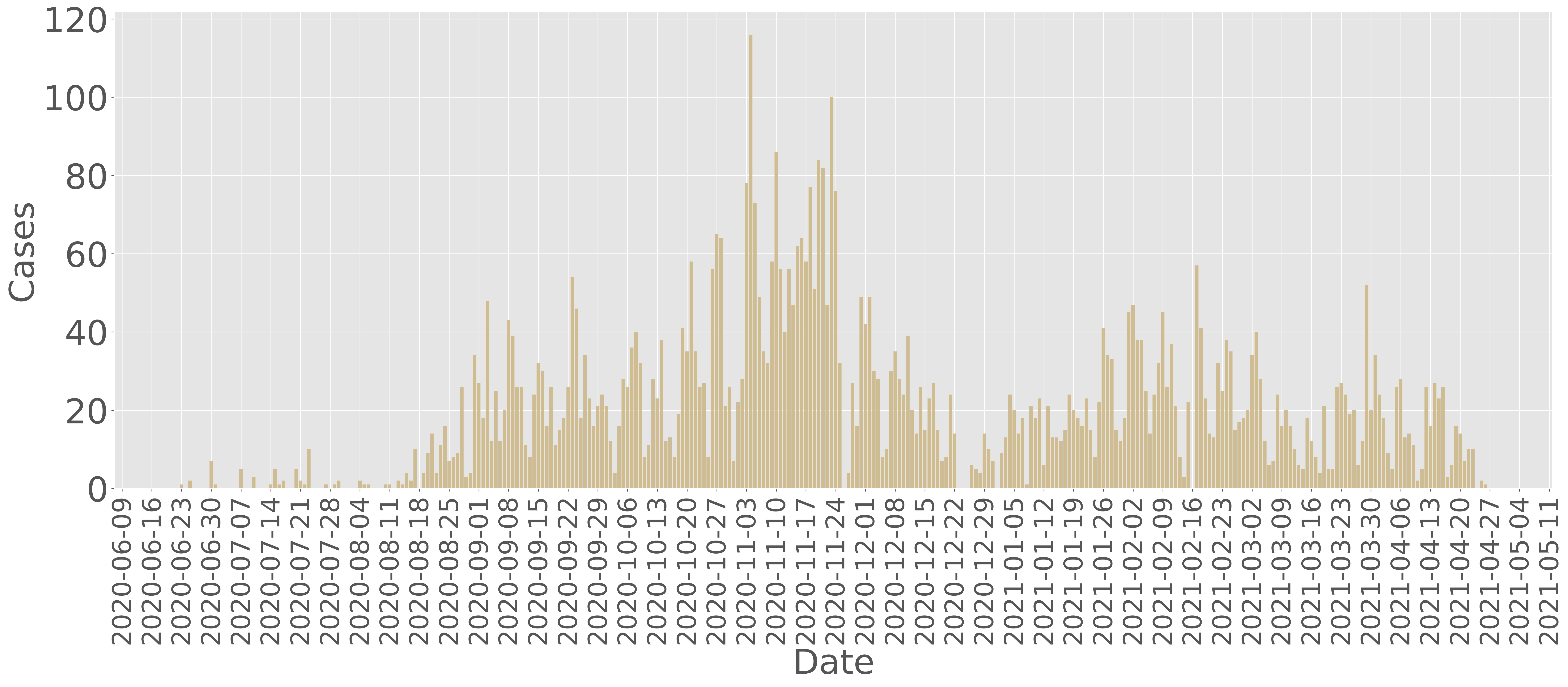}
\caption{Daily confirmed positive cases at Purdue from Fall 2020 to Spring 2021 captured by the IDA+A team at Purdue.}
\label{fig:Purdue_P}
\end{figure} 
\begin{figure}[htbp]
\centering
\includegraphics[width=1\linewidth]{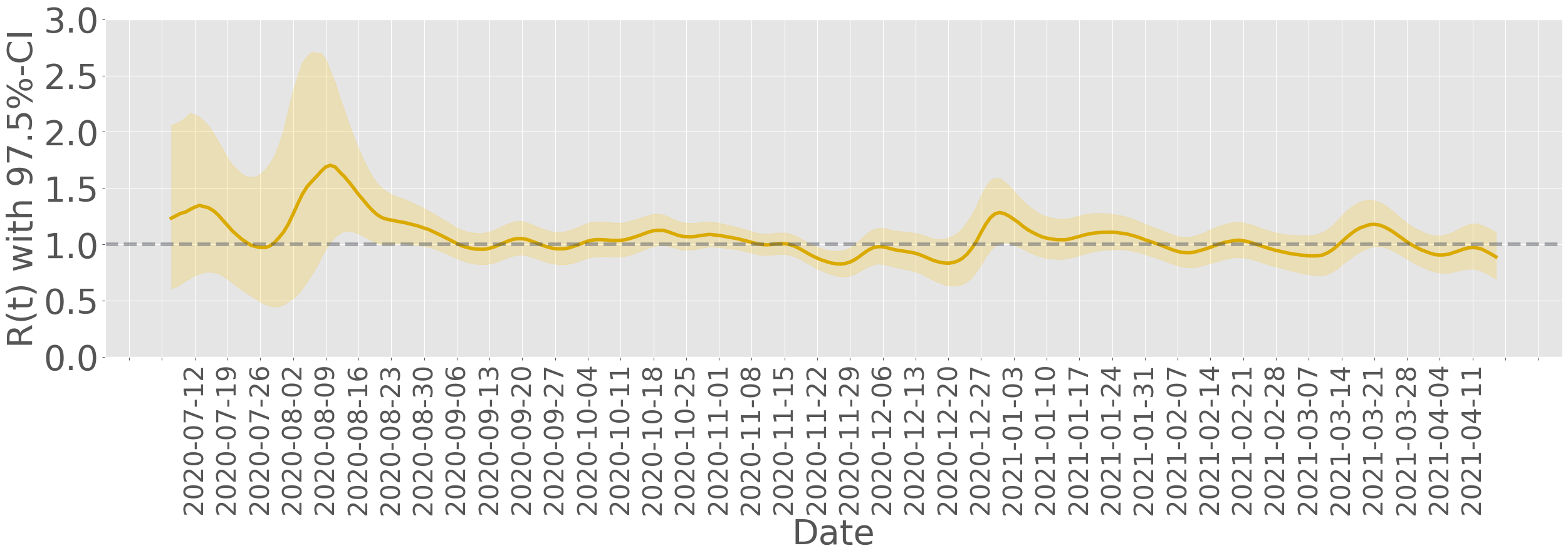}
\caption{Estimated reproduction number at Purdue from Fall 2020 to Spring 2021.}
\label{fig:Purdue_R}
\end{figure} 




Similar to UIUC, we leverage confirmed cases from Fall 2020 to Spring 2021 at Purdue University via the IDA+A team, as shown in Figure~\ref{fig:Purdue_P}. The confirmed cases include the total number of confirmed symptomatic and asymptomatic cases. The daily confirmed cases through voluntary testing and surveillance testing can be found in SI-1-B and SI-1-C. 
Figure~\ref{fig:Purdue_P} shows four main spikes: from August 18th 2020 to September 10th, from October 13th to November 15th, from January 16th to February 1st, and from March 16th to March 30th.  These four spikes correspond to the entry screening of the Fall 2020 semester, the increasing number of gatherings and activities at the start of the football season, the entry screening of the Spring 2021 semester, and the return from Spring break, respectively. We leverage the total confirmed positive cases shown in Figure~\ref{fig:Purdue_P}, to estimate the reproduction number, as illustrated in Figure~\ref{fig:Purdue_R}. 
From the estimation,
we  observe four periods where the reproduction number is higher than 1. The result aligns with the four spikes in the confirmed positive cases in Figure~\ref{fig:Purdue_P}.
Additionally, the two entry screenings can be captured by the two huge spikes of the reproduction number being much higher than 1 in Figure~\ref{fig:Purdue_R}. Note that although UIUC and Purdue implemented different testing-for-isolation strategies, the estimated reproduction number reflects similar spreading trends in terms of entry-screening and in-semester spikes. One main reason for this observation is that both universities have a similar size, culture, \baike{and location.}

In addition to leveraging the estimated reproduction number to analyze the epidemic spread, as illustrated in Figures~\ref{fig:UIUC_R} and~\ref{fig:Purdue_R}, we find that the estimated reproduction number fluctuated around $1$ 
at both universities.
Since we propose a mechanism to measure the impact of the isolation rate on the reproduction number, we can leverage the isolation rate as a control variable to manipulate the reproduction number directly. 
The effective testing-for-isolation strategies implemented by both universities inspire the feedback control strategy that we propose in the framework, where the mitigation goal is to maintain the reproduction number at a desired threshold, i.e., less than or equal to 1. 
Through the estimated reproduction number, we simulate the spreading process and further propose a methodology to evaluate the effectiveness of the implemented strategies by reverse engineering the reproduction, as shown in the Reconstruction framework in~Figure~\ref{fig:framework}.
The impact on the selection of sliding window and step size when estimating the reproduction number can be found in  SI-2-D-4 and~\cite{cori2013new,gostic2020practical}. 
\subsection{Reconstruction of Epidemic Spread at the UIUC and Purdue} 
\label{Sec-C}
One challenge faced by researchers in modeling, analyzing, and predicting epidemic spreading processes is the fact that such processes are irreversible. We cannot experience the exact same epidemic spreading process under the exact same conditions twice. 
Therefore, in order to evaluate the effectiveness of the existing intervention strategy and to create a testing environment to assess the impact of different intervention strategies on the epidemic spread, we introduce a methodology to reconstruct the spreading process~\cite{huisman2022estimation,Epyestim_Python_2020}.

We leverage the estimated reproduction number to reconstruct the spread.
Reconstructing the epidemic spreading data through the estimated reproduction number can be formulated as the inverse process of estimating the reproduction number through the confirmed cases. We first generate the infected cases and then add delays from the incubation period and testing-to-confirmation delay, in order to obtain the simulated confirmed cases. 
We use a Poisson random process to reconstruct the new daily infected cases, where the mean of the Poisson process is determined by the estimated reproduction number on that day~\cite{cori2013new}. \baike{Recall from \eqref{eq:serial_int} that $w(\alpha)$ represents the serial interval  distribution of the spreading under the isolation rate $\alpha$.}
The new generated infected cases $I_t$ at time step $t$ is Poisson distributed with the mean~\cite{cori2013new},
\begin{equation}
   \mathbb{E}(I_t)= \mathcal{R}_t\sum_{s=1}^{t}I_{t-s}w_s(\alpha).
\end{equation}
The probability of \baike{$k$} new infected cases on day \baike{$t$} is given by 
\begin{equation}
\label{eq:Pois}
    \mathbb{P}(I_t=k)= \frac{(\mathcal{R}_t\Lambda_t)^{k}e^{(-\mathcal{R}_t\Lambda_t)}}{k!},
\end{equation}
where $\Lambda_t=\sum_{s=1}^{t}I_{t-s}w_s(\alpha)$, i.e., we have $I_t\sim Pois(\mathcal{R}_t\Lambda_t)$, where $Pois(\lambda)$ denotes the Poisson distribution with mean $\lambda$. \eqref{eq:Pois} indicates that the new generated cases at time step $t$ are determined by the serial interval distribution $w(\alpha)$, the past infected cases $I_t$, and the reproduction number $\mathcal{R}_t$. 
The mechanism from~\eqref{eq:Pois} generates infection-to-infection data, where $I_t$ captures the infected cases on day $t$. 
However, 
recall that the data we collected from UIUC (\ref{fig:UIUC_P}) and Purdue (\ref{fig:Purdue_P})
consist of daily confirmed cases, which are the infected cases with the incubation period and testing-to-confirmation delays, rather than the directly-measured infected cases~\cite{cori2013new,Cori2022, gostic2020practical,huisman2022estimation,Epyestim_Python_2020}.
Therefore, to simulate confirmed cases that align with the confirmed data, we first generate daily infected cases using~\eqref{eq:Pois}, and then incorporate the incubation period and testing-to-confirmation delays to the simulated infected cases. We utilize the same incubation period distribution and the testing-to-confirmation distribution as the distributions that we leverage for reproduction number estimation~\cite{brauner2020effectiveness,huisman2022estimation,tariq2020real,brauner2021inferring,Epyestim_Python_2020}.  More detailed discussions on how to reconstruct the spreading process using the given reproduction number, the serial interval distribution, and delay distributions can be found in SI-2-B.



\begin{figure}[htbp]
\centering
\includegraphics[width=1\linewidth]{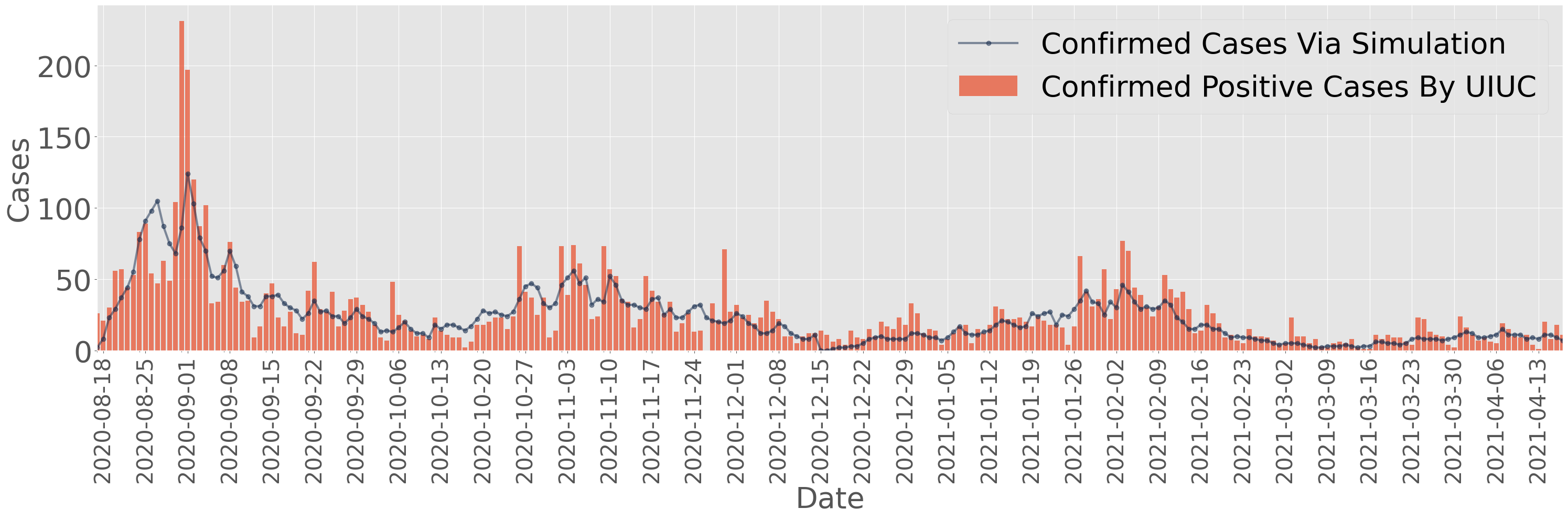}
\caption{Daily confirmed positive cases at the UIUC during Fall 2020 and Spring 2021.}
\label{fig:UIUC_Rec}
\end{figure}
\begin{figure}[htbp]
\centering
\includegraphics[width=1\linewidth]{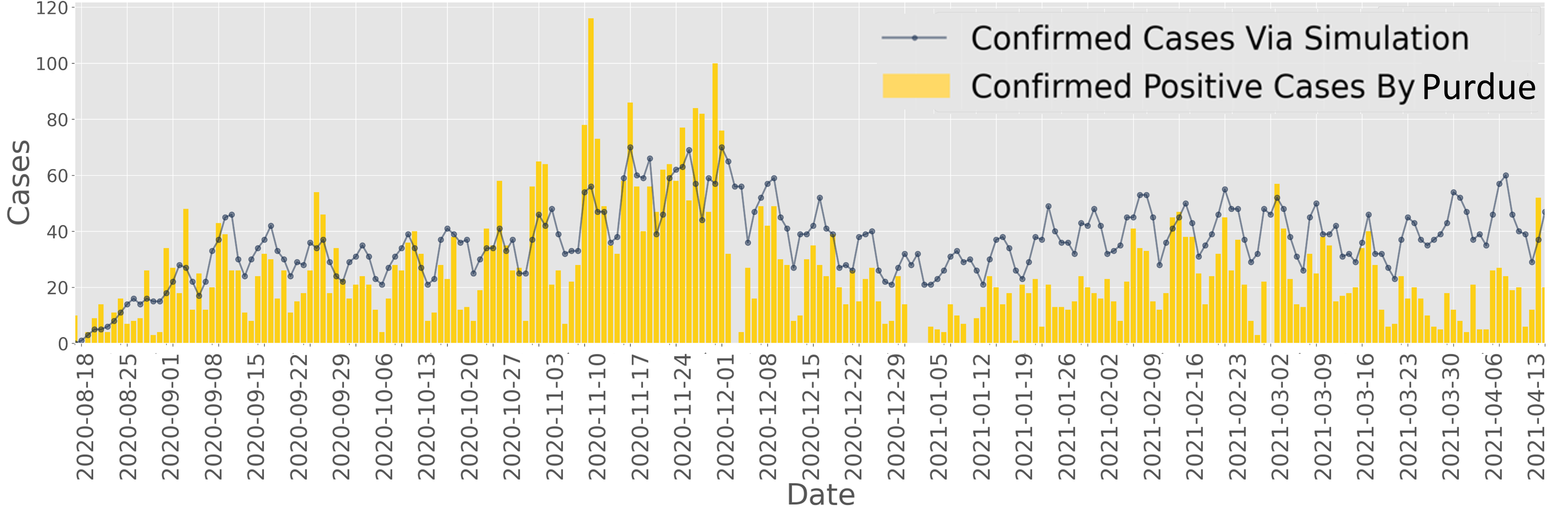}
\caption{Daily confirmed positive cases at Purdue during Fall 2020 and Spring 2021.}
\label{fig:Purdue_Rec}
\end{figure}

We obtain the reconstructed spreading processes on both UIUC and Purdue campuses in the form of daily confirmed cases, as shown in Figures~\ref{fig:UIUC_Rec} and~\ref{fig:Purdue_Rec}, respectively. The reconstructed spreading processes can successfully capture the epidemic spreading trend over both campuses, especially the spikes (see SI-2-E-1). 
We use the reconstructed spreading methodology to construct a testing environment to evaluate the effectiveness of the implemented testing rates and to test different potential testing strategies.
\subsection{Evaluation of The Testing-for-Isolation Strategies by Purdue and UIUC}
We perform counterfactual analysis to study what would have happened 
if UIUC and Purdue had not implemented their testing-for-isolation strategies. \baike{We analyze Fall 2020 semester for both campuses.}
We propose a novel method to reverse engineer the reproduction number through
methodologies proposed in  
the previous two sections that quantify the impact of the isolation rate on the reproduction number. 
\baike{Recall that although Purdue  tested approximately 10\% of the population to catch asymptomatic infections, 
it was possible that the isolation rate for the asymptomatic infected population 
was higher than 10\%. Hence, we leverage the same conditions used in estimating the reproduction number and reconstructing the spread at UIUC and Purdue to investigate the possible spreading scenarios without any implemented testing-for-isolation strategies.}

Consider that the only difference between the spread without any testing-for-isolation strategies and the historical spreading process is the isolation rate.
In order to generate a spreading process without any isolation strategy, we need to reverse engineer the reproduction number affected under an isolation rate back to the reproduction number without the isolation rate, as indicated by~\eqref{eq:S_F}. 
However, with a finite number of population on university campuses,
the reproduction number is also affected by the existing susceptible population. Specifically, we consider cases where the population on both campuses is fixed during the semester. Therefore, when reverse engineering the reproduction number at any given moment, we should not only consider the isolation rate at that moment but also take into account the impact of the susceptible population.

We define the reconstructed reproduction number under reverse engineering, without any isolation strategy, at any time step $t$ ($t\geq 0$) as $\hat{\mathcal{R}}_t$,
\begin{equation}
\label{eq:R_F}
   \hat{\mathcal{R}}_t = \frac{\mathcal{R}_t \hat{S}(t)}{\mathcal{F} S(t)}.
\end{equation}
We use $\hat{S}(t)$ to denote 
the total susceptible population of the reconstructed environment without any isolation strategies. Meanwhile, we use ${S}(t)$ to denote the total historical susceptible population at any given time step $t$. 
We use $\mathcal{R}_t$ to represent the estimated reproduction number from the confirmed cases at $t$, and $\mathcal{F}$ is defined in~\eqref{eq:S_F}.
\eqref{eq:R_F} indicates that it is critical to consider the two factors to reverse engineer the reproduction number: 
1) The scaling factor $\mathcal{F}$ and 2) the ratio of the susceptible populations $\frac{\hat{S}(t)}{{S}(t)}$. The scaling factor $\mathcal{F}$ will be lower if we have a higher isolation rate, and vice versa. Hence, it is natural to think that without the higher implemented isolation rate, the outbreak could be worse. 
Meanwhile, a higher ratio between the susceptible population $\frac{\hat{S}(t)}{{S}(t)}$ will result in higher scaling of  $\hat{\mathcal{R}}_t$. 


We use the reverse engineering method on the reproduction number to first study the spread on the UIUC campus.
When studying the spread on the UIUC campus without any implemented isolation strategies, we consider the worst-case scenario. In this worst-case scenario, the implemented isolation rate is treated as the testing rate, implying that UIUC has successfully isolated all confirmed cases. Consequently, without isolation, everyone caught by UIUC's testing strategy would not have been isolated. 
Further, without the testing-for-isolation strategy, both symptomatic and asymptomatic cases will behave normally and will not isolate themselves from the population. This worst-case scenario creates a situation where every individual on campus does not take actions against the pandemic. 
The detailed process on reconstructing the spreading process on the UIUC campus without their testing-for-isolation strategy can be found in SI-2-E-2.

\begin{figure}[htbp]
\centering
\includegraphics[width=1\linewidth]{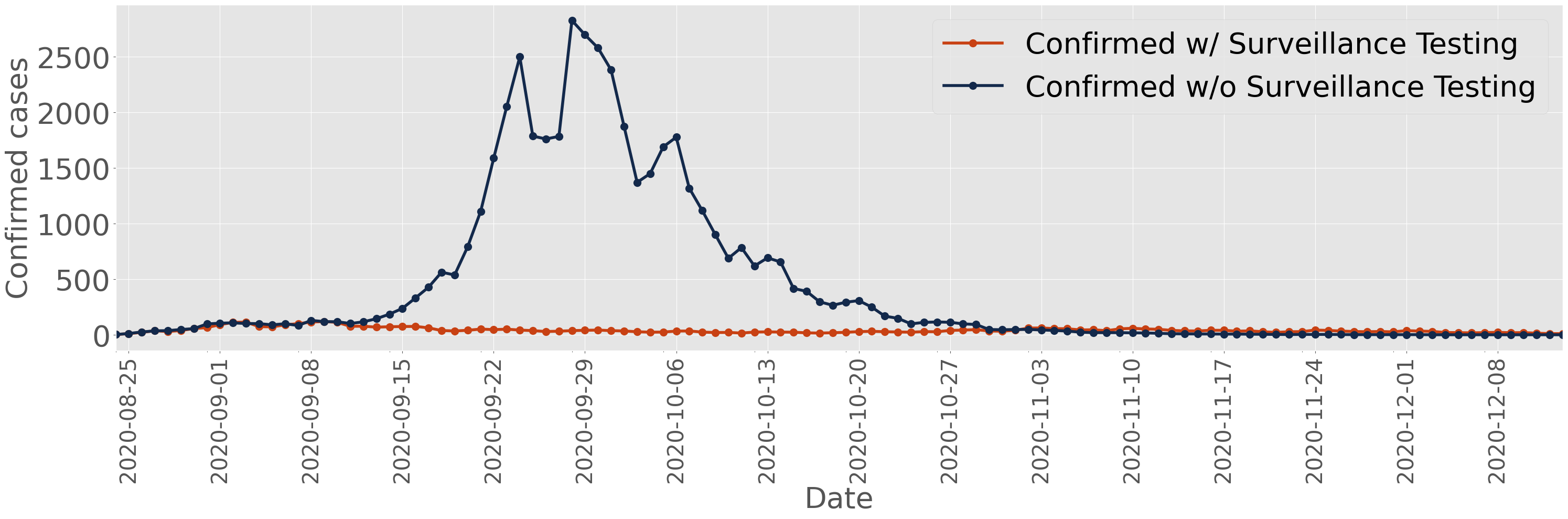}
\caption{Simulated Daily confirmed positive cases at UIUC during Fall 2020.}
\label{fig:UIUC_w_o_1}
\end{figure}

Note that although without any isolation, we can still record the confirmed cases through the same testing strategy UIUC implemented.
Figure~\ref{fig:UIUC_w_o_1} shows that, without any isolation and further with everyone takes no actions against the virus,
there would have been a significant outbreak on the UIUC campus during Fall 2020. Around 90\% of the total population on campus will be infected during the Fall 2020 semester.
Starting from September 2020,
the confirmed cases start to grow slowly, since the strict entry-screen caught most of the infected cases. However, due to the return of the Big Ten football season around the beginning of October, the violation of the other implemented intervention policies further increase the transmission rates, consequently elevating the reproduction number. Hence, the confirmed cases continue to increase and eventually reach their peak around the end of October.
Later on, the confirmed cases start to decrease. The decrease in confirmed cases is caused by the fixed population size on campus during Fall 2020, resulting in a reduced $\hat{S}(t)$. Therefore, the campus reaches the herd immunity threshold~\cite{randolph2020herd,cobey2020modeling}, where the epidemic begins to fade away after a certain portion of the population becomes infected and gains immunity against the virus. 
Note that the peak value is enormous because we consider no isolation and intervention for the infected cases, which is essentially the worst-case scenario. This scenario can be demonstrated by a similar large infection peak in China during Spring 2023 when most COVID-19 interventions were suddenly lifted, allowing the virus to spread freely~\cite{normile2023china}. Further reconstructed situations are explored by considering different isolation rates on the UIUC campus in SI-2-E-3.

\begin{figure}[htbp]
\centering
\includegraphics[width=1\linewidth]{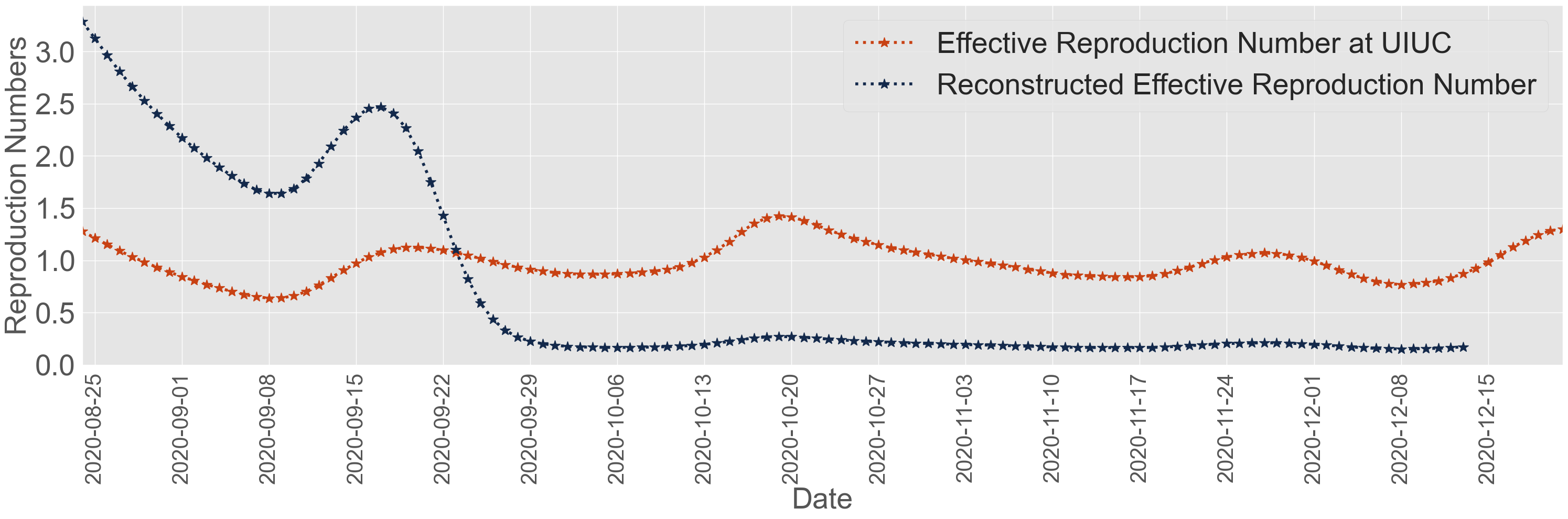}
\caption{Reverse engineering the reproduction number at UIUC.}
\label{fig:UIUC_Rt_Comp}
\end{figure}

After discussing what would have happened without the implemented testing-for-isolation strategies, we will compare  the reconstructed reproduction number under 
reverse engineering
with the historical reproduction number to validate the observation.
Figure~\ref{fig:UIUC_Rt_Comp}
shows that starting from the beginning of the Fall 2020 semester until late September 2020, the reconstructed reproduction number $\hat{\mathcal{R}}_t$ is higher than the estimated reproduction number $\mathcal{R}_t$, reflected by the exponential growth before the end of September.
This phenomenon is primarily determined by the scaling factor $\mathcal{F}$ in \eqref{eq:R_F}.
As explained in \eqref{eq:R_F},
another determining factor for reverse engineering the reproduction number is the ratio between the susceptible populations. 
Figure~\ref{fig:UIUC_w_o_1} illustrates that after a large amount of the population on campus is infected, the infected population starts to decrease. Therefore, from late September 2020, the reconstructed reproduction number is lower than the estimated reproduction number.


Using the same methodology, we reconstruct a possible scenario for Purdue campus without its implemented testing strategies. For Purdue University, we consider a different population behavior compared to the situation at the UIUC.  
As discussed regarding the impact of Purdue’s testing strategy on the infection profile, all confirmed symptomatic cases will self-report and isolate themselves from the population when they test positive, as they are cautious and willing to be tested. 
Based on the testing data from Purdue, we have $\theta$ = 55\%. 
In contrast to UIUC, we focus on the impact of the surveillance test strategy on asymptomatic cases at the Purdue campus. 
Recall that we consider a 30\% isolation rate when simulating the spreading process over the Purdue campus.
\baike{For further discussion on the impact of choosing the isolation rate and the symptomatic ratio, refer to SI-2-E-3.}

\begin{figure}[htbp]
\centering
\includegraphics[width=1\linewidth]{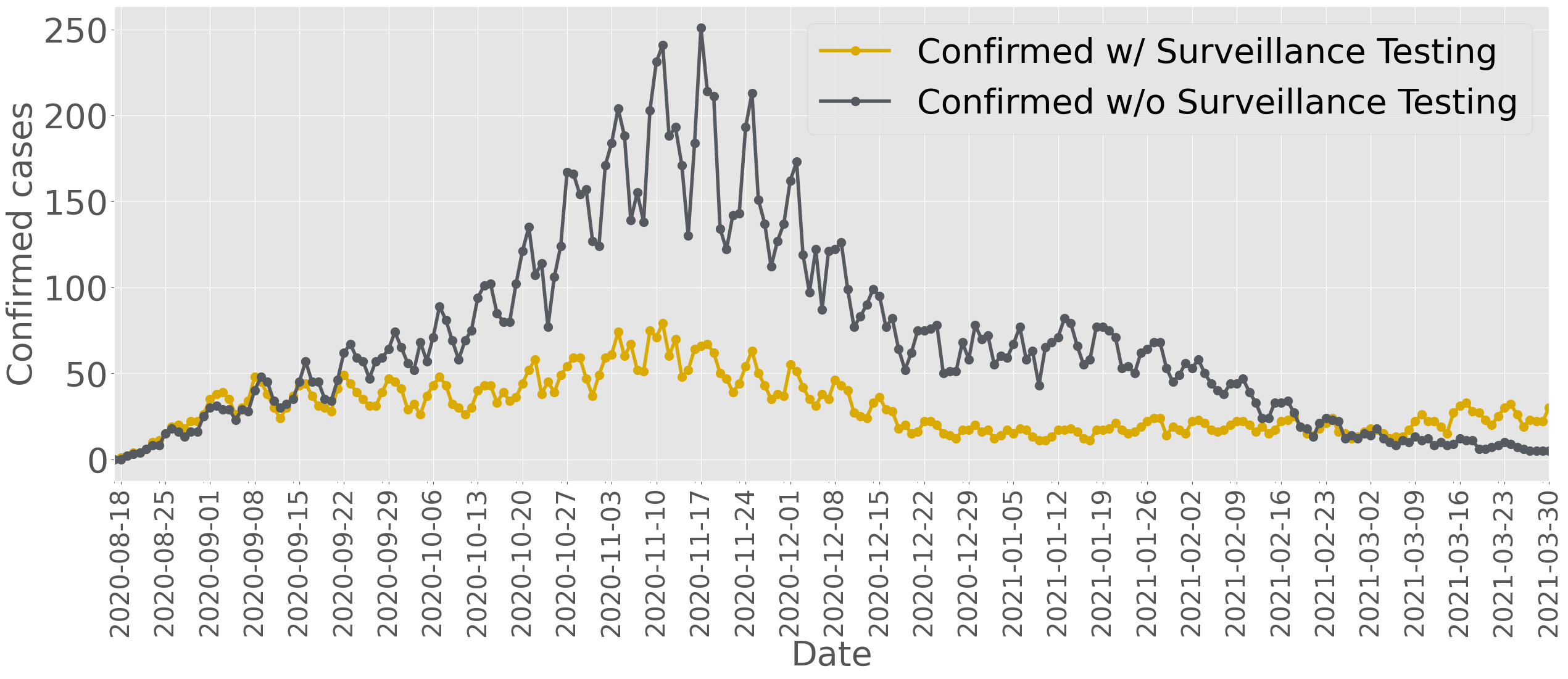}
\caption{Reconstructed daily infected cases at Purdue.}
\label{fig:Purdue_0.3}
\end{figure}

\begin{figure}[tbhp]
\centering
\includegraphics[width=1\linewidth]{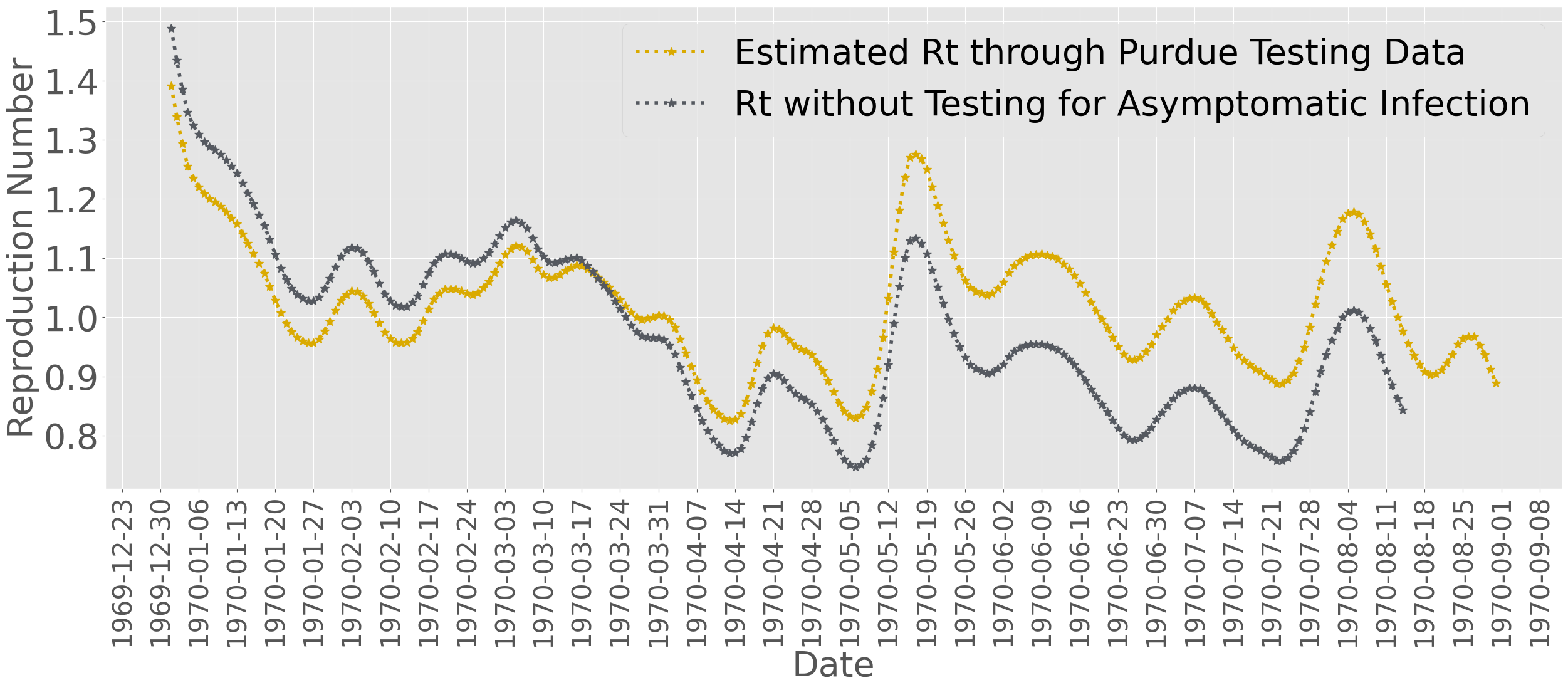}
\caption{Reverse engineering the reproduction number at Purdue.}
\label{fig:Purdue_0.3_Rt}
\end{figure}

We reconstruct the spreading over the Purdue campus during Fall 2020, as illustrated in Figure~\ref{fig:Purdue_0.3}. 
Figure~\ref{fig:Purdue_0.3} shows that without the testing-for-isolation strategy that isolates the $30\%$ asymptomatic infected population, and under the condition that all symptomatic cases will self-report and isolate themselves from the population, there would be a larger outbreak. In particular,
the confirmed cases would start to surpass the historical confirmed cases beginning in October. Furthermore, due to the existence of $\theta = 55\%$ symptomatic population being isolated, the reconstructed spreading process regarding the outbreak caused by the return of the BIG Ten football season is much milder than that of UIUC. Additionally, Figure~\ref{fig:Purdue_0.3_Rt}indicates that the reconstructed reproduction number is slightly higher than the historical reproduction number from the beginning of the Fall 2020 semester until the end of October. We can explain this phenomenon by the absence of the testing-for-isolation strategy for asymptomatic cases.

We perform counterfactual analysis that centers around reverse engineering the reproduction number to assess the testing-for-isolation strategies implemented by UIUC and Purdue. The evaluation shows that the testing-for-isolation is crucial for epidemic mitigation. Without testing-for-isolation, there would have been a huge outbreak, as illustrated by the analysis of the spread over the UIUC campus. Even under the ideal situation where all symptomatic cases are tested voluntarily and isolate themselves, there still would have been an outbreak due to the existence of asymptomatic cases, as illustrated by the evaluation process on the Purdue campus. We further discuss additional evaluations of the implemented testing-for-isolation strategies over the UIUC and Purdue campuses under different scenarios in SI-2-E-2 and SI-2-E-3.

\subsection{Open-Loop Epidemic Control}
\label{Sec: Control}
We have evaluated the spreading processes over both the UIUC and Purdue campuses through reverse engineering the reproduction number, which completes the Epidemic Reconstruction and Strategy Evaluation part in Figure~\ref{fig:framework}.
Now, to execute the Feedback Control step illustrated in Figure~\ref{fig:framework}, we leverage the previously reconstructed spreading environment of both campuses to adjust various isolation strategies for epidemic mitigation. Our approach involves counterfactual analysis on implementing different fixed isolation rates for UIUC or Purdue. The foundation of this analysis is also based on reverse engineering the reproduction number through the intensity of the intervention strategy, specifically the isolation rate.

First, we implement different isolation rates for the reconstructed worst-case scenario over the UIUC campus, as shown in Figure~\ref{fig:UIUC_w_o_1}. In the worst-case scenario, if an infected case is not isolated after testing, the individual will behave as uninfected until recovery. We compare the outcomes if we had implemented different fixed isolation rates that are less than $200\%$ weekly on the UIUC campus during Fall 2020. 
Under the condition that the weekly testing rate is $200\%$, 
the fixed isolation rates are drawn from $\{0\%, 10\%, 20\%, \dots, 90\%, 100\%, 120\%,\dots 180\%, 200\%\}$ in order. Meanwhile, the testing and isolation process does not distinguish between symptomatic and asymptomatic cases. 

\begin{figure} [h]
\centering
\includegraphics[width=1\linewidth]{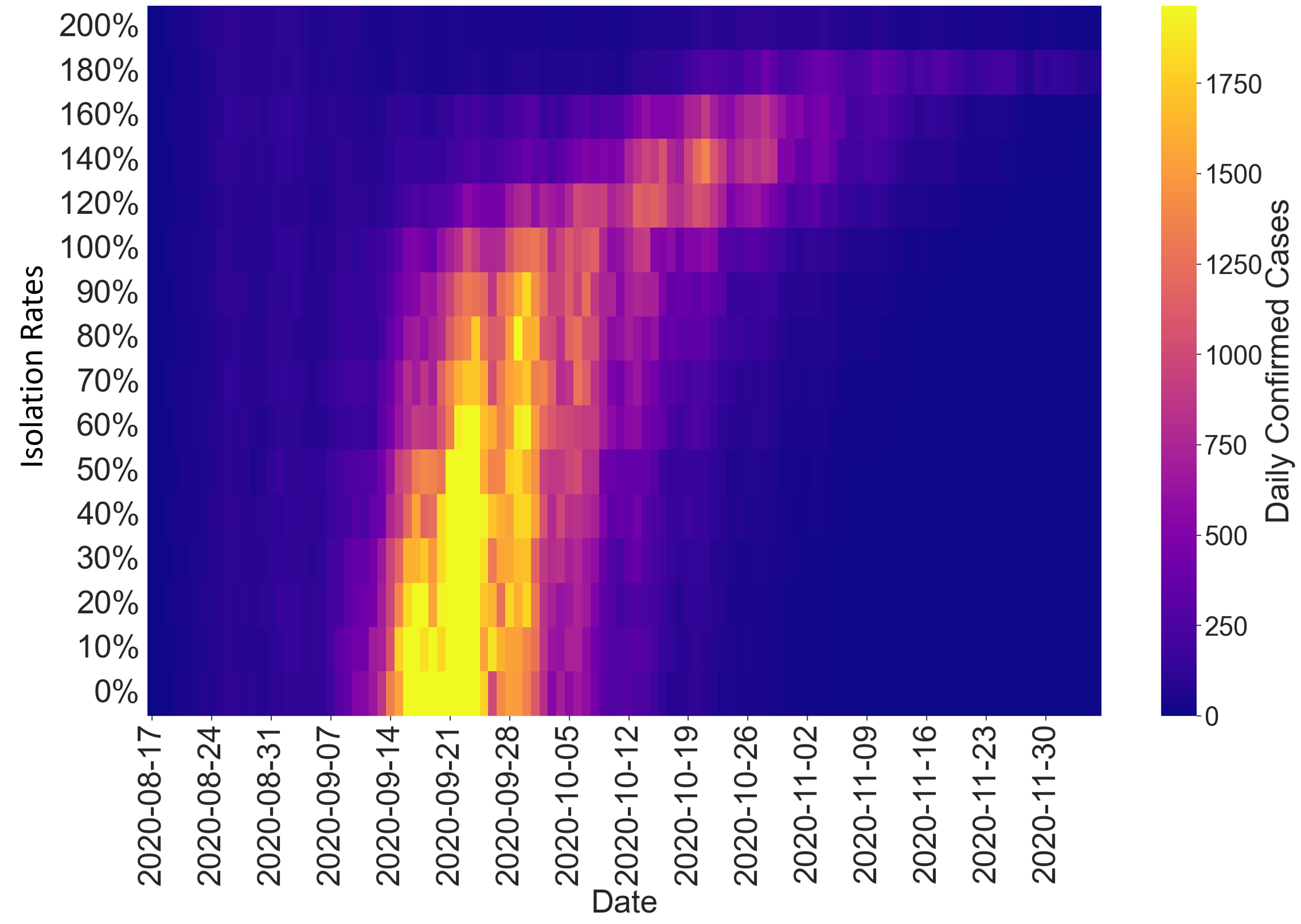}
\caption{Daily confirmed cases over the UIUC campus with different isolation rates.}
\label{fig:UIUC_Control_Fixed_Heat}
\end{figure}

We capture the daily confirmed cases using the heatmap in Figure~\ref{fig:UIUC_Control_Fixed_Heat}, 
which indicates the higher peak infection levels with brighter colors. 
%
Figure~\ref{fig:UIUC_Control_Fixed_Heat} implies that higher isolation rates will result in relatively smoother and flatter curves in terms of confirmed cases. The shape of the brighter area in Figure~\ref{fig:UIUC_Control_Fixed_Heat} also indicates that a higher isolation rate will lead to lower spikes, while these lower spikes will also be further delayed.
Note that all the analyses are based on the worst-case scenario situation proposed in the previous section. Testing these fixed rates in a different reconstructed testing environment at UIUC would yield significantly different results.  Therefore, when evaluating intervention strategies such as the fixed isolation rate, it is critical to consider the conditions about population and virus spreading behavior. We also present the impact of the isolation rate on the cumulative confirmed cases in SI-2-F. 

\begin{figure} [h]
\centering
\includegraphics[width=1\linewidth]{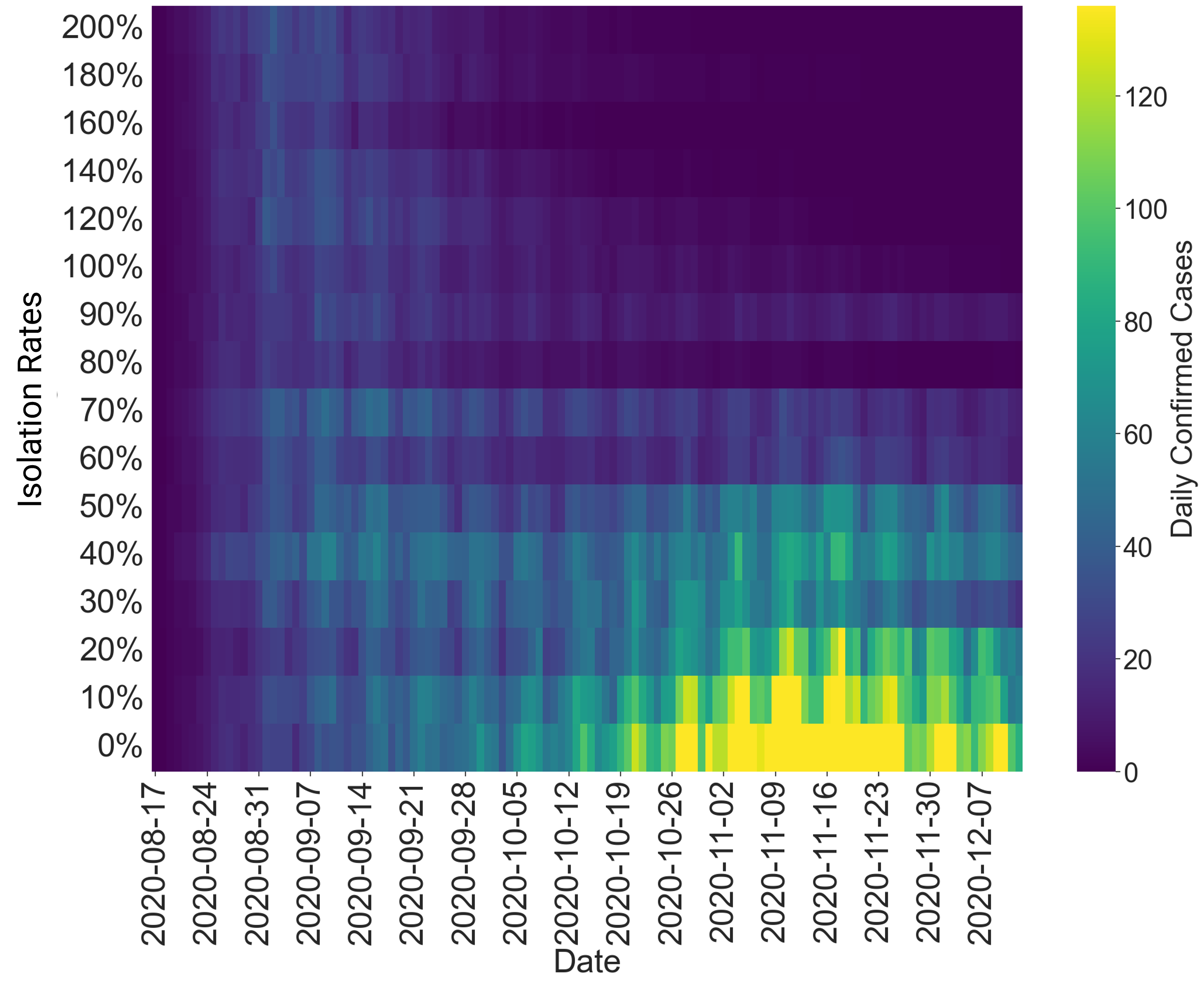}
\caption{Daily confirmed cases over the Purdue campus with different isolation rates.}
\label{fig:Purdue_Control_Daily_Heat}
\end{figure}

Compared to UIUC, we study the change of the isolation rate under surveillance testing on the Purdue campus.
We use the reconstructed environment at Purdue during Fall 2020, as shown in Figure~\ref{fig:Purdue_0.3},
where positive cases confirmed by voluntary testing would isolate themselves from others. 
Further, based on the data from Purdue, $\theta = 55\%$ of the population was symptomatic. 
We vary the isolation rate from the surveillance testing for asymptomatic cases from $\{0\%, 10\%, 20\%, \dots, 90\%, 100\%, 120\%,\dots 180\%, 200\%\}$. We capture  daily confirmed cases using the heatmap in Figure~\ref{fig:Purdue_Control_Daily_Heat}.

As expected, the daily confirmed cases at Purdue were much lower than the daily confirmed cases on the UIUC campus, given the different population behavior.
Higher isolation rates generate relatively smoother and flatter curves in terms of confirmed cases. 
Additionally, Figure~\ref{fig:Purdue_Control_Daily_Heat} indicates that higher isolation rates result in lower spikes during the return of the BIG Ten football season.  Figure~\ref{fig:Purdue_Control_Daily_Heat} also shows that 
there is a noticeable difference between isolation rates for asymptomatic cases below $50\%$ per week and 
isolation rates higher than $50\%$ of the asymptomatic cases per week. Based on the  analysis in this specific example, when the isolation rate for asymptomatic cases is higher than $50\%$ per week, there would not be any significant outbreak during the Fall 2020 semester at the Purdue campus. \baike{The same as the UIUC study, the simulation results are based on certain conditions. Changing simulation conditions will generate different conclusions.}
Further detailed analyses and confirmed cases for Purdue during Fall 2020 under different isolation rates can be found in SI-2-F.

To validate the efficacy of leveraging the framework through reverse engineering of the reproduction number to design an open-loop mitigation strategy, such as a fixed isolation rate, in this section, we explore hypothetical scenarios for UIUC and Purdue campuses by introducing different fixed isolation rates. Specifically, we identify how the isolation rate influences the peak infection value and time. Moreover, we investigate certain threshold conditions associated with the isolation rates that aid in avoiding potential outbreaks. Note that conditions set in reconstructing the testing environment will influence simulation results. While we cannot alter the historical spreading process with the implemented testing-for-isolation strategies, our counterfactual analysis, based on comprehensive spreading information, can yield valuable conclusions on setting isolation rates and threshold conditions. These insights are vital for real-time policy-making in the context of epidemic mitigation. Following the validation of open-loop control strategies, the next section introduces a data-driven feedback control mechanism in our framework, adjusting the control input (i.e., isolation rates) based on the epidemic's severity.

\subsection{Data-Driven Feedback Epidemic Control}
After reconstructing the spread across the UIUC and Purdue campuses and evaluating the impact of different isolation rates, we will proceed to design and validate our data-driven feedback control framework.
If all the conditions from~\eqref{eq:R_total} were perfectly known, it would be possible to generate an isolation rate that maintains the reproduction number exactly at the desired value $\mathcal{R}^*$
with a single computation. However, the complex nature of the spread introduces uncertainty, making it difficult to design such a rate with perfect settings. To address this challenge, 
instead of relying on a single computed isolation rate from~\eqref{eq:R_total}, we propose a data-driven feedback control mechanism to adjust the isolation rate according to the severity of the spread. Note that it is natural to consider that the isolation rate is proportional to the testing rate. For the sake of simplicity, we do not differentiate between the testing rate and the isolation rate in this subsection.

The feedback control strategy is straightforward: to save the total number of tests conducted while maintaining the reproduction number at a certain level $\mathcal{R}^*\in(0,1]$, we increase testing to isolate more infected individuals when the outbreak is severe and decrease testing to isolate fewer infected individuals when the spread is less severe. We leverage the estimated reproduction number to indicate the severity of the outbreak. Utilizing the testing environment previously described, the feedback control framework that incorporates the reproduction number completes the loop in the data-driven framework illustrated in Figure~\ref{fig:framework}.
We further discuss in SI-2-G how the goal of maintaining the daily infected population at an acceptable level aligns with the optimal mitigation strategy~\cite{acemoglu2020optimal,she2022optimal}. 

We then introduce the necessary information for implementing the data-driven feedback control mechanism. 
For an epidemic spread, with any implemented testing-for-isolation strategies, we can obtain spreading data through testing. Illustrated in Figure~\ref{fig:framework}, under the condition that we know 1) the infection profiles ($v$) of the virus through the public health or the contact tracing data, 2) testing rates (isolation rates) ($\underline{\alpha}$ and  $\overline{\alpha}$)  implemented by the authorities, and 3) the ratio of symptomatic ($\theta$) and asymptomatic cases ($1- \theta$) from the testing data, 
we can estimate the reproduction number $\mathcal{R}_t$
at any given time window. 
If the estimated reproduction number is not equal to the expected value $\mathcal{R}^*$,
we need to update the testing rate.
We integrate the methodologies developed in \eqref{eq:S_F}, \eqref{eq:serial_int}, and \eqref{eq:R_F} 
to propose a novel mechanism to adjust the testing rate based on the reproduction number.

We introduce and then validate the mechanism by considering the spread over the UIUC campus, without distinguishing between symptomatic and asymptomatic infections.
Although $\mathcal{R}_t$ is continuous, in reality, we can estimate only a finite number of the reproduction number. 
Hence, we use $\mathcal{R}_k(\alpha_k)$ to represent the
estimated reproduction number 
under the testing rate $\alpha_k$ at the $k^{th}$ step, $k\in \{1,2,\dots\}$. Based on the estimated reproduction number at step $k$, we propose the following mechanism to update the testing rate at the $(k+1)^{th}$ step. We first compute the modified infection profile through the estimated reproduction number $\mathcal{R}_k (\alpha_k)$ under the testing rate $\alpha_k$,  
$v(\alpha_k) = \sum_{i=1}^n w_i(\alpha_k)\mathcal{R}_k(\alpha_k),$
where $w_i(\alpha_k)$ is defined in~\eqref{eq:serial_int}.
If $\mathcal{R}_k(\alpha_k)\neq \mathcal{R}_k^*$, we will update the testing rate $\alpha_{k+1}$ for the $(k+1)^{th}$ step. Specifically, we establish the following methodology to update the testing rate \baike{in order to control the reproduction number at $\mathcal{R}_k^*\in(0,1]$}:
\begin{equation}
\label{Eq:Testing_Rate}
\sum_{i=1}^n v_i(\alpha_k)(1-\alpha_{k+1})^i =\mathcal{R}_k(\alpha_k)\sum_{i=1}^n w_i(\alpha_k)(1-\alpha_{k+1})^i =\mathcal{R}^*.
\end{equation}
In \eqref{Eq:Testing_Rate}, $\mathcal{R}_k(\alpha_k)$ is the estimated reproduction number at step $k^{th}$, under the implemented testing rate $\alpha_k$. Recall that $w(\alpha_k)$ 
is the modified serial interval distribution under the testing rate $\alpha_k$. 
The only unknown in~\eqref{Eq:Testing_Rate} is the testing rate to be updated, $\alpha_{k+1}$. 
Therefore, by solving~\eqref{Eq:Testing_Rate}, we compute the updated testing rate at the next time step directly with the feedback information from the estimated effective reproduction number $\mathcal{R}_k(\alpha_k)$ and the modified serial interval distribution $w(\alpha_k)$ under the previous implemented testing rate $\alpha_k$, \baike{making it a data-driven control policy.} More details can be found in SI-2-G.

Consider implementing the data-driven feedback control framework in the reconstructed testing environment in
Figure~\ref{fig:UIUC_w_o_1}, 
where the feedback control strategy will adjust the testing rate every two weeks. 
We estimate the reproduction number over the past two weeks and update the testing rate for the following two weeks. This mechanism indicates that we use the estimated reproduction number of the past 14-day period as the indicator of the reproduction number for the subsequent two weeks. This mechanism considers that the reproduction number will remain the same if the implemented testing rate does not change. 
Therefore, there are no prediction mechanisms when updating the future testing rate. 

\begin{figure}[htbp]
\centering
\includegraphics[width=1\linewidth]{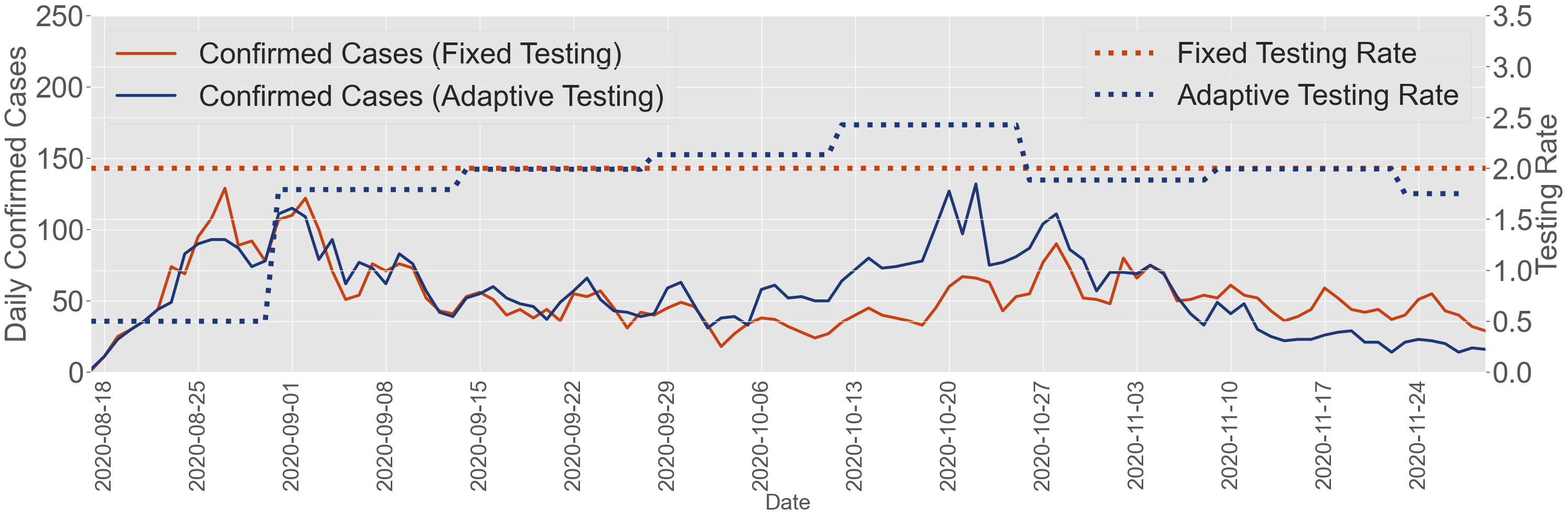}
\caption{Data-driven feedback control for the spread at UIUC.}
\label{fig:UIUC_Control}
\end{figure}

We implement the proposed feedback control framework in the testing environment and compare it to the testing strategy implemented by UIUC, which involved testing the entire campus twice a week.  Our goal is to control the target reproduction number at $\mathcal{R}^*=0.95$. The target reproduction number, slightly smaller than 1, ensures that the epidemic can gradually fade away with sufficient testing resource.
The feedback control framework, as depicted in Figure~\ref{fig:UIUC_Control}, demonstrates that it can achieve a similar number of daily and total confirmed cases (both around 6300) compared to the testing policy implemented by UIUC.
Furthermore, we find that the implemented testing strategy by UIUC required testing/isolating everyone 16 times in total, while our proposed feedback control strategy only requires testing/isolating everyone 14 times in total. 
Additionally, for most of the Fall 2020 period, the feedback control framework implemented a lower testing rate compared to UIUC's implemented 200\% testing rate. However, during October, the feedback control framework employs higher testing rates to mitigate the potential outbreak associated with the return of the BIG Ten football season. This adjustment is based on the real-world confirmed data and the estimated reproduction number during Fall 2020, as depicted in Figures~\ref{fig:UIUC_P} and~\ref{fig:UIUC_R}.


Overall, the feedback control strategy demonstrates the core idea of reverse engineering the reproduction number for data-driven feedback pandemic mitigation shown in Figure~\ref{fig:framework}. It adapts the testing/isolation rate based on the risk of outbreaks, implementing fewer tests when the risk is low and increasing the rate when there is a potential spike. The example highlights the effectiveness and flexibility of the feedback control mechanism in pandemic mitigation. 
We generalize the methodology in~\eqref{Eq:Testing_Rate} for the epidemic spread, considering both symptomatic and asymptomatic cases in SI-2-G, to study the feedback control mechanism over the Purdue campus.

\section{Discussions}
\subsection{Data-Driven Counterfactual
Analysis, Strategy Evaluation, and Feedback 
Control} We propose a framework for data-driven counterfactual analysis, strategy evaluation, and feedback control of epidemics, incorporating the reverse engineering of the reproduction number. Without assuming a specific model of the spreading process, our approach utilizes statistical information, including the infection profile of the spread, the infection-to-confirmation delay, and the testing/isolation rate, to guide the evaluation and control of spreading behavior. We validate the proposed framework by leveraging testing-for-isolation data from UIUC and Purdue. Through analysis, we evaluate the implemented strategies at both universities during the early stage of the COVID-19 pandemic.

We design a data-driven control mechanism that relies solely on infection data to generate intervention strategies. To validate the effectiveness of our framework, we compare it with the intervention strategies implemented by UIUC and Purdue. Our data-driven feedback control framework effectively manages the spreading process and can adapt to changing conditions. This feedback control mechanism not only serves as a foundation for designing future data-driven frameworks to allocate resources for epidemic mitigation but also lays the groundwork for incorporating control analysis in societal-scale challenges in the absence of complex models.
\subsection{Interventions Beyond Testing-For-Isolation} 
We leverage COVID-19 data from universities to introduce and demonstrate our proposed framework, as shown in Figure~\ref{fig:framework}. However, our framework is adaptable to other scenarios involving different intervention strategies. 
The concept of using the reproduction number as feedback necessitates an examination of the relationship between the intervention strategy and the reproduction number, grounded in the infection profile.
For example, we consider a different intervention strategy, such as varying vaccination percentages. To implement the entire framework outlined in Figure~\ref{fig:framework}, it is crucial to quantify the impact of vaccination on altering the reproduction number. This connection is essential to effectively reverse engineer the reproduction number considering various vaccination rates. Hence, by exploring different intervention strategies' effects on the reproduction number, we can generalize the framework.
\subsection{Future Works}
We acknowledge several limitations of the proposed framework and provide potential avenues for improvement through future work (For more details, see SI-4). 
First, When estimating the reproduction number, we use existing infection profiles from the literature. However, it is essential to update these profiles with contact tracing data from testing-for-isolation strategies to account for varying spreading behaviors. Then, we utilize the past reproduction number to project future values in our control design, without considering any predictions. Since the reproduction number can fluctuate due to various factors, it is crucial to incorporate predictive control mechanisms and machine learning techniques to improve the framework. 
Additionally, as for the mitigation goal, 
with a substantial initial infected population, 
maintaining the reproduction number ($\mathcal{R}^*$) slightly below 1 may still lead to a large number of infections.
Hence, adjusting the control goal of the framework becomes highly significant~\cite{vegvari2022commentary, parag2022epidemic}. 

Furthermore, while the feedback control framework can potentially save mitigation resources, frequent policy changes may be impractical.  Meanwhile, the policy generated by the feedback control design could exceed the resource capacity during a certain period. Thus, exploring constrained optimization on the mechanism  is necessary. Last,
we propose the framework using aggregated data. However, the framework can be improved with spatial and heterogeneous spreading data, where we can estimate the reproduction number for sub-regions and adjust mitigation strategies accordingly. Future work can explore a high-resolution distributed data-driven strategy evaluation and feedback control framework, by leveraging machine learning techniques like graph learning and causal inference to infer connections between sub-regions.

Nonetheless, we aim for this work to provoke discussions about the role and limitations of data-driven counterfactual analysis in pandemic mitigation, considering both analytical and computational perspectives. We firmly believe that harnessing statistical information and inference in the domain of epidemic spreading processes can inspire and significantly benefit the development of rigorous data-driven strategy evaluation and feedback control across various research fields.

\section*{Acknowledgments}
We thank Professor Nigel Goldenfeld at the University of California San Diego for taking the time to review our paper and for providing invaluable insights and feedback. We thank The SHIELD: Target, Test, Tell team at the University of Illinois Urbana-Champaign for collecting and providing data, and for helpful discussion for this research.
The IRB for this research was ruled exempt by UIUC (protocol \#21216). 
We thank the Institutional Data Analytics + Assessment team at Purdue University for collecting and providing data, and for helpful discussion for this research. 
The IRB for this research was ruled exempt by Purdue IRB-2020-1683. We thank Humphrey C. H. Leung for plotting several figures based on testing data from Purdue.
\bibliographystyle{unsrt}  
\bibliography{references}  
\newpage
\section*{\Huge Supporting Information}
\section*{Reverse Engineering the Reproduction Number: A
Framework for Model-Free Counterfactual
Analysis, Strategy Evaluation, and Feedback Control of Epidemics}
\label{secSI:overview}
\begin{figure}[p]
\centering
\includegraphics[width=\textwidth]{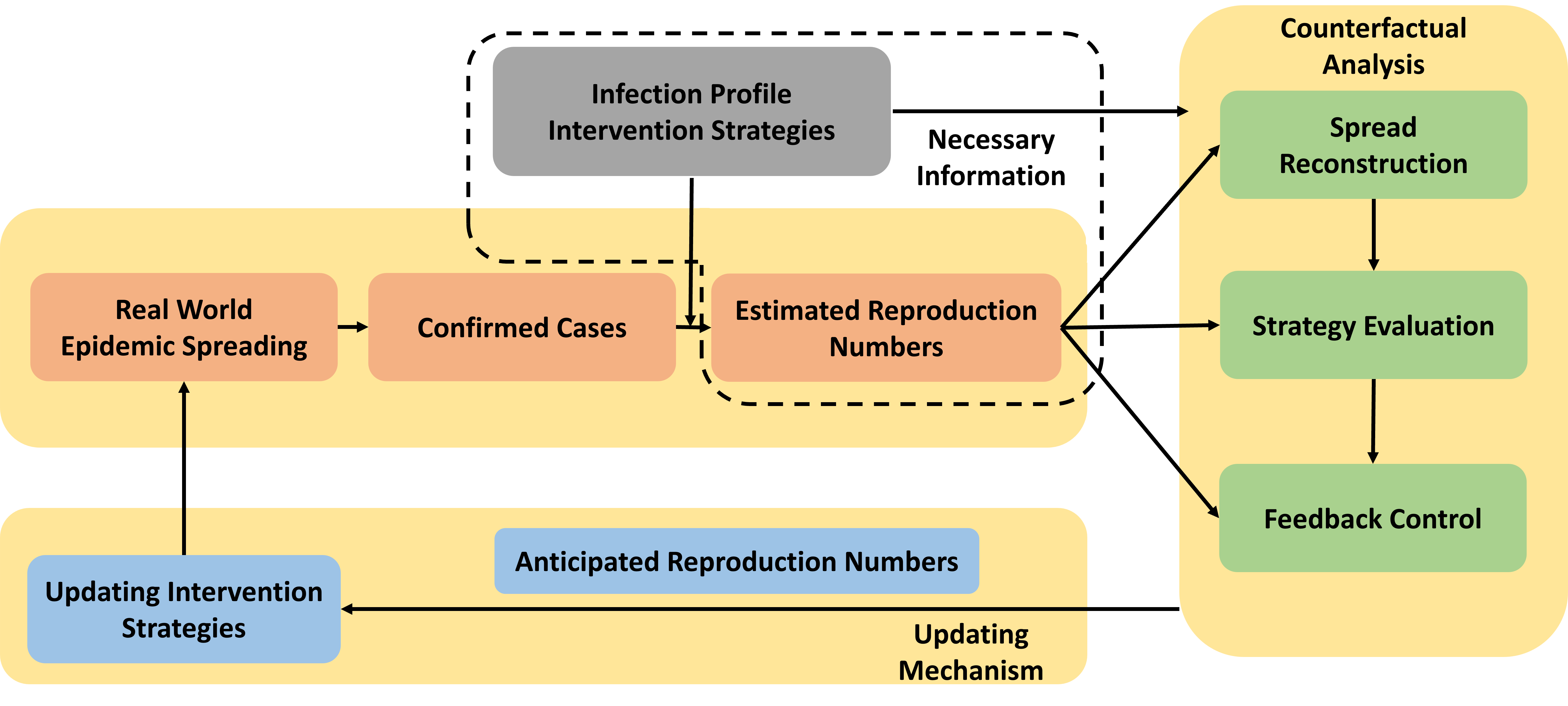}
\caption{SI-A
framework for model-free counterfactual
analysis, strategy evaluation, and feedback
control of epidemics.}
\label{fig:Control_Framework_SI}
\end{figure}

We present a general overview of the 
framework for model-free counterfactual
analysis, strategy evaluation, and feedback
control of epidemics
framework in Figure~\ref{fig:Control_Framework_SI}. 
In any testing-based epidemic spreading process, we can obtain confirmed positive cases and  utilize these confirmed cases to estimate the reproduction number of the spreading process~\cite{cori2013new,huisman2022estimation, gostic2020practical}.We propose
methodologies to reverse engineer the reproduction number in order to evaluate the implemented intervention strategies and
update the strategy through leveraging the reproduction number as feedback information.
This Supporting Information will provide a detailed introduction to the entire framework, including the real-world data utilized, the methodologies proposed, sensitivity analysis, and the goals associated with updating these strategies.

In particular, we validate the framework through leveraging COVID-19 data from the University of Illinois Urbana-Champaign (UIUC) and Purdue University (Purdue), where the intervention strategy is the implemented testing-for-isolation strategy.  During the COVID-19 pandemic, in order to safely operate university campuses, the University of Illinois Urbana-Champaign and Purdue University  implemented testing-for-isolation strategies. By testing a proportion of the total population on campus, including students, faculty, and staff, both universities isolated confirmed positive cases\footnote{\baike{Quarantine and isolation processes are not forced. Confirmed positive cases are encouraged to isolate themselves from the population.}}, aiming to keep the infected population below acceptable thresholds and prevent potential large-scale outbreaks. Both universities successfully maintained the infected population under their acceptable level. Inspired by these successful implementations, we are motivated to validate our \baike{framework that incorporates} 
reverse engineering of the reproduction number, counterfactual analysis of the intervention strategy, and 
feedback control mechanism to further improve the existing intervention strategies for spread reconstruction, strategy evaluation, and pandemic mitigation.

To summarize, in this Supporting Information, we will explain the following perspectives in detail:
 \begin{itemize}
   \item What type of data from UIUC and Purdue do we leverage to analyze the effectiveness of the testing-for-isolation strategies implemented by both universities? 
   \item How do we quantify the impact of the intervention strategy (testing-for-isolation in this work) on the infection profile, reproduction number, and the overall spreading process?
    \item What methodology do we propose for reverse engineering the reproduction number?
    \item How do we utilize the reverse engineering of the reproduction number to conduct counterfactual analysis of the spread?
    \item How can the developed framework serve as a foundation for model-free strategy evaluation and feedback control design based on the severity of the spread?
\end{itemize}
We organize the material by answering these questions, which will provide further information for our work.
\section*{SI-1. Intervention Strategies and Data from UIUC and Purdue}
\label{secSI:Data}
\subsection*{SI-1-A. Background}
First, we present the data sets. All the real-world data \baike{from Purdue and UIUC} were collected by Institutional Data Analytics + Assessment at Purdue University and the SHIELD team at the University of Illinois Urbana-Champaign, \baike{respectively}. 
During the COVID-19 pandemic, the SHIELD team at UIUC relied on guidance from reputable sources such as the Centers for Disease Control and Prevention, the Illinois Department of Public Health, and the Champaign Urbana Public Health District~\cite{UIUC_SHIELD}. They actively monitored the pandemic on and around campus and adapted their decisions in response to the evolving understanding of COVID-19. The IDA+A team at Purdue leveraged COVID-19 data specifically from the Purdue campus to perform statistical analysis and construct sampling methodologies~\cite{Purdue_Idata}. Their findings and reports were provided to campus leaders and decision-makers to assist in evaluating the spread of the virus on campus and the effectiveness of intervention strategies implemented at Purdue. Both teams adopted testing-for-isolation strategies  to assess the severity of the pandemic and made necessary adjustments to their plans.

At the early stage of the COVID-19 pandemic, when the availability of an efficient vaccine was \baike{absent}, significant challenges arose in mitigating the spread of the virus. Smart communities with dense populations, such as technology companies and research laboratories, began adopting remote working practices. In order to ensure the safe operation of densely populated areas, large universities like the University of Illinois Urbana-Champaign and Purdue University implemented testing-for-isolation strategies, along with other interventions. 
These strategies involved testing a proportion of the total campus population, enabling the identification of infected cases. Subsequently, infected and high-risk individuals were encouraged to isolate themselves within their own residences or be quarantined in designated quarantine centers. As a result, the severity of the epidemic on these campuses decreased since a portion of the infected population was no longer actively transmitting the virus.

Compared to UIUC and Purdue, in order to better support the testing-for-isolation strategy, different universities implemented their unique methodologies. 
For example, the team at UIUC proposed an agent-based model to capture the spreading on campus~\cite{ranoa2022mitigation}, while researchers at UCSD and Harvard studied the impact of testing using a network model~\cite{goyal2021evaluation}. Emory University developed a \baike{compartmental} model to study SARS-CoV-2 spreading among disparate populations of students, faculty, and staff~\cite{lopman2020model}. In addition, 
the researchers in~\cite{ghaffarzadegan2021diverse} proposed a group of models to simulate the spread over campuses after reopening. To determine whether in-person instruction could safely continue during the pandemic and evaluate the necessity of various interventions,~\cite{gressman2020simulating} utilized a stochastic agent-based model to study the spread over campuses. A customized susceptible, exposed, infected, and recovered compartmental model was presented in~\cite{brown2021simple} to describe the control of asymptomatic spread of COVID-19 infections on Boston University campus. 
Furthermore, an agent-based model on a network aimed at capturing unique features of COVID-19 spread through small residential colleges was proposed in~\cite{bahl2021modeling}.
Moreover,~\cite{borowiak2020controlling} used mathematical models to evaluate the impact of class sizes on the reproduction number to suppress the spread of the virus on campus, specifically in classrooms.
\subsection*{SI-1-B. Testing-For-Isolation at UIUC and Purdue}
\label{secSIt_Testing_UIUC_Purdue}
\begin{figure}[p]
\centering
\includegraphics[width=\textwidth]{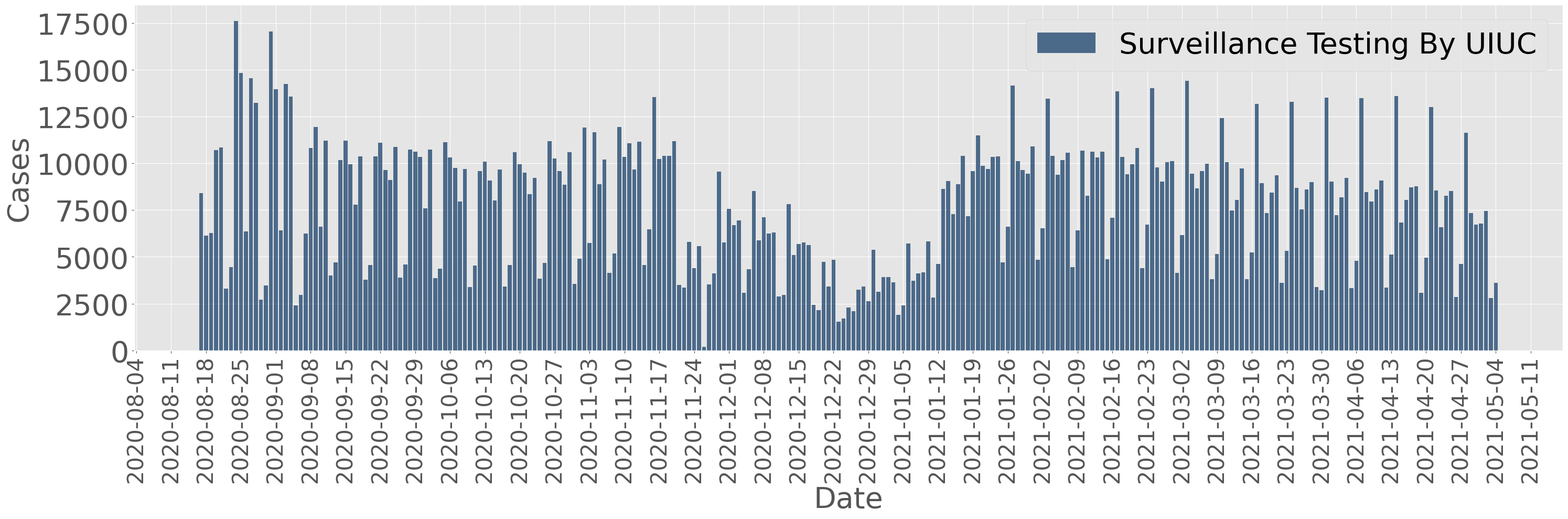}
\caption{Daily surveillance testing at UIUC during Fall 2020 and Spring 2021 implemented by the SHIELD team at UIUC.}
\label{fig:UIUC_T_SI}
\end{figure}

\begin{figure}[p]
\centering
\includegraphics[width=\textwidth]{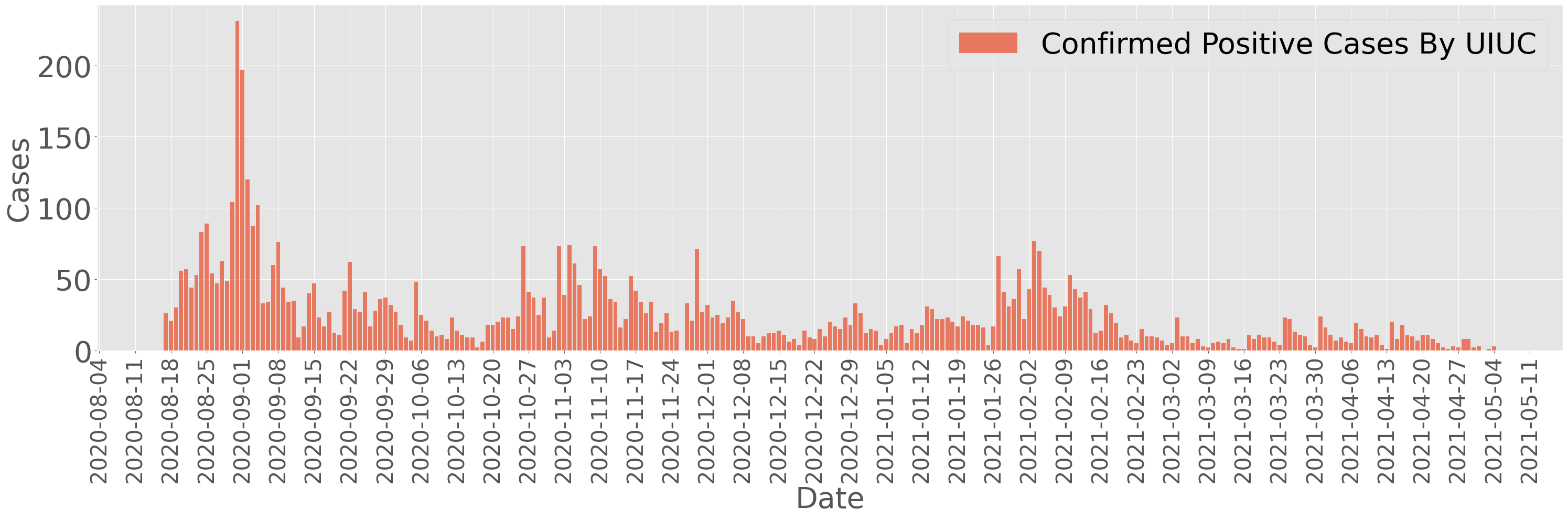}
\caption{Daily confirmed positive cases at UIUC during Fall 2020 \baike{and Spring 2021} captured by the SHIELD team at UIUC.}
\label{fig:UIUC_P_SI}
\end{figure}

We first introduce the aggregated data from UIUC. The University of Illinois Urbana-Champaign conducted campus-wide surveillance testing twice a week during Fall 2020 and three times a week during Spring 2023, illustrated by Figure~\ref{fig:UIUC_T_SI}. The surveillance testing was supported through spatial analyses in order to target high-risk regions such as the fraternity house.
The daily confirmed cases are presented in Figure~\ref{fig:UIUC_P_SI}. In the Methodology Section, we provide a more detailed explanation of the distinction between confirmed cases and infected cases. 
From Figure \ref{fig:UIUC_P_SI}, we observe multiple spikes during Fall 2020 and Spring 2021 on the UIUC campus. We highlight two significant spikes. The first spike occurred around the middle of August 2020 when UIUC implemented an entry-screening to identify the infected population returning to campus. The second spike occurred around the middle of October 2020, which we attribute to gathering events associated with the return of the college football season in Fall 2020. 
In addition to the entry-screening, UIUC managed to maintain a consistently low level of daily confirmed cases with mild fluctuations under the high surveillance testing frequency.

\begin{figure}[p]
\centering
\includegraphics[width=\textwidth]{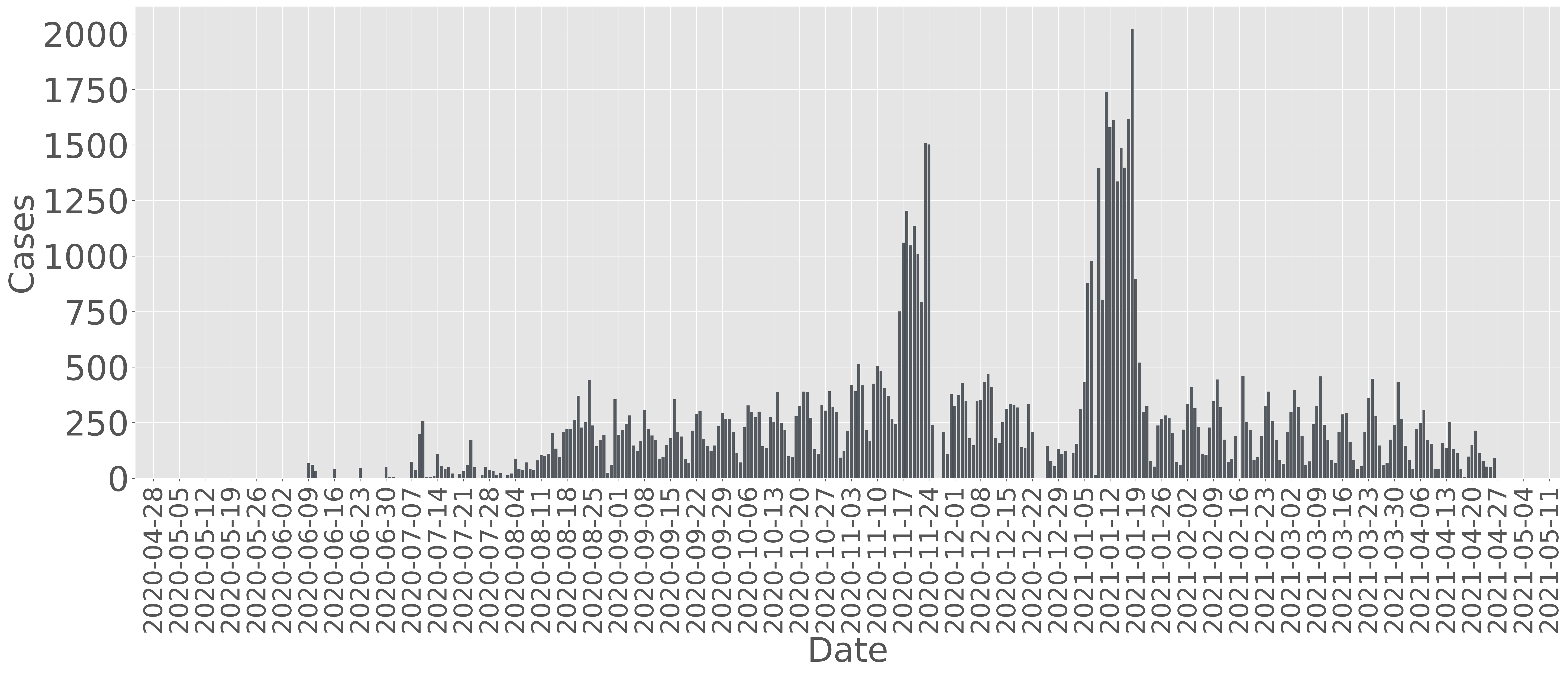}
\caption{Daily voluntary tests at Purdue University during Fall 2020 and Spring 2021 captured by the IDA+A team at Purdue.}
\label{fig:Purdue_SR_T_SI}
\end{figure}

\begin{figure} [!h]
\centering
\includegraphics[width=\textwidth]{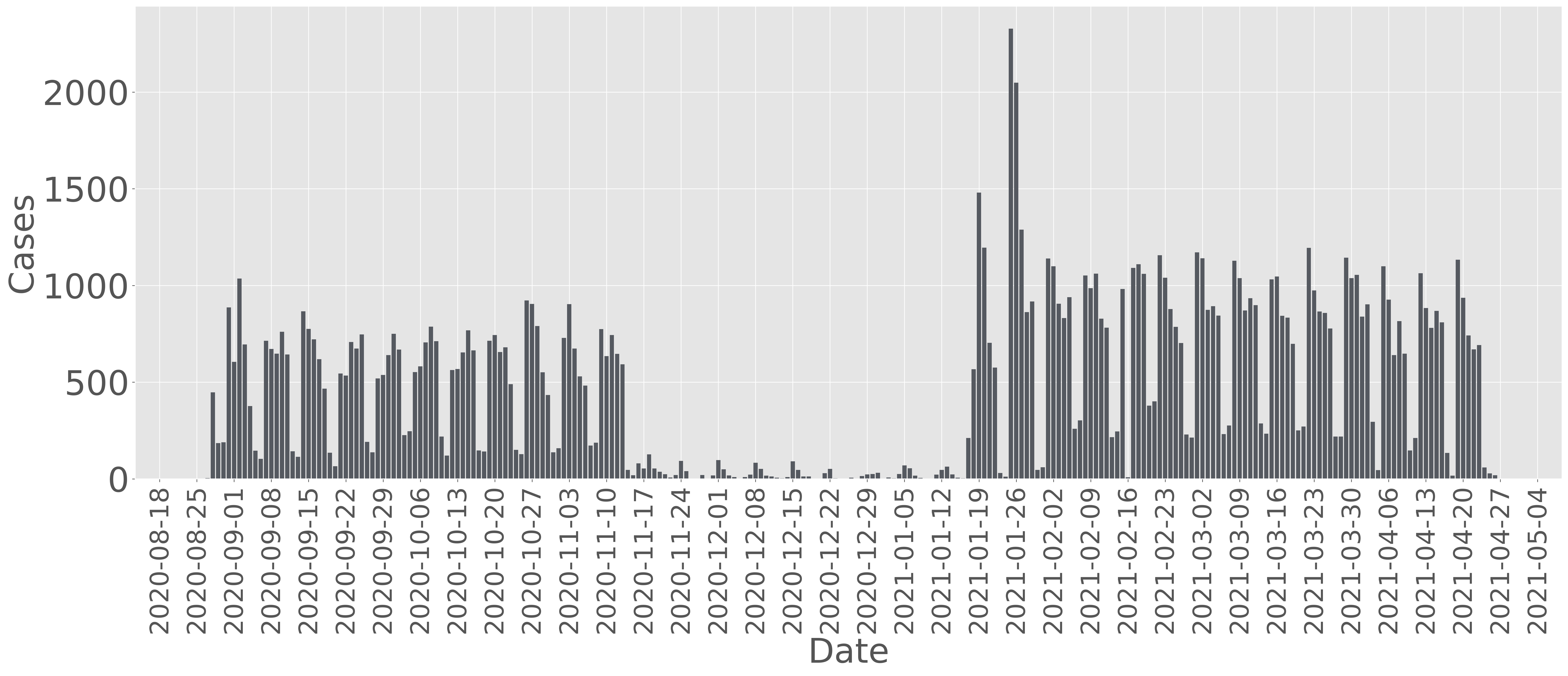}
\caption{Daily surveillance tests at Purdue University during Fall 2020 and Spring 2021 captured by the IDA+A team at Purdue.}
\label{fig:Purdue_SVL_T_SI}
\end{figure}

Distinct from UIUC, which relies on surveillance testing, Purdue University implemented a different testing-for-isolation strategy. Purdue allocated testing resources into two categories. The first category included symptomatic cases and cases identified through contact-tracing analysis, requiring these individuals to undergo testing. In this work, we refer to this type of testing as voluntary testing. The allocation of voluntary testing resources related to this category is illustrated in Figure~\ref{fig:Purdue_SR_T_SI}.
However, 
COVID-19 also involves asymptomatic infections~\cite{kronbichler2020asymptomatic}. To capture asymptomatic cases, Purdue designated approximately 5000 tests per week (approximately 8\% to  12\% of the total campus population) for surveillance testing during Fall 2020. Additionally, Purdue increased the surveillance testing rate during Spring 2021 to target more active spreading areas, such as the fraternity houses. As shown in Figure~\ref{fig:Purdue_SVL_T_SI}, there was an increase in the amount of surveillance testing resources used in Spring 2021.

In detail, Purdue's strategy involved randomly sampling and testing a proportion of the campus population. We refer to this type of testing-for-isolation strategy as surveillance testing, and Purdue's allocation of testing resources is shown in Figure~\ref{fig:Purdue_SVL_T_SI}. The absence of the surveillance testing dataset from the middle of November 2020 to the beginning of January 2021, as depicted in Figure~\ref{fig:Purdue_SVL_T_SI}, reflects the fact that Purdue University sent students back home after the Thanksgiving break in Fall 2020, and students did not return to campus until the Spring 2021 semester.

Furthermore, for residential students, Purdue randomly sampled 8\% of each residence hall floor based on their analyses. For off-campus students, Purdue utilized various contact tracing elements that the IDA+A team constructed in consultation with experts. The IDA+A team created a series of clusters representing the off-campus student body. Each cluster connected students via different contact tracing metrics such as dining swipes and network logs. Each student in the cluster was assigned centrality and connectedness features, which were then summed across each cluster to give each student an aggregate 'connectedness' value. Purdue then performed a weighted random sample, weighting by connectedness, from the entire off-campus student body. The amount of this sampling varied week-to-week depending on test availability.


\begin{figure}[p]
\centering
\includegraphics[width=\textwidth]{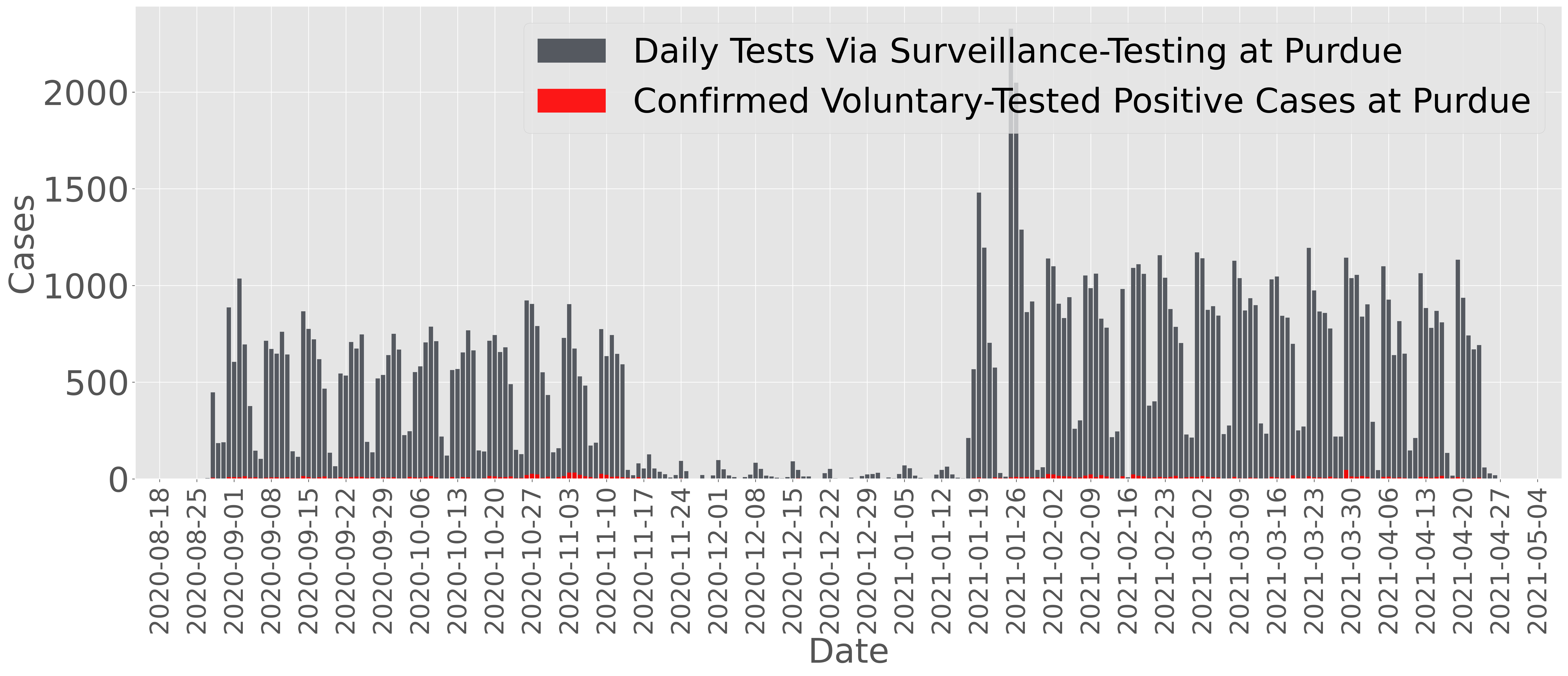}
\caption{Daily surveillance tests and 
confirmed cases through surveillance testing
at Purdue University during Fall 2020 and Spring 2021 captured by the IDA+A team at Purdue.}
\end{figure}

\begin{figure} [!h]
\centering
\includegraphics[width=\textwidth]{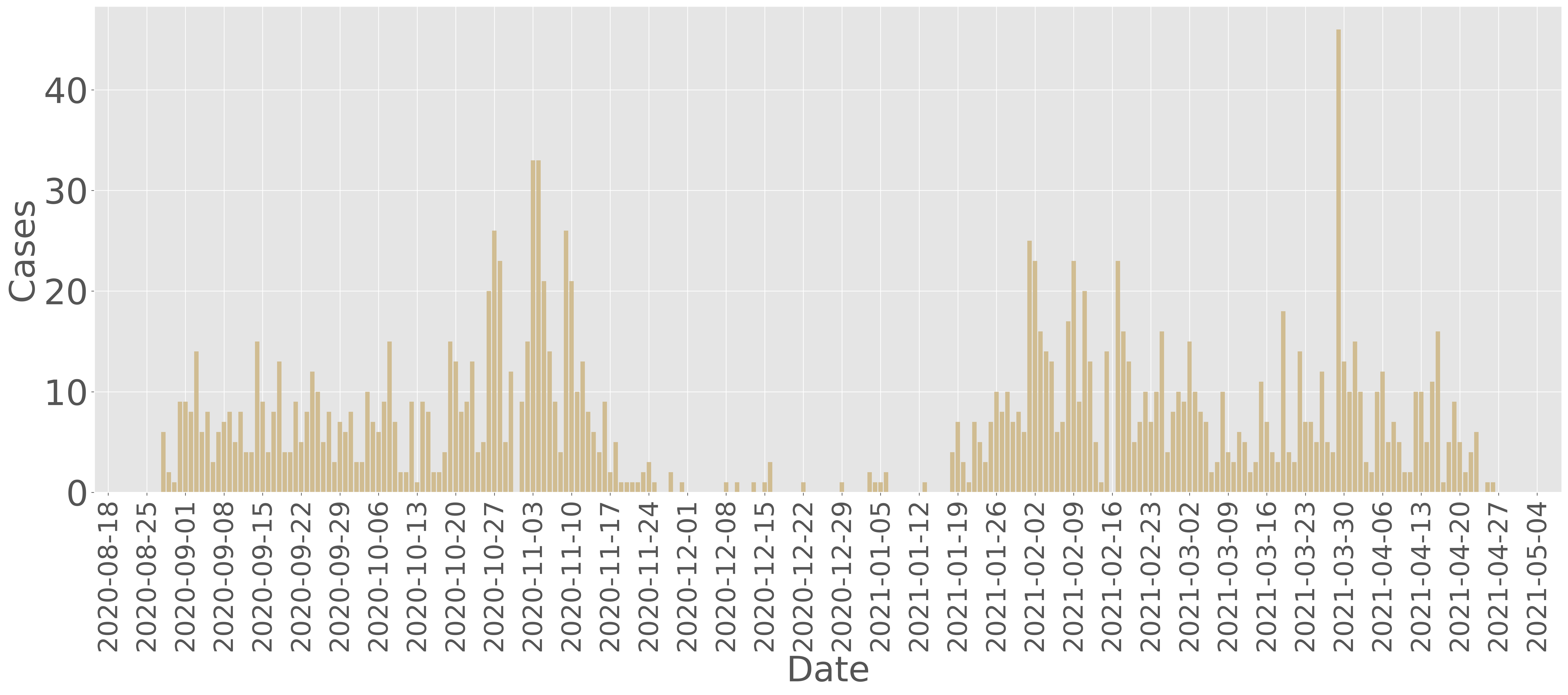}
\caption{Daily confirmed positive cases at Purdue University during Fall 2020 and Spring 2021 captured by the surveillance testing.}
\label{fig:Purdue_SVL_P_SI}
\end{figure}

\begin{figure}[p]
\centering
\includegraphics[width=\textwidth]{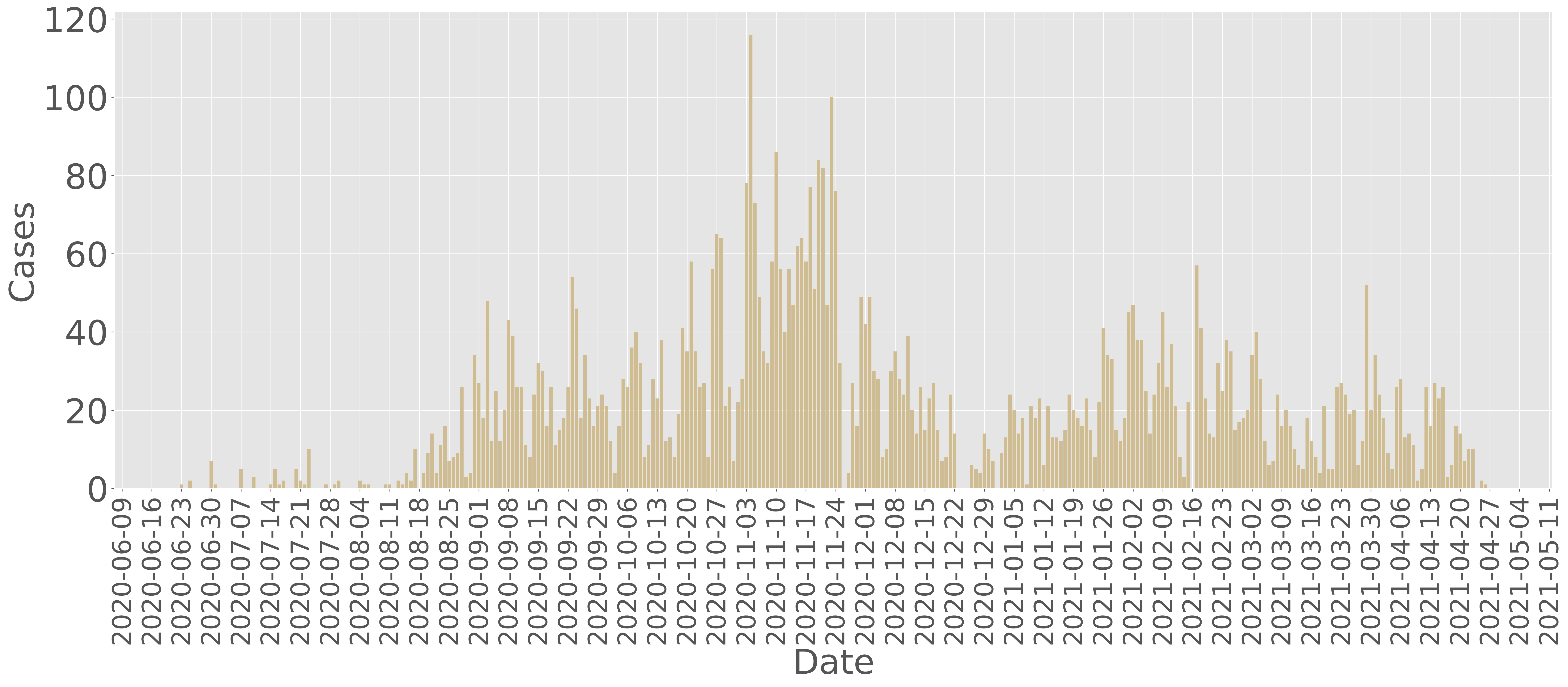}
\caption{Total daily confirmed positive cases at Purdue University during Fall 2020 and Spring 2021.}
\label{fig:Purdue_Total_SI}
\end{figure}

We additionally present the confirmed cases through the surveillance testing-for-isolation strategy and the total confirmed cases in Figure~\ref{fig:Purdue_SVL_P_SI} and Figure~\ref{fig:Purdue_Total_SI} during Fall 2020 and Spring 2021, respectively. From Figure~\ref{fig:Purdue_SVL_P_SI}, we conclude that the surveillance testing-for-isolation strategy successfully captured asymptomatic cases. Similar to UIUC during Fall 2020, Figure~\ref{fig:Purdue_Total_SI} exhibits two significant spikes. The first spike, occurring around the middle of August, was related to the entry-screening conducted at the beginning of the semester. The other spike was associated with gathering events resulting from the return of the college football season towards the end of October.

By comparing the total confirmed cases between UIUC (Figure~\ref{fig:UIUC_P_SI}) and Purdue (Figure~\ref{fig:Purdue_Total_SI}) during Fall 2020, we observe that Purdue University had a higher average daily number of confirmed cases and experienced greater fluctuations, despite allocating fewer testing resources. We can intuitively conclude that higher testing rates at UIUC enabled the timely identification and isolation of infected cases, thereby reducing the spread of the virus. Consequently, compared to Purdue, UIUC had fewer confirmed cases during the Fall 2020 semester. We primarily leverage the total confirmed cases from both universities, under their respective testing-for-isolation strategies, to validate the proposed  pandemic mitigation framework for counterfactual analysis, strategy evaluation and feedback control.

\subsection*{SI-1-C. Difference Between Isolation and Quarantine}
\begin{figure} [!h]
\centering
\includegraphics[width=\textwidth]{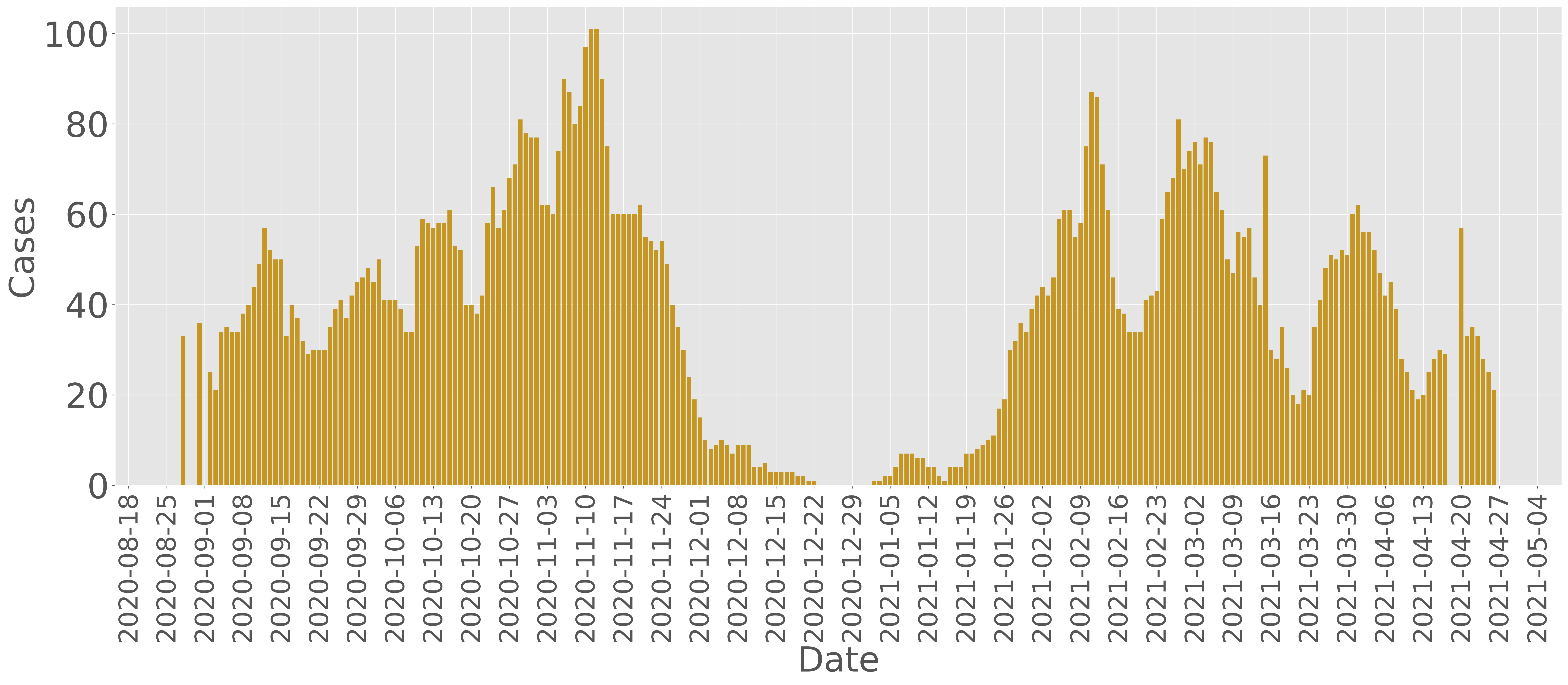}
\caption{Daily isolated cases at Purdue University during Fall 2020 and Spring 2021.}
\label{fig:Purdue_iso_SI}
\end{figure}

\begin{figure} [!h]
\centering
\includegraphics[width=\textwidth]{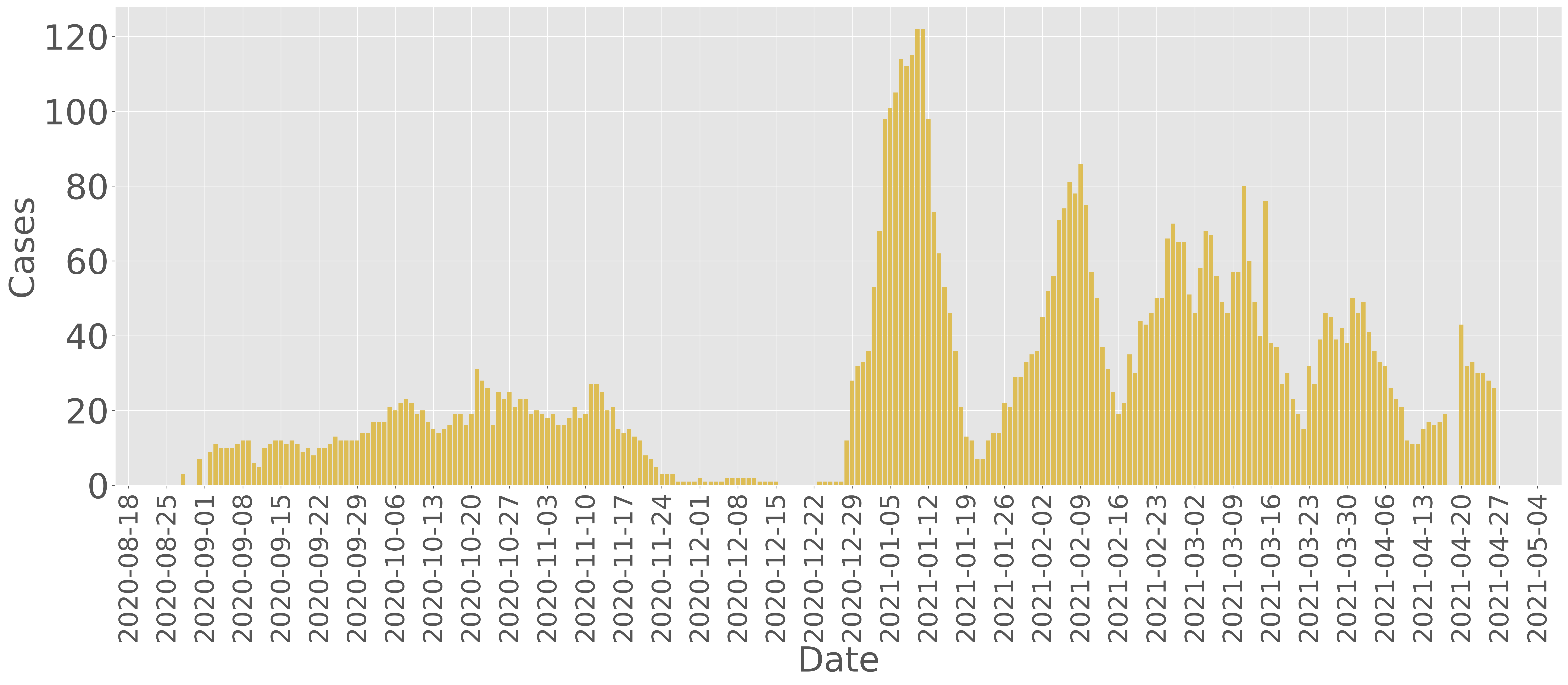}
\caption{Daily quarantined cases at Purdue University during Fall 2020 and Spring 2021.}
\label{fig:Purdue_qua_SI}
\end{figure}

In addition to presenting testing resources and confirmed cases from Purdue University, we further illustrate the implemented testing-for-isolation strategy at Purdue through the provided isolation data in Figure~\ref{fig:Purdue_iso_SI} and quarantine data in Figure~\ref{fig:Purdue_qua_SI}. When a case is confirmed to be infected through testing, Purdue University encourages the case to isolate from others and designates it as an isolated case. The termination of the isolation period is determined by whether the case tests negative or not. Therefore, the isolation data in Figure~\ref{fig:Purdue_iso_SI} can be considered as the cumulative number of infected cases with an average recovery time of one to two weeks. It is worth noting that during the Winter break, most students were not on campus, resulting in only a few recorded isolated cases during Winter 2020.


Unlike isolated cases, where all individuals were confirmed infected, Purdue also implemented contact tracing strategies to identify and quarantine individuals who had close contact with confirmed cases, as shown in Figure~\ref{fig:Purdue_qua_SI}. During the Fall 2020 semester, Purdue implemented a contact-tracing-based testing-for-isolation strategy by targeting a few closely contacted cases and encouraging them to quarantine. To enhance the effectiveness of testing, Purdue established a more complicated contact-tracing network, as described before. By constructing these contact-tracing networks, Purdue improved its contact-tracing policy during Spring 2021, resulting in more closely contacted cases being traced and subsequently quarantined. Consequently, compared to Fall 2020, we observe a significant increase in the number of daily quarantined cases, as depicted in Figure~\ref{fig:Purdue_qua_SI}.


Note that both UIUC and Purdue distinguish between isolation and quarantine. Quarantine keeps someone who has been in close contact with someone who has COVID-19 away from others, while isolation keeps someone who is sick or has tested positive for COVID-19 away from others, even within their own home. In this work, we refer to isolation as a measure to prevent those who tested positive from further spreading the virus. We do not consider quarantine. Additionally,  testing at both universities was not conducted through uniformly random sampling in reality. Both universities leveraged their spatial data to target and enhance the testing rates in high-risk areas.
\section*{SI-2. Methodology}
We introduce the core methodologies that we leverage and develop for the framework in Figure~\ref{fig:Control_Framework_SI}. We first introduce a testing environment to validate our proposed framework.
We present the mechanism used to generate synthetic data to simulate the spreading processes. This procedure includes explaining how to estimate the reproduction number based on the confirmed cases under the impact of the implemented intervention strategy, specifically the testing-for-isolation strategies, as indicated within the dashed circle in Figure~\ref{fig:Control_Framework_SI}.

\subsection*{SI-2-A. Epidemic Infection Profiles}
For an epidemic spreading process such as COVID-19, it is impossible for us to compare the effectiveness of different strength of the implemented intervention strategy under the exact spreading conditions at the exact same time in reality. This inherent characteristic of spreading processes presents challenges when proposing epidemic evaluation and feedback control design. However, it is important and necessary to evaluate different intervention strategies in order to prepare for future outbreaks.
Therefore, counterfactual analysis, a method used to evaluate what might have happened in a situation that did not occur, is critical. We first introduce a 
methodology that leverages the reproduction number and random processes to
generate daily confirmed cases, which serves as a foundation for reconstructing the spread over the UIUC and Purdue campuses with their implemented interventions in the next section. This approach
facilitates the counterfactual analysis of different intensity of the intervention strategy. 

\subsubsection*{SI-2-A-1. Infection Profile and The Reproduction Number}
The infection profile represents the average time between the onset of infections in a primary case and the onset of infections in its secondary cases. Different infectious diseases exhibit distinct spreading behaviors, leading to varying infection profiles. Moreover, the infection profile of the same virus can differ due to changes in the spreading environment. Factors such as the age structure of the population can influence the infection profile. Additionally, symptomatic and asymptomatic infections may generate different infection profiles within the same population.

\begin{figure}[p]
\centering
\includegraphics[width=\textwidth]{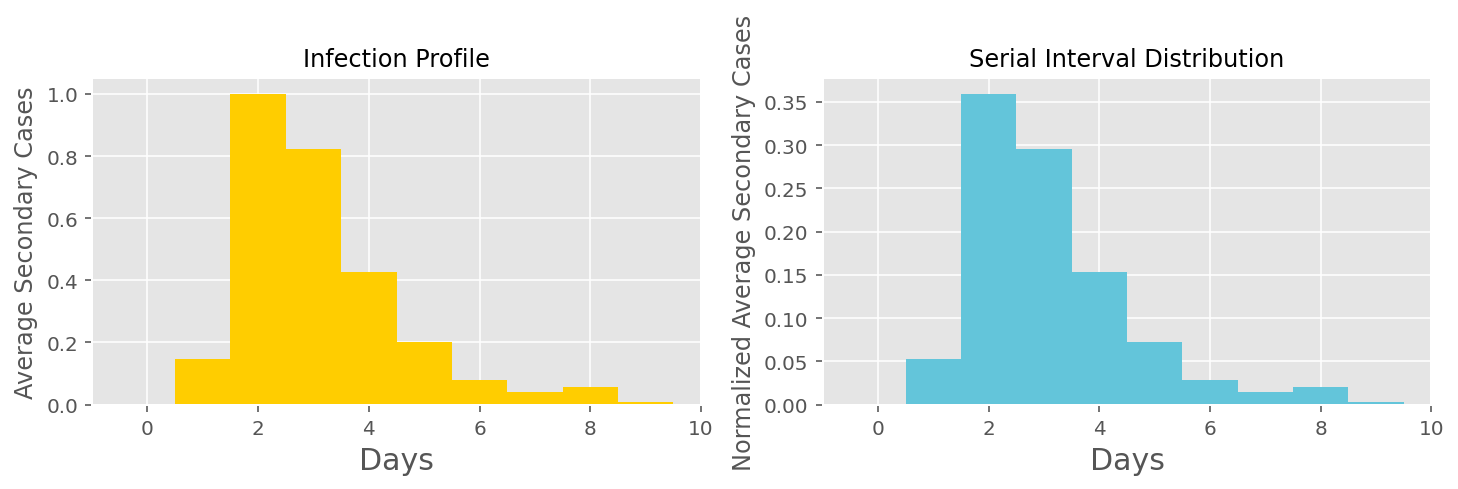}
\caption{One COVID-19 infection profile (Left); Serial interval distribution (Right) \cite{goyal2021viral, he2020temporal}.}
\label{fig:COVID19_In_Pro_SI}
\end{figure}

Hence, to generalize the methodology developed in this work, we denote the infection profile of symptomatic infections as follows:
\begin{equation*}
    \underline{v}=[\underline{v}_1, \underline{v}_2, \cdots, \underline{v}_n],
\end{equation*}
where $\underline{v}\in \mathbb{R}^n_{>0}$ and $n$ is the number of days during which a symptomatic case is infectious. We use $\underline{v}_i\in\mathbb{R}_{\geq0}$ to represent the average number of infected cases that a symptomatic case can generate on day $i$. Similarly, 
we define infection profile of asymptomatic cases as a vector 
\begin{equation*}
  \overline{v}=[\overline{v}_1, \overline{v}_2, \cdots, \overline{v}_m], 
\end{equation*} where $v\in \mathbb{R}^m_{>0}$ and $m$ is the number of the days  during which an asymptomatic case is infectious. We use $\overline{v}_i\in \mathbb{R}_{>0}$ to represent the average number of infected cases that an asymptomatic case can generate on day $i$. The  reproduction number can capture the average number of infected cases generated by one infected individual in a  susceptible population~\cite{delamater2019complexity}. Hence, it is naturally to bridge the gap between the infection profile and the reproduction number through the following equations:
\begin{equation}
\label{eq_SI: Reproduction_Num}
\underline{\mathcal{R}} = \sum_{i=1}^n \underline{v}_i, \ \ \ \ \overline{\mathcal{R}}= \sum_{i=1}^m \overline{v}_i,
\end{equation}
where $\underline{\mathcal{R}}$ and $\overline{\mathcal{R}}$ are the reproduction number of the symptomatic and asymptomatic infections, respectively. Further, for an infectious disease that can generate both symptomatic and asymptomatic infections, if symptomatic cases are  $\theta$ percent, the reproduction number of the spreading process is defined as 
\begin{equation}
\label{eq_SI: R_0}
    \mathcal{R} = \theta \sum_{i=1}^n \underline{v}_i+ (1-\theta)\sum_{i=1}^m\overline{v}_i=\theta \underline{\mathcal{R}}+ (1-\theta)\overline{\mathcal{R}}.
\end{equation}
Specifically, in order to facilitate comparisons of the implemented intervention strategy, we utilize the same infection profile of the COVID-19 pandemic for both UIUC and Purdue University, given by~\cite{goyal2021viral, he2020temporal}
\begin{equation}
\label{eq_SI:SI_Dist}
v=\underline{v}=\overline{v} = [0.148, 1.0, 0.823, 0.426, 0.202, 0.078, 0.042, 0.057, 0.009].
\end{equation}
\eqref{eq_SI:SI_Dist} were leveraged by the research team at UIUC~\cite{ranoa2022mitigation}. We illustrate the infection profile in~\eqref{eq_SI:SI_Dist} by Figure~\ref{fig:COVID19_In_Pro_SI} (Left).
Additionally, based on \eqref{eq_SI: R_0}, the reproduction number we leverage for COVID-19 is $\mathcal{R}= 2.785$. Note that we consider the reproduction number $\mathcal{R}$ as the basic reproduction number.
In epidemic mitigation and prediction, researchers not only rely on the basic reproduction number but also focus on the effective reproduction number. Unlike the basic reproduction number, the effective reproduction number captures the average number of infected cases generated by one infected case in a mixed population, including susceptible, recovered, and infected individuals, etc. When the basic reproduction number of an infectious disease spreading over a population is known, it is common to approximate the effective reproduction number $\mathcal{R}_t$ by scaling the basic reproduction number using the following equation:
\begin{equation*}
\mathcal{R}_t=\frac{S(t)}{N}\mathcal{R},
\end{equation*}
where $N$ represents the total population and $S(t)\in [0,N]$ represents the  susceptible population at time step $t\in [0, +\infty]$. Meanwhile, we can also approximate the effective reproduction number $\mathcal{R}_t$ using the estimated reproduction number. Researchers proposed various methods to estimate the reproduction number: utilizing data on infected cases, confirmed cases, hospitalized cases, and more. In this work, we employ a popular method based on Bayesian inference to estimate the reproduction number using confirmed cases~\cite{cori2013new,gostic2020practical,huisman2022estimation}. To do so, we first need to understand the concept of the serial interval distribution. 

\subsubsection*{SI-2-A-2. Serial Interval Distributions}
A serial interval distribution represents the normalized infection profile. \baike{If an infected case can transmit the infection to another person, it defines the probability distribution of the duration between the infection of the primary case and the infection of the secondary case.} Based on this definition, we can define the serial interval distributions of symptomatic and asymptomatic infections as 
\begin{align*}
    \underline{w}=\underline{v}/\underline{\mathcal{R}}=[\underline{v}_1/\underline{\mathcal{R}}, \underline{v}_2/\underline{\mathcal{R}}, \cdots, \underline{v}_n/\underline{\mathcal{R}}],\\
    \overline{w}= \overline{v}/\overline{\mathcal{R}}=[\overline{v}_1/\overline{\mathcal{R}}, \overline{v}_2/\overline{\mathcal{R}}, \cdots, \overline{v}_m/\overline{\mathcal{R}}],
\end{align*}
respectively. \baike{Note that $\underline{w}\in [0,1]^n$ and $\overline{w}\in [0,1]^m$}.
According to~\eqref{eq_SI: Reproduction_Num} and~\eqref{eq_SI: R_0},
the serial interval distribution of a spreading process with both symptomatic and asymptomatic infections is given by 
\begin{equation}
\label{eq_SI:SI_Def}
    w=[(\theta\underline{v}_1+(1-\theta)\overline{v}_1)/\mathcal{R}, (\theta\underline{v}_2+(1-\theta)\overline{v}_2)/\mathcal{R}, \dots, (\theta\underline{v}_m+(1-\theta)\overline{v}_m)/\mathcal{R}, \theta\underline{v}_{m+1}/\mathcal{R}, \dots, \theta\underline{v}_n/\mathcal{R}].
\end{equation}
where $n\geq m$.
Note that, $\sum_{i=1}^n{\underline{w}}=\sum_{i=1}^m{\overline{w}} = \sum_{i=1}^n{w} =1$ by definition, since the serial interval distribution is a probability distribution. In this work, we consider 
\begin{equation}
\label{eq_SI:SI_Normal}
    w=\underline{w}=\overline{w}= [0.053, 0.36, 0.29, 0.153,  0.078,
 0.028, 0.015, 0.02, 0.003],
\end{equation}
where \eqref{eq_SI:SI_Normal} is obtained by normalizing \eqref{eq_SI:SI_Dist}. We show how to leverage the serial interval distribution to generate spreading data, which will lay a foundation for estimating the reproduction number. 
\subsection*{SI-2-B. Generating Spreading Data}
\label{secSI:Gen_data}
We introduce a mechanism for generating spreading data that matches real-world spreading processes~\cite{Epyestim_Python_2020,huisman2022estimation}. We utilize this data-generation method to generate confirmed cases that mimic the confirmed cases from UIUC and Purdue. The data generation method is based on a widely used model, which can be found in~\cite{cori2013new, huisman2022estimation}and in the Python package~\textit{Epyestim}~\cite{Epyestim_Python_2020}.
We consider an epidemic spreading process with the serial interval distribution, given by $w$, follows a Poisson process,
such that the number of new generated infected cases $I_t$ at time $t$ is Poisson distributed with mean 
~\cite{cori2013new}
\begin{equation}
\label{eq_SI:Pos_mean}
   \mathbb{E}(I_t)= \mathcal{R}_t\sum_{s=1}^{t}I_{t-s}w_s.   
\end{equation}
Further, we define $\Lambda_t=\sum_{s=1}^{t}I_{t-s}w_s$. Then, the probability distribution of \baike{$k$} infected cases on day $t$ is denoted by~\cite{cori2013new} 
\begin{equation}
\label{eq_SI:Pos_dis}
    \mathbb{P}(I_t=k)= \frac{(\mathcal{R}_t\Lambda_t)^{k}e^{(-\mathcal{R}_t\Lambda_t)}}{k!}.
\end{equation}
Therefore, we have $I_t\sim Pois(\lambda=\mathcal{R}_t\Lambda_t)$, where $Pois(\lambda)$ denotes the Poisson distribution with mean $\lambda$. \eqref{eq_SI:Pos_dis} implies that the new infected cases at time step $t$ are determined by the serial interval distribution $w$, the existing
daily infected cases $I_t$, $t\in \{t-1, t-2,\dots, 0\}$,
and the reproduction number at time step $t$, i.e., $\mathcal{R}_t$. By further investigating~\eqref{eq_SI:Pos_mean},
the term $\Lambda_t$ represents the convolution of the daily infected cases and the serial interval distribution of the corresponding disease. If $t-s$ is greater than the length of the serial interval distribution $w$,  \baike{the average number of new daily infected cases} on day $t-s$ cannot generate new infections on day $t$, meaning that the infected cases \baike{becoming infectious on day $0$}  are already non-infectious on day $t-s$.  Furthermore, $\Lambda_t$ can only generate "scaled" new infected cases because $w$ is a normalized infection profile. To obtain the average number of new infected cases based on~\eqref{eq_SI:Pos_mean}, we need to scale $\Lambda_t$ by the current reproduction number $\mathcal{R}_t$. Hence, we obtain~\eqref{eq_SI:Pos_mean} 
to describe the average daily number of new infected cases. 
A more detailed introduction and discussion on how to leverage this technique to capture new infected cases can be found in~\cite{cori2013new, huisman2022estimation,Epyestim_Python_2020}.

\begin{figure}[p]
\centering
\includegraphics[width=\textwidth]{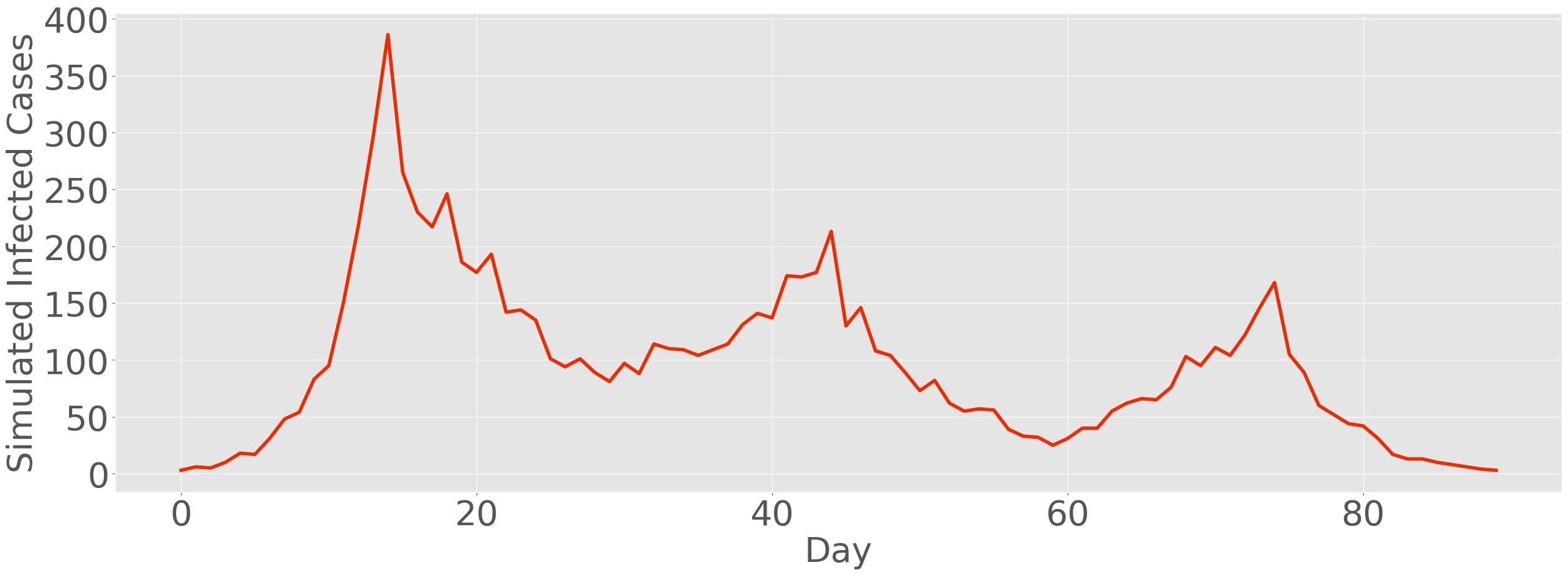}
\caption{Simulated daily infected cases.}
\label{fig:Simulated_I_SI}
\end{figure}

After introducing the mechanism to generate infected cases, we use the following example to illustrate the process of generating new infected cases using~\eqref{eq_SI:Pos_mean} and~\eqref{eq_SI:Pos_dis}.
The data generation process is built upon the Python package to generate infection data and estimate the reproduction number~\cite{Epyestim_Python_2020}.
We consider an epidemic spreading process \baike{over a sufficiently large population} with a fixed serial interval distribution, as given in~\eqref{eq_SI:SI_Dist}.
The reproduction number of the spreading process is illustrated in Figure~\ref{fig:Simulated_I_SI}. We consider the initial infected cases from day 1 to day 7 to be~$[1, 0, 1, 0, 2, 1, 3]$, i.e., the initial condition.
By implementing the mechanism described by~\eqref{eq_SI:Pos_mean} and~\eqref{eq_SI:Pos_dis}, we generate the infected cases using the Poisson process. Figure~\ref{fig:Simulated_I_SI} shows one typical simulation of the daily infected cases. We can see from Figure~\ref{fig:Simulated_I_SI} that the trend of the infected cases matches the changes in the reproduction number. A higher reproduction number results in larger spikes in the number of infected cases, while the quantity of infected cases begins to decrease when the reproduction number is less than 1. 
\begin{figure}[p]
\centering
\includegraphics[width=\textwidth]{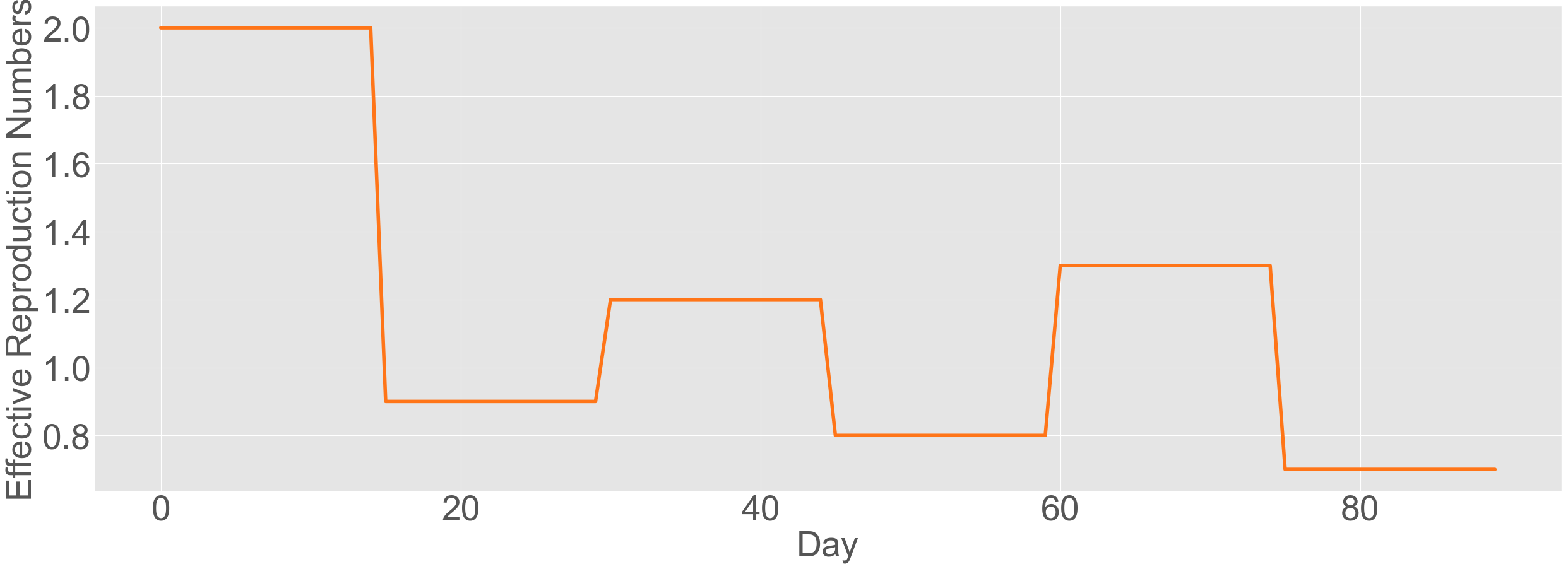}
\caption{\baike{Simulated} reproduction number.}
\label{fig:Simulated_R_SI}
\end{figure}

In reality, directly obtaining  the infected cases described in Figure~\ref{fig:Simulated_I_SI} is extremely challenging. Instead, it is more common to have access to confirmed cases, which include a delay between a case being infected and being confirmed. 
This period is influenced by factors such as the incubation period, testing delays, and report delays. For instance, all the cases we collected in Figure~\ref{fig:UIUC_P_SI} and Figure~\ref{fig:Purdue_Total_SI} were confirmed cases rather than infected cases. To simulate confirmed cases in the modeling process, we need to generate confirmed cases based on the infected cases. \baike{The idea is to add factors such as the incubation period, testing delays, and report delays to the simulated infected cases.}

The incubation period captures the time from when an infected case becomes infected to when it becomes infectious~\cite{lauer2020incubation}. Testing and reporting delays capture the delays from becoming infectious to being reported/confirmed. In this work, we mainly consider these two delays. We consider the incubation period of the infections during Fall 2020 and 2021 follows the distribution given by~\cite{brauner2021inferring, brauner2020effectiveness}, where the incubation period of an infected case follows a discrete Gamma distribution with shape parameter 1.35 and scale parameter 3.77. Meanwhile, we leverage another discrete Gamma distribution, with shape parameter 2 and scale parameter 3.2 to capture the testing-to-confirmation delay~\cite{tariq2020real}. We generate the combined delay distribution through convolution between the incubation period distribution and the testing-to-confirmation delay distribution~\cite{Epyestim_Python_2020}.
The combined distribution is defined as the infection-to-confirmation delay distribution, denoted as $\Delta$. The mean of the infection-to-confirmation delay distribution $\Delta$ in this work is 10.3 day~\cite{huisman2022estimation,brauner2021inferring,Epyestim_Python_2020}. 
Through the convolution between the infection-to-confirmation delay distribution and the infected cases generated by~\eqref{eq_SI:Pos_mean} and~\eqref{eq_SI:Pos_dis} , we obtain the synthetic data to describe confirmed cases~\cite{huisman2022estimation}, given by
\begin{equation}
    C_t = \sum_{s=1}^t I_{t-s}\Delta_s,
\end{equation}
where $C_t$ represents confirmed cases on day $t$ from the infected cases with incubation period and testing-to-confirmation delays.

\begin{figure}[p]
\centering
\includegraphics[width=\textwidth]{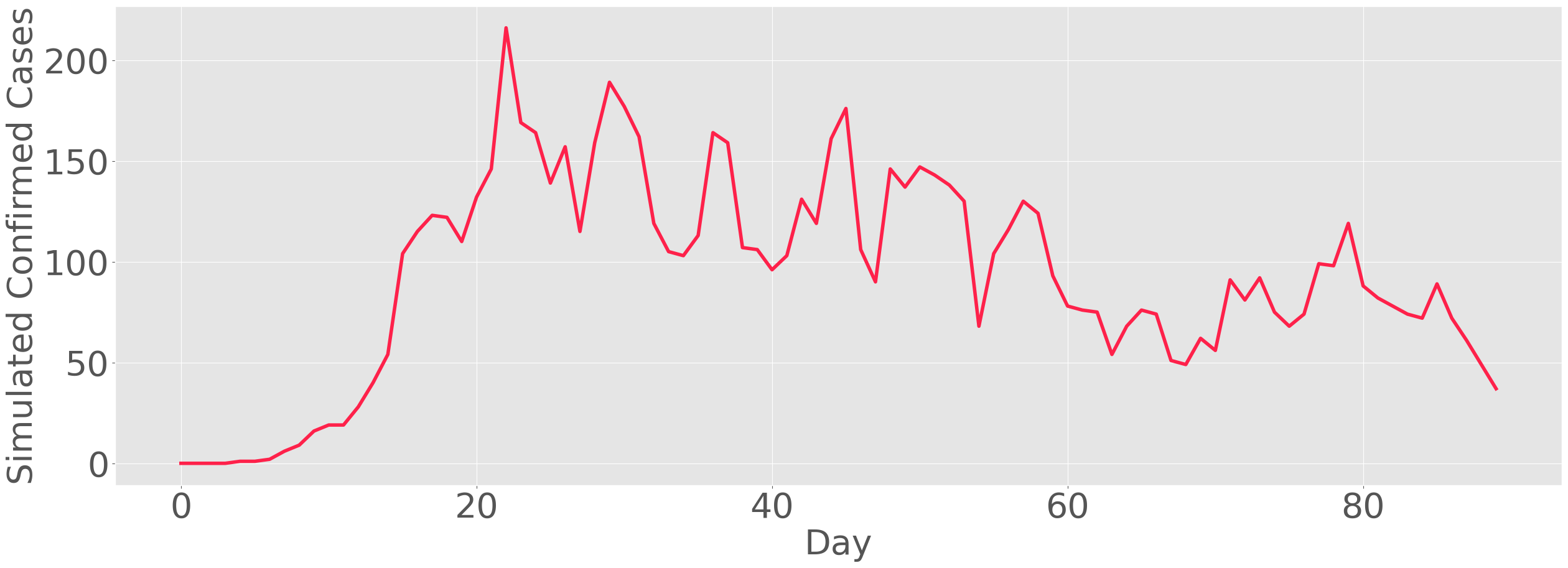}
\caption{Simulated daily confirmed cases.}
\label{fig:Simulated_C_SI}
\end{figure}

Additionally, to better align with the weekly testing and reporting patterns implemented by UIUC and Purdue, as shown in Figure~\ref{fig:Purdue_SVL_T_SI},
we introduce noise and weekly patterns to the confirmed cases
\baike{through the methodology in~\cite{huisman2022estimation}}. We utilize a sinusoidal function to generate the weekly testing and reporting pattern. By incorporating noise and weekly patterns, we can generate confirmed cases that reflect the infected data shown in Figure~\ref{fig:Simulated_I_SI}, as illustrated in Figure~\ref{fig:Simulated_C_SI}. In comparison to Figure~\ref{fig:Simulated_I_SI}, the confirmed cases in Figure~\ref{fig:Simulated_C_SI} exhibit delayed infection with weekly patterns. More detailed techniques about generating synthetic confirmed data can be found in~\cite{Epyestim_Python_2020,huisman2022estimation}. 

We utilize the techniques introduced in this section to generate confirmed data for both UIUC and Purdue, aiming to match the confirmed cases observed on both campuses. To achieve \baike{the data generation process}, we need to estimate the  
 reproduction number \baike{$\mathcal{R}_t$} through the confirmed cases, taking into account the impact of the implemented testing-for-isolation strategies. As a result, we propose a methodology to quantify the impact of the testing-for-isolation intervention on spreading processes.
\subsection*{SI-2-C. The Impact of Testing-For-Isolation On Spreading Processes}
\label{secSI:Testing}
The spreading processes in this work follow the serial interval distribution normalized from the infection profile in~\eqref{eq_SI:SI_Dist}.  Additionally,
we explore the impact of testing-for-isolation strategies on 
the serial interval distribution, since the testing-for-isolation strategy is the only control intervention we consider~\cite{ali2020serial}. 
Typically, researchers qualitatively study the impact of interventions on serial interval distributions by leveraging real-world spreading data~\cite{brauner2021inferring}. However, one disadvantage of studying the impact of the intervention strategy at different intensity is the limited amount of data from the irreversible feature of the spread. Hence, it is  challenging to directly quantify the influence of all possible intervention strategies on the spread using real-world spreading data.

To address this problem, we propose a novel mechanism to quantify the impact of the testing-for-isolation strategy on the infection profile, and thus, the reproduction number and spread. 
In particular, the mechanism we propose presents a new way to compute how the isolation rate affects the serial interval distribution and subsequently alters the spreading process. 
Based on this mechanism, we can reverse engineer the reproduction number based on the isolation rate for counterfactual analysis.
To illustrate the mechanism, we first examine the impact of the isolation rate on the infection profile and reproduction number. Then, we map the influence to the serial interval distribution.  This mechanism provides valuable insights into understanding the relationship between  the isolation rate and the spreading process. By quantifying the impact of the testing-for-isolation strategy on the serial interval distribution, we can better assess the effectiveness of different isolation rates in evaluating and controlling the spread of infectious diseases.

We consider implementing testing-for-isolation strategies for a mix of symptomatic and asymptomatic cases to simplify the introduction and validation. In our approach, the symptomatic and asymptomatic cases have the exact same infection profile. However, the same mechanism can be applied to symptomatic infections and asymptomatic infections with different infection profiles through~\eqref{eq_SI: Reproduction_Num} and~\eqref{eq_SI: R_0}, or even a disease with multiple infection profiles based on the host. 
If we uniformly randomly sample a proportion $\alpha$, $\alpha\in (0, 1]$, of the population on campus to test daily, we can confirm and isolate $\alpha$ times the total infected population daily, assuming the ideal situation where we can isolate all confirmed cases. Consequently, the isolated cases cannot infect others. Therefore, we consider the removal of an average of $\alpha$ of the total infected cases each day. In this context, we consider that the infection profile of a spreading process is represented by the vector $v=[v_1, v_2, \dots, v_n]$. Then, we propose the modified infection profile under the daily isolation rate $\alpha$ as  
\begin{equation}
\label{eq_SI: Inf_Prof_Test}
    v(\alpha)=[v_1(1-\alpha), v_2(1-\alpha)^2, \cdots, v_n(1-\alpha)^n].
\end{equation}
We  explain \eqref{eq_SI: Inf_Prof_Test} step by step: 
If we have $k\in\mathbb{N}_{>0}$
infectious cases on day 1,
without testing-for-isolation strategies, 
these $k$ infectious cases will generate an average number of $kv_i$ cases on day $i$, $i\in \{1,2,\dots, n\}$. However, consider the same number of $k$ infectious cases under a testing-for-isolation strategy. 
If we test and then isolate $k\overline{\alpha}$ cases on day $1$ from the $k$ cases, 
there will be $kv_1(1-\overline{\alpha})$ new infected cases that are generated by the $k(1-\overline{\alpha})$ cases. On day $2$, there will be $kv_2(1-\overline{\alpha})^2$ cases generated by $k(1-\overline{\alpha})^2$ infectious cases. Consequently, the new infected cases caused by the original $k$ infectious cases are $kv_i(1-\overline{\alpha})^i$ on day $i$, $i\in\{1,2,\dots, n\}$. Thus, the average number of infected cases generated by a single infectious individual on day $i$, $i\in\{1,2,\dots, n\}$, is given by $kv_i(1-\overline{\alpha})^i$. 
Consequently, we can obtain the infection profile of the original $k$ infectious cases as shown in~\eqref{eq_SI: Inf_Prof_Test}. 
Note that~\eqref{eq_SI: Inf_Prof_Test} provides a mechanism to quantify the impact of the daily isolation rate on the initial infection profile. Given a certain spreading process, we can study the impact of the isolation rate on the infection profile and further analyze its effect on the reproduction number.

In order to study the impact of the daily isolation rate $\alpha$ on the reproduction number, we define the reproduction number under  the impact of the isolation rate $\alpha$ as 
\begin{equation}
    \mathcal{R}(\alpha) = \sum_{i=1}^n {v(\alpha)}.
\end{equation}
Then, we define the scaling factor of the  reproduction number under the isolation rate $\alpha$ as $\mathcal{F}(\alpha)$, where 
\begin{equation}
\label{eq_SI:F}
      \mathcal{F}(\alpha) = \frac{\mathcal{R}(\alpha)}{\mathcal{R}}.  
\end{equation}
Based on~\eqref{eq_SI:F}, we can quantify how the isolate rate $\alpha$ scales down the reproduction number. From the definition of the scaling factor, $\mathcal{F}(\alpha)\in [0,1]$.
Additionally, we further explore the impact of the isolation rate on the serial interval distribution. Based on the definition of the serial interval distribution in~\eqref{eq_SI:SI_Def}, we define the serial interval distribution under the isolation rate $\alpha$ as
\begin{equation}
\label{eq_SI:SI_Und_Test}
    w(\alpha) = \frac{ v(\alpha)}{\mathcal{R}(\alpha)}
    = \frac{ v(\alpha)}{\mathcal{F}(\alpha) \mathcal{R}}.
\end{equation}
Note that, both $w$ and $w(\alpha)$ are probability distribution. Hence, the isolation rate $\alpha$ only changes the shape of the distribution but not the summation. 

\begin{figure}[p]
\label{Fig:Inf_Prof_Comp}
\centering
\includegraphics[width=\textwidth]{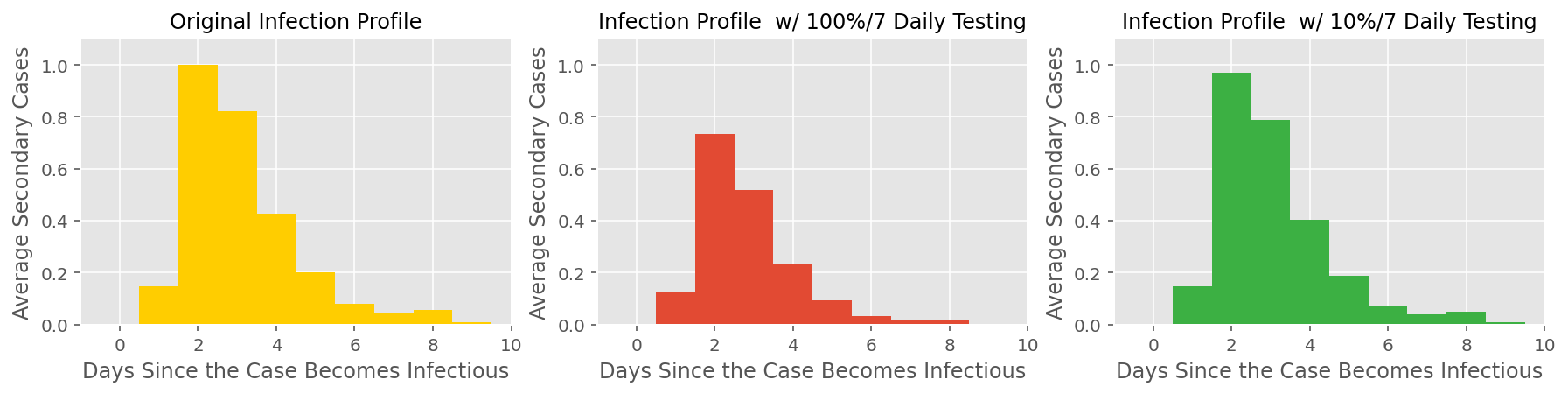}
\caption{Infection Profiles: (Left) Original infection profile, (Middle) Infection profile under $\alpha_1 = 1/7$ testing/isolation rate, (Right) Infection profile under $\alpha_2 = 0.1/7$ testing/isolation rate.}
\label{fig:In_Prof_Compa_SI}
\end{figure}

We use an example to illustrate the defined scaling factor $\mathcal{F}(\alpha)$. Consider an epidemic follows the given infection profile $v = [0.148, 1.0,0.823, 0.426, 0.202, 0.078, 0.042, 0.057, 0.009]$, with the reproduction number $\mathcal{R} = 2.785$. We consider two testing-for-isolation strategies: testing and then isolating $100\%$ of the total population weekly, uniformly split over seven days\footnote{Testing and isolating the whole population on one day can prevent the virus from spreading. However, it is challenging to implement such strategy due to limited testing capacity. For instance, UIUC implemented a high testing rate, such that everyone was tested two or three times a week. Nevertheless, these tests at UIUC were still split over the week, as shown in Figure~\ref{fig:UIUC_T_SI}.
}, which gives us $\alpha_1 = 1/7$;  and testing and isolating $10\%$ of the total population weekly, uniformly split over seven days, which gives us 
$\alpha_2 = 0.1/7$. 
We apply these isolation rates on the infection profile $v$ using the proposed mechanism in~\eqref{eq_SI: Inf_Prof_Test}, and as a result, we obtain the following profiles
\begin{align*}
  v(\alpha_1)= [0.127, 0.735, 0.518, 0.223, 0.0935,
 0.031, 0.014,  0.017, 0.002],\\ v(\alpha_2)= [0.146, 0.972, 0.7882, 0.402, 0.188,
 0.072, 0.038, 0.051, 0.008]. 
\end{align*}
We plot the original infection profile and the two infection profiles under the isolation rates $\alpha_1 = 1/7$ and $\alpha_2 = 0.1/7$ in Figure~\ref{fig:In_Prof_Compa_SI}.
Additionally, we compute the reproduction number under the two different isolation rates,  resulting in $\mathcal{R}(\alpha_1) = 1.77$, and $\mathcal{R}(\alpha_2) = 2.66$. The corresponding scaling factors are  $\mathcal{F}(\alpha_1) = 0.64$, $\mathcal{F}(\alpha_2) = 0.95$, respectively. Consequently, testing and isolating the entire population weekly can reduce the reproduction number of the spreading process, as captured by the infection profile in~\eqref{eq_SI:SI_Dist} by $36\%$.  Similarly, a lower isolation rate, such as isolating $10\%$ of the population weekly, can reduce the  reproduction number of the spreading process by $5\%$. These findings demonstrate the impact of the isolation rate on the reproduction number. Next, we utilize the proposed mechanism to generate synthetic data, estimate reproduction number, evaluate the impact of different isolation rates, and further design feedback control mechanism, through taking into account the influence of the isolation rate.

\subsection*{SI-2-D. Estimation of The Reproduction Number}
\label{secSI:Reproduction_Num}
The key metric we utilize to generate synthetic data is the reproduction number. To accomplish our objective of reconstructing spreading processes at the UIUC and Purdue campuses, we present the methodology we employ to estimate the reproduction number from the confirmed cases on both campuses. The fundamental methodology we employ for estimating the  reproduction number is based on the approach outlined in~\cite{cori2013new,huisman2022estimation,Cori2022} using the \textit{Epyestim} package~\cite{Epyestim_Python_2020}
\footnote{https://github.com/lo-hfk/epyestim}, 
which is a Python package that builds upon the R package EpiEstim~\cite{cori2013new}.
As the development of methodologies for estimating the reproduction number is not the main focus of our work, we provide only a brief introduction of the core techniques developed in reference~\cite{cori2013new}.
Moreover, the techniques presented in~\cite{cori2013new} can be applied to infected cases without considering the incubation period and test delays that we have introduced. Therefore, to implement the estimation techniques described in~\cite{cori2013new}, the frameworks presented in~\cite{ huisman2022estimation} provide a means to preprocess confirmed cases to fit into the techniques described in~\cite{cori2013new}. 
Here, we introduce the general concepts of preprocessing confirmed cases for the estimation of the reproduction number from~\cite{huisman2022estimation} and~\cite{Epyestim_Python_2020}. For more detailed information and discussions on methods and challenges in estimating reproduction number, refer to references~\cite{cori2013new,thompson2019improved, nash2022real, bhatia2021generic, nash2022estimating,huisman2022estimation,Epyestim_Python_2020}.

\subsubsection*{SI-2-D-1. Pre-processing Confirmed Data}
One essential step for estimating the reproduction number via EpiEstim~\cite{Cori2022} involves preprocessing the confirmed data.  We can consider the estimation of the reproduction number from the confirmed cases (e.g., in Figures~\ref{fig:UIUC_P_SI} and ~\ref{fig:Purdue_Total_SI}) as a reverse process of generating synthetic data in Section~SI-2-B. In contrast to Section~SI-2-B, where we leverage the known reproduction number to generate synthetic confirmed data, we use real-world confirmed cases to estimate the reproduction number that generated such confirmed data.
Recall that, The Poisson process in~\eqref{eq_SI:Pos_mean} and~\eqref{eq_SI:Pos_dis} can generate infected data. However, However, real-world spreading data, including data from Figure~\ref{fig:UIUC_P_SI} and Figure~\ref{fig:Purdue_Total_SI} are reported as confirmed cases. Therefore, in order to obtain the reproduction number through~\eqref{eq_SI:Pos_mean} and \eqref{eq_SI:Pos_dis}, we need to preprocess the real-world spreading data to obtain the infected cases. As introduced in~\cite{huisman2022estimation}, we summarize the core data preprocessing procedures in terms of data smoothing and deconvolution.
To smooth the confirmed cases,~\cite{huisman2022estimation} implemented a local polynomial regression (LOESS) with first-order polynomials and tricubic weights. In our settings, we included a 21-day window of confirmed cases in the local neighborhood of each point as our smoothing parameter.  Further,~\cite{huisman2022estimation}extended the deconvolution method of~\cite{cori2013new, Cori2022, cauchemez2006estimating}, which is itself an adaptation of the Richardson-Lucy algorithm~\cite{richardson1972bayesian,lucy1974iterative}, to obtain the infected cases based on the confirmed cases under the delays from the incubation period and infection-to-confirmation delays. A more detailed discussion about data preprocessing can be found in~\cite{huisman2022estimation}.
\subsubsection*{SI-2-D-2. Bayesian Inference}
We utilize Bayesian inference to estimate the reproduction number, as proposed by~\cite{cori2013new, huisman2022estimation,gostic2020practical}. Again, we leverage the Python package~\textit{Epyestim}~\cite{Epyestim_Python_2020}\footnote{https://github.com/lo-hfk/epyestim, which is a realization of the R package \textit{EpiEstim CRAN} package~\cite{cori2013new}~(https://cran.r-project.org/web/packages/EpiEstim/index.html).} inspired  by~\cite{huisman2022estimation,cori2013new,Cori2022} to estimate the reproduction number from confirmed cases.
In this section, we introduce some key notations for the Bayesian inference framework. For more comprehensive details, please refer to~\cite{cori2013new, huisman2022estimation,gostic2020practical}. 
\baike{Since we consider no other intervention strategies other than the testing-for-isolation strategy,} the infection profile, which we employ in this work, is independent of calendar time and is solely influenced by the testing-for-isolation strategy. In Section~SI-2-B, the new infected cases are modeled as a Poisson process, where the number of new infected cases at time $t$ follows a Poisson distribution with a mean of 
$\mathcal{R}_t\sum_{s=1}^{t}I_{t-s}w_s$.
Hence, if the number of new infected cases at time $t$ is given by  $I_t$, the likelihood of the incidence $I_t$ given the unknown reproduction number $\mathcal{R}_t$, conditional on the previous infected cases $\{I_0,\dots, I_{t-1}\}$ is
\begin{equation}
P(I_t|I_0,\dots,I_{t-1},w,\mathcal{R}_t)=\frac{(\mathcal{R}_t\Lambda_t)^{I_t}e^{(-\mathcal{R}_t\Lambda_t)}}{I_t!},
\end{equation}
with $\Lambda_t=I_{t-s}w_s$, $s=1,\dots, t$~\cite{cori2013new}.
Note that in real-world spreading processes, the reproduction number usually varies over time. To describe the spreading process within a short time period, we can leverage the reproduction number during a sliding window of a certain time period.
Hence, we \baike{leverage an estimated} reproduction number $\mathcal{R}_{t,\tau}$ to represent the average reproduction number over a time period $[t-\tau+1, t]$. 
The likelihood of the new infected cases during the time period $[t-\tau+1, t]$, $\{I_{t-\tau+1},\dots, I_t\}$, given the reproduction number $\mathcal{R}_{t,\tau}$, conditional on the previous incidences $\{I_0,\dots,I_{t-\tau}\}$, is 
\begin{equation}
\label{eq_SI:Bay_Likely}
P(I_{t-\tau+1},\dots,I_t,|I_0,\dots,I_{t-1},w,\mathcal{R}_{t,\tau})=\prod^{t}_{s=t-\tau+1}\frac{(\mathcal{R}_t\Lambda_s)^{I_s}e^{(-\mathcal{R}_t\Lambda_s)}}{I_s!}.
\end{equation}
In the framework of Bayesian inference, the likelihood function in~\eqref{eq_SI:Bay_Likely} follows a Poisson distribution. Therefore, it is natural to consider choosing the prior distribution of $\mathcal{R}_t$  as the conjugate prior of a Poisson distribution, which is a Gamma distribution with parameters $(a,b)$. Using a posterior joint distribution of $\mathcal{R}_{t,\tau}$, the posterior joint distribution of $\mathcal{R}_{t,\tau}$ is given by~\cite{cori2013new, Cori2022}:
\begin{align*}
P(I_{t-\tau+1},\dots,I_t,\mathcal{R}_{t,\tau}|I_0,\dots,I_{t-1},w) =
\mathcal{R}_{t,\tau}^{a+\sum_{s=t-\tau+1}^t (I_s-1)}e^{-\mathcal{R}_{t,\tau}(\sum_{s=t-\tau+1}^t\Lambda_s+\frac{1}{b})}\prod^{t}_{s=t-\tau+1}\frac{\Lambda_s^{I_s}}{I_s!}\frac{1}{\Gamma(a)b^a},
\end{align*}
which is proportional to $ \mathcal{R}_{t,\tau}^{a+\sum_{s=t-\tau+1}^t (I_s-1)}e^{-\mathcal{R}_{t,\tau}(\sum_{s=t-\tau+1}^t\Lambda_s+\frac{1}{b})}\prod^{t}_{s=t-\tau+1}\frac{\Lambda_s^{I_s}}{I_s!}$.
Therefore, the posterior distribution of $\mathcal{R}_{t,\tau}$ is a Gamma distribution with parameters 
$(a+\sum_{s=t-\tau+1}^t I_s,\frac{1}{\sum_{s=t-\tau+1}^t\Lambda_s+\frac{1}{b}})$.
Further, the posterior mean of $\mathcal{R}_{t,\tau}$ is $(\frac{a+\sum_{s=t-\tau+1}^t I_s}{\frac{1}{\sum_{s=t-\tau+1}^t\Lambda_s+\frac{1}{b}}})$.  Again, for a detailed discussion of the Bayesian inference for the  reproduction number, we refer readers to~\cite{cori2013new, Cori2022, huisman2022estimation}. Additionally,~\cite{gostic2020practical} and~\cite{nash2022real} discuss the advantages and limitations of current widely-used estimation methods for the reproduction number. 
\subsubsection*{SI-2-D-3. Estimation of The Reproduction Number Over Campuses}
In this subsection, we combine the data pre-processing method we introduced from~\cite{huisman2022estimation} 
with Bayesian inference techniques to estimate the  reproduction number using confirmed cases from UIUC and Purdue. Both UIUC and Purdue implemented testing-for-isolation strategies, which require modifications to the initial serial interval distribution given in~\eqref{eq_SI:SI_Normal}. Specifically, for UIUC, where the daily isolation rate equals the testing rate, i.e., $\alpha_I = 2/7$  of the population (surveillance testing),  we can calculate the modified serial interval distribution for the spreading process at UIUC using the default infection profile from~\eqref{eq_SI:SI_Dist},~\eqref{eq_SI: Inf_Prof_Test}, and~\eqref{eq_SI:SI_Und_Test}. 
We denote this modified serial interval distribution as $w(\alpha_I)$. We use $w(\alpha_I)$ as the serial interval distribution to estimate the reproduction number at the UIUC campus. Compared to UIUC, Purdue encouraged symptomatic cases to be self-reported and tested (voluntary testing), while focused on testing around $10\%$ of the total population to identify asymptomatic cases (surveillance testing). Based on the testing data from Purdue,  \baike{we consider the following conditions when estimating the reproduction number\footnote{\baike{We perform sensitivity analysis on the ratio of the symptomatic infection $\theta$ and the isolation rate on our proposed methodologies.}}}:
\begin{itemize}
    \item The ratio of the symptomatic infection, $\theta = 0.55$.
    \item Symptomatic cases are expected to report and get tested and then to be isolated within a week of showing symptoms. Therefore, the isolation rate for symptomatic cases is $100\%/7$.
    \item For asymptomatic cases, around 30\% of the total asymptomatic cases were tested and isolated through the 10\% weekly non-uniformly testing. \baike{This condition is based on the fact that the 10\% testing excludes symptomatic and suspicious cases, and the contact-tracing network at Purdue makes the testing not uniformly random.}
    \item Both symptomatic and asymptomatic infections have the same infection profile given by~\eqref{eq_SI:SI_Dist}.  
\end{itemize}
Based on these conditions, the overall weekly isolation rate at Purdue is calculated as 
$100\%\times 0.55 + 30\%\times 0.45 = 68.5\%$ per week. 
Therefore, the daily isolation rate at Purdue is $\alpha_P=0.685/7$. Similarly, we can calculate the serial interval distribution of the spreading process at Purdue using~\eqref{eq_SI: Inf_Prof_Test} and~\eqref{eq_SI:SI_Und_Test}, denoted as~$w(\alpha_P)$. We use $w(\alpha_P)$ as the serial interval distribution to estimate the reproduction number at the Purdue campus.

\begin{figure}[p]
\centering
\includegraphics[width=\textwidth]{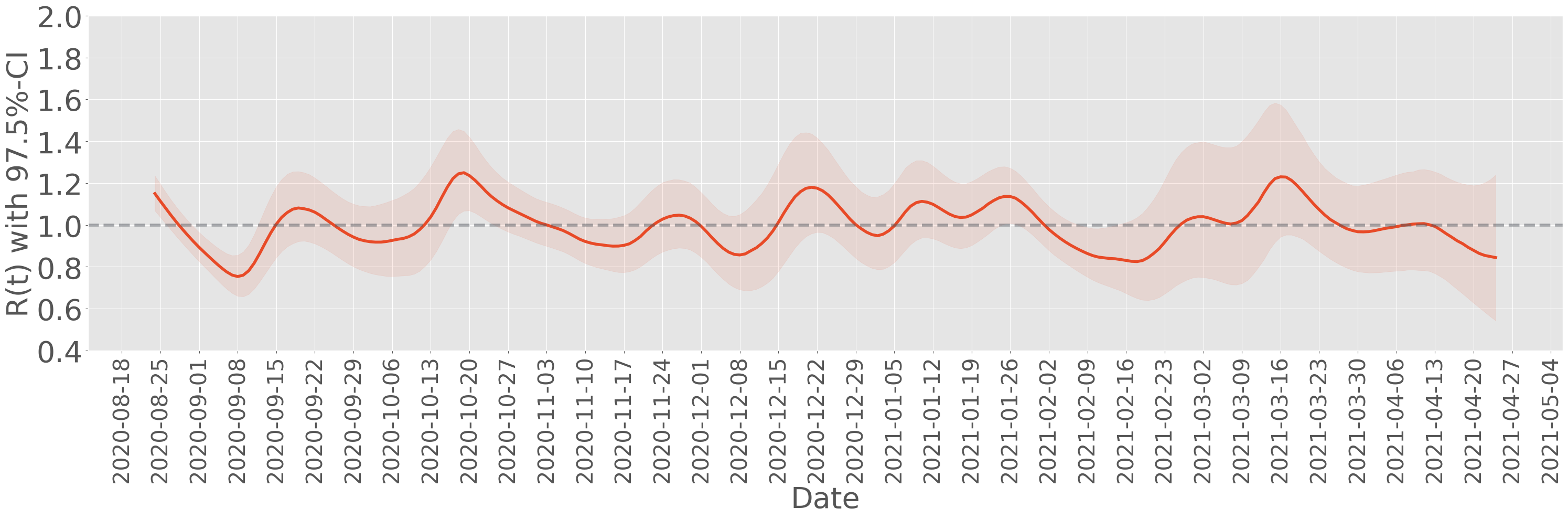}
\caption{The estimated reproduction number of the spreading process over the UIUC campus.}
\label{fig:Estimated_R_UIUC_SI}
\end{figure}
\begin{figure}[p]
\centering
\includegraphics[width=\textwidth]{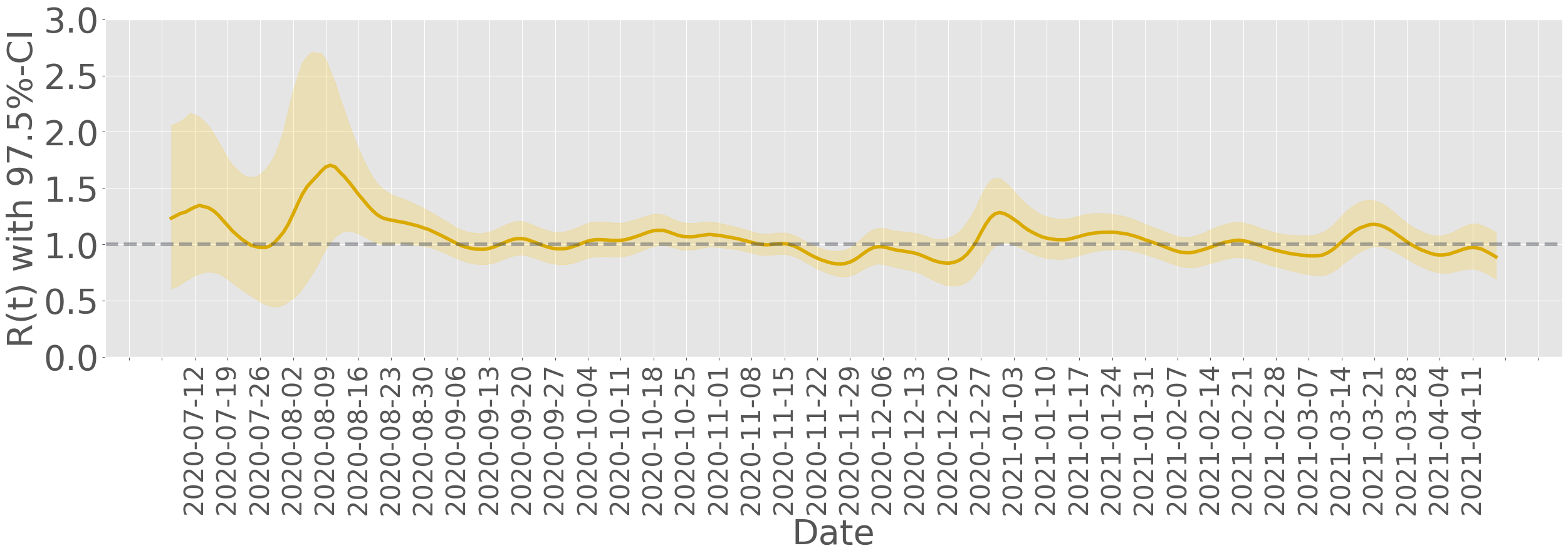}
\caption{The estimated reproduction number of the spreading process over the Purdue campus.}
\label{fig:Estimated_R_Purdue_SI}
\end{figure}

In the estimation process, we use a sliding window of 21 days to smooth the confirmed cases when estimating the reproduction number. We consider that the estimated reproduction number can affect new infections in the next 7 days, denoted as $\tau = 7$ in~\eqref{eq_SI:Bay_Likely}, \baike{such that the policy and population behavior remain unchanged within a short period}.
The estimated reproduction number of UIUC and Purdue are illustrated in Figure~\ref{fig:Estimated_R_UIUC_SI} and Figure~\ref{fig:Estimated_R_Purdue_SI}, respectively.  
In Figure~\ref{fig:Estimated_R_UIUC_SI} and Figure~\ref{fig:Estimated_R_Purdue_SI}, the solid lines represent the mean values of the estimated reproduction number, while the shadow areas represent the $97.5\%$ confidence intervals of the estimated reproduction number. Note that \baike{since we use an average $10.3$-day} infection-to-confirmation delay distribution from being infected to confirmed, there are lags between the estimated reproduction number and the confirmed cases. These lags capture the delays between the infected cases and the corresponding confirmed cases. 

For UIUC, the estimated reproduction number are greater than 1 for multiple periods, particularly at the beginning of Fall 2020, around the middle of October 2020, and the middle of Spring 2021. This estimation aligns with the observations from the confirmed cases at UIUC during Fall 2020 \baike{and Spring 2021}, where multiple mild spikes were observed. The larger spikes during Fall 2020 were caused by the entry-screening at the beginning of the semester and the gathering events during the return of the football season around the middle of October. \baike{The larger spike during Spring 2021 was caused by the entry-screening for students returning from the Spring break.}
Similarly, for Purdue, the estimated reproduction number are greater than 1 for multiple periods during Fall 2020 and Spring 2021. 
\baike{There were two major spikes observed in the estimated reproduction number at Purdue. One occurred around the beginning of August, and the other around the beginning of January. These two spikes describe the infection process during the summer and the Christmas break. Additionally, these spikes were reflected in the confirmed cases around 10 days later, corresponding to the beginning of the Fall 2020 and Spring 2021 semesters, respectively.}

These estimation results further highlight the importance of the infection-to-confirmation delay distribution. The spikes observed during the entry-screening at the beginning of the Fall 2020 and Spring 2021 semesters were due to infections that occurred \baike{during breaks off-campus.}
In addition to these spikes at Purdue, the estimated reproduction number also indicate that the reproduction number was greater than 1 during October 2020 and the middle of March 2021. The return of the college football season in October 2020 affected students at Purdue, leading to increased gatherings. Furthermore, the spike observed around the middle of March 2021 was caused by the Spring break at Purdue. After the Spring break, Purdue conducted another round of entry-screening for those who left campus.
In summary, the estimated reproduction number can reflect the spikes observed in the confirmed cases, and we can connect these spikes to real-world events. \baike{These analyses inspire us to use the reproduction number as an indicator (\baike{feedback}) to facilitate the design of the model-free epidemic mitigation framework that incorporates intervention strategy evaluation and
closed-loop feedback control.}

\subsubsection*{SI-2-D-4. Sensitivity Analysis on Estimating The Reproduction Number}
\begin{figure}[p]
\centering
\includegraphics[width=\textwidth]{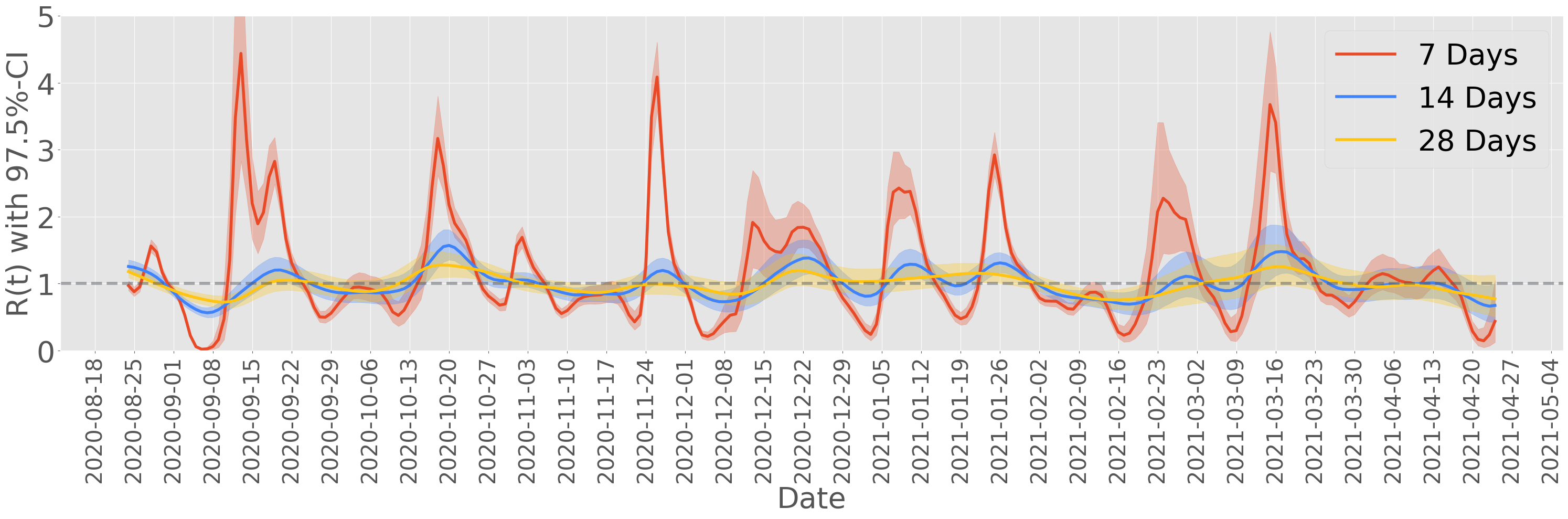}
\caption{Estimated reproduction number of the spreading process over the UIUC campus (\baike{7-, 14-, 28-day average data)}.}
\label{fig:Estimated_R_UIUC_SI_7}
\end{figure}

\begin{figure}[p]
\centering
\includegraphics[width=\textwidth]{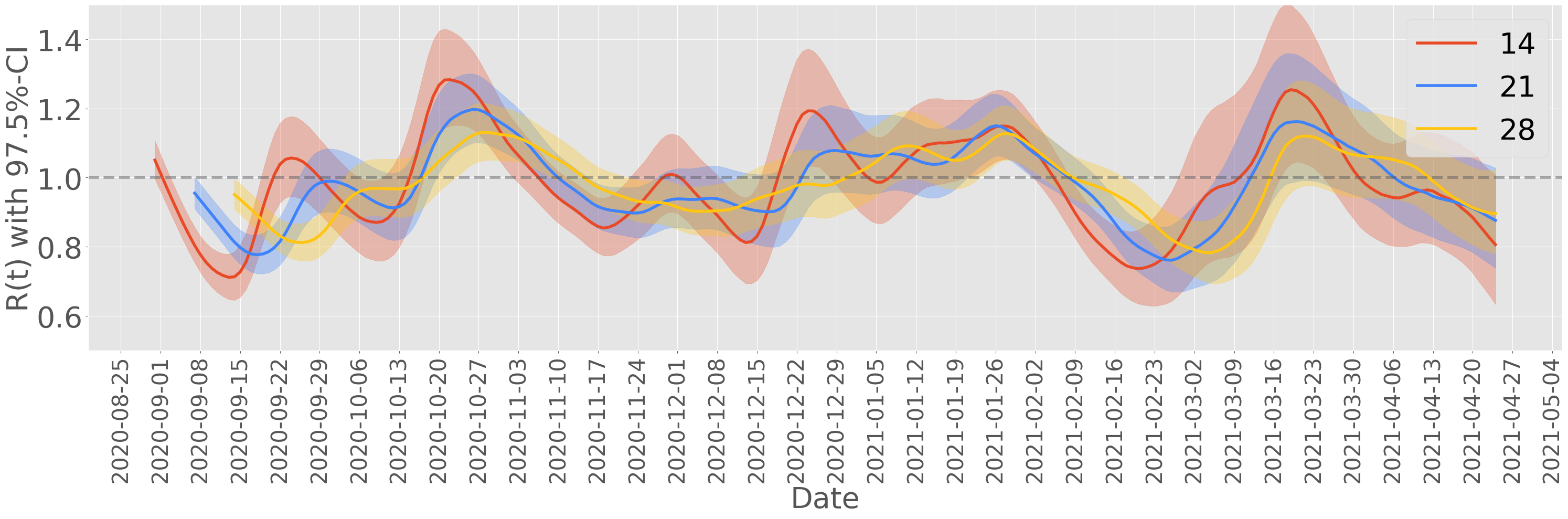}
\caption{Estimated reproduction number of the spreading process over the UIUC campus ($R_{t,\tau=14,21,28}$).}
\label{fig:Estimated_R_UIUC_SI_r_14}
\end{figure}

To further analyze the impact of the length of the smoothing windows on the estimation results, we use the confirmed cases from UIUC (Figure~\ref{fig:UIUC_P_SI}) as an example. \baike{Fixing $\tau = 7$,} we consider smoothing windows of  $7$, $14$, and $28$ days and plot the corresponding estimated reproduction numbers in Figure~\ref{fig:Estimated_R_UIUC_SI_7}.
From Figure~\ref{fig:Estimated_R_UIUC_SI_7},
we observe that a longer smoothing window generates a smoother estimated reproduction number, as it averages out more noise and variations in data collection. All three plots can capture the trend of the epidemic spreading in terms of spikes and plateaus. However, with longer smoothing windows, the estimated reproduction numbers become less sensitive to changes in the spreading behavior. 
For instance, when using a relatively shorter smoothing window, such as 7 days (\baike{the red line in} Figure~\ref{fig:Estimated_R_UIUC_SI_7}),  the estimated reproduction number is highly sensitive to changes in the confirmed cases.  In contrast, when using a longer smoothing window, such as 28 days (\baike{the golden line in} Figure~\ref{fig:Estimated_R_UIUC_SI_7}), the estimated reproduction number is less sensitive to changes in the confirmed cases. It is crucial to find a balance between sensitivity and responsiveness in the estimation process to ensure reliable pandemic evaluation and decision-making.  Therefore, in this work, we choose a 21-day smoothing window  (as shown in~\ref{fig:Estimated_R_UIUC_SI}) and $\tau=7$ in our estimation to capture the epidemic trend while maintaining sensitivity to changes in the spreading behavior.
In addition to analyzing the impact of the data smoothing window, we also study the impact of $\tau$ on the estimation results. We maintain a 21-day smoothing window and vary the value of $\tau$  to observe its effect. Specifically, we set $\tau$ as  14, 21, and 28, respectively, and plot the corresponding estimated reproduction number over the UIUC campus in Figure~\ref{fig:Estimated_R_UIUC_SI_r_14}.
From these plots, it is evident that as we increase the value of  $\tau$, representing a longer window for estimating the reproduction number, the estimation becomes less sensitive to changes. In other words, a longer $\tau$  results in a smoother estimated reproduction number with reduced sensitivity to short-term fluctuations. Additionally, considering the weekly nature of the confirmed testing data in~\ref{fig:UIUC_P_SI}, it is reasonable  to consider that the estimated reproduction number can exhibit weekly \baike{changes in} behavior rather than monthly. Hence, to capture the changing trends in the spread of the epidemic, we choose $\tau=7$ as the value for estimating the reproduction number. Note that the estimated reproduction number's accuracy is influenced by several factors, such as the estimation algorithm, the choice of distributions, and the data collection process, among others. More detailed discussions on this topic are given  in~\cite{cori2013new,Cori2022,gostic2020practical,parag2021improved, linka2020reproduction, abbott2020epinow2}.

\subsection*{SI-2-E. Reconstruction of Spreading Processes Over Campuses}
\label{secSI:Reconstruction}
In order to create an environment for validating the proposed framework, we reconstruct the spreading processes over both campuses using the mechanisms introduced in Section~SI-2-C  and the estimated reproduction number from both campuses in Section~SI-2-D. We consider two reconstruction scenarios. First, we reconstruct the spreading processes over both campuses under their implemented testing-for-isolation strategies using the methodologies introduced and proposed in Sections~SI-2-B, SI-2-C, and~SI-2-D. This approach integrates the estimation techniques and testing-for-isolation strategies discussed earlier. Next, we reconstruct the spreading process over both campuses under the hypothesis that the universities had not implemented their testing-for-isolation strategies during Fall 2020, aiming to conduct the counterfactual analysis of the potential outbreak. By comparing the reconstruction results between the two scenarios, we aim to evaluate the impact of the testing-for-isolation strategies on the dynamics of the spreading processes and emphasize the significance of implementing these strategies in mitigating the spread of the epidemic.
\subsubsection*{SI-2-E-1. Reconstructing the Original Spread}
\begin{figure}[p]
\centering
\includegraphics[width=\textwidth]{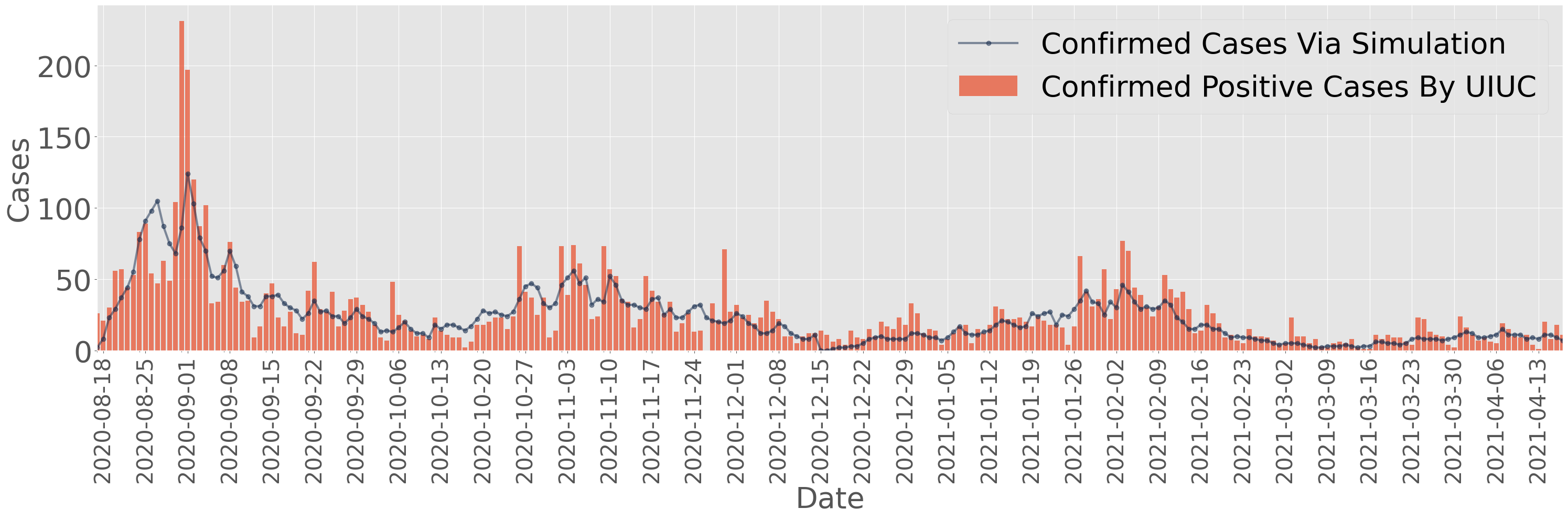}
\caption{Reconstructed spreading process over the UIUC campus.}
\label{fig:UIUC_Rec_SI}
\end{figure}

\begin{figure}[p]
\centering
\includegraphics[width=\textwidth]{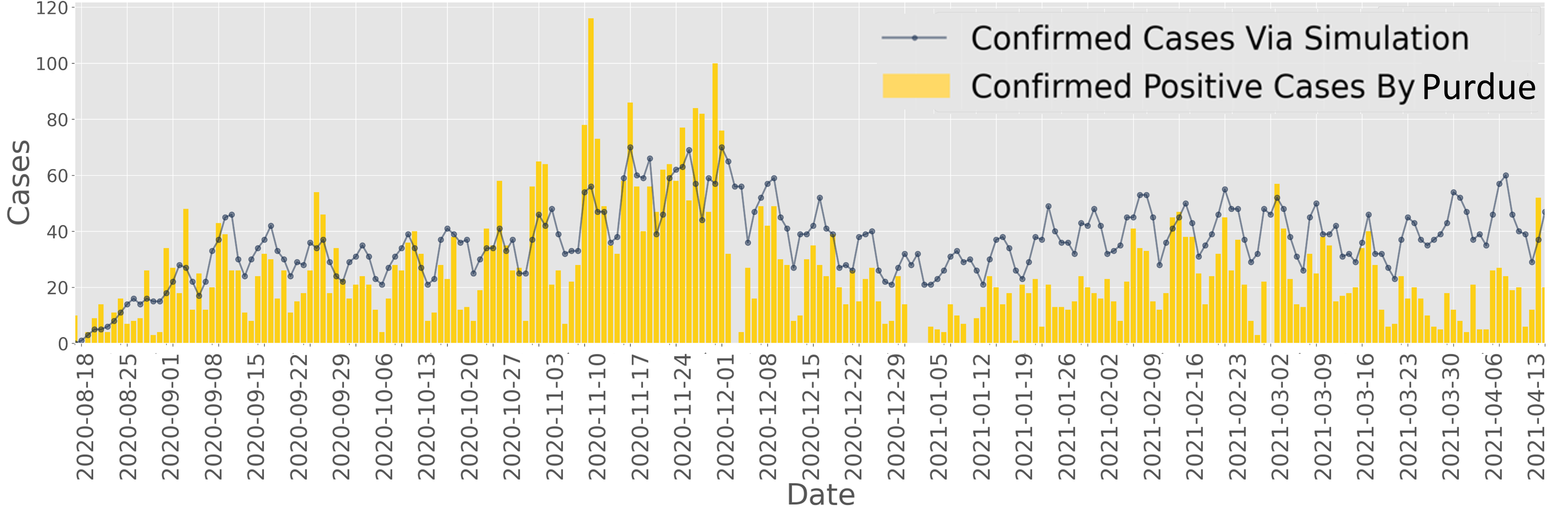}
\caption{Reconstructed spreading process over the Purdue campus.}
\label{fig:Purdue_Rec_SI}
\end{figure}

\baike{We utilize} the estimated reproduction number in \baike{Figure~\ref{fig:Estimated_R_UIUC_SI} and Figure~\ref{fig:Estimated_R_Purdue_SI}} in Section~SI-2-D along with the reconstruction mechanism introduced in Section~SI-2-B to generate synthetic data that aligns with the real-world confirmed cases on both campuses, as shown in Figures~\ref{fig:UIUC_P_SI} and~\ref{fig:Purdue_Total_SI}. 
We begin by reconstructing the spreading process over the UIUC campus.
We leverage the estimated  reproduction number shown in Figure~\ref{fig:Estimated_R_UIUC_SI}, 
 along with the modified serial interval distribution $w(\alpha_I)$ under the isolation rate $\alpha_I$, and the infection-to-confirmation delay distribution $\Delta$.   The resulting reconstruction is illustrated in Figure~\ref{fig:UIUC_Rec_SI} with a dotted solid line, which closely matches the confirmed cases observed over the UIUC campus during Fall 2020 and Spring 2021. Notably, the reconstruction accurately captures the spreading trend, including spikes and weekly confirmed patterns\footnote{\baike{In order to eliminate the impact of the winter break, when most people were not on campus, 
 we simulate the spreading process over Fall 2020 and Spring 2021 separately, since at the beginning of each semester, the entry-screening resets the spreading process. In addition, UIUC implemented different testing rates during Fall 2020 (two times a week) and Spring 2021 (three times a week).}}
We apply the same techniques to reconstruct the spreading process over the Purdue  campus. However, we utilize a different modified serial interval distribution, $w(\alpha_P)$, which is modified based on Purdue's testing-for-isolation strategy. We continue to use the same infection-to-confirmation delay distribution, $\Delta$, used for UIUC. Figure~\ref{fig:Purdue_Rec_SI}  illustrates the reconstruction (dotted solid line) that matches the confirmed cases observed on the Purdue campus during Fall 2020 and Spring 2021. Further, the reconstructed spreading process accurately captures the major spikes observed in the confirmed cases. Note that, in contrast to UIUC where we reset the initial condition at the beginning of the Spring 2021 semester, we did not reset the initial condition for Purdue during the same period. Therefore, the impact of the entry-screening at the beginning of Spring 2021 is not considered, resulting in an overestimation of the daily infected cases during the Spring 2021 semester at Purdue.

In summary, the reconstruction process relies on the estimated reproduction number, distributions, and parameters utilized during the estimation process. Different distributions and parameters may result in varied estimations of the reproduction number. However, when the conditions for estimation remain constant, the generation of synthetic data for reconstruction acts as an inverse process of the estimation, aligning the reconstructed spreading with the real-world spreading data.
\subsubsection*{SI-2-E-2. Evaluating the Spreading Process without the Isolation}
After reconstructing the spreading process under the implemented testing-for-isolation strategies over both campuses, we are motivated to study the scenario in which both universities had not implemented their testing-for-isolation strategies. This type of epidemic reconstruction process will also provide a foundation for evaluating 
different strength of interventions, including varying the isolation rate.
First, we propose a novel mechanism, i.e., reverse engineering of the reproduction number, to facilitate the reconstruction of the spread over campuses without the implemented testing-for-isolation strategies. \baike{Consider} that all the conditions remain exactly the same during the spreading period of interest, such as the spreading environment, other intervention policies, and population behaviors, among others. The only difference is that the spreading processes will be affected by a different isolation rate, including the zero isolation rate. 
To account for this difference, we can extend the reproduction number scaling mechanism we proposed in~\eqref{eq_SI:F} to compute $\mathcal{R}_t(\alpha=0)$ based on the estimated  reproduction number $\mathcal{R}_t(\alpha)$ under the implemented isolation rate $\alpha=\alpha_I$ or $\alpha=\alpha_P$. Note that besides using~\eqref{eq_SI:F} to compute the scaling factor, we can also compute $\mathcal{F}(\alpha)$ by using the serial interval distribution.
Recall that we denote the serial interval distribution of a spreading process as $w\in\mathbb{R}^n$. Then, we can define the pseudo serial interval distribution under the isolation rate $\alpha$ as
\begin{equation}
\label{eq_SI:Pseudo_SI_Dis}
   w^*(\alpha)=[w_1(1-\alpha), w_2(1-\alpha)^2, \cdots, w_n(1-\alpha)^n].  
\end{equation}
Note that $\sum_{i=1}^n {w_i^*(\alpha)} \leq 1$, since 
$\sum_{i=1}^n {w_i} = 1$. Hence, we call $ w^*(\alpha)$ a pseudo serial interval distribution instead of a serial interval distribution. we show that \eqref{eq_SI:Pseudo_SI_Dis} provides a way of
using $w^*(\alpha)$ to compute the impact of the isolation rate on the serial interval distribution directly, through the scaling factor $\mathcal{F}(\alpha)$. The connection is given by 
\begin{equation}
\label{eq_SI:F_w}
\mathcal{F}(\alpha) =\frac{\mathcal{R}(\alpha)}{\mathcal{R}}= \frac{\mathcal{R}\sum_{i=1}^n {w_i^*(\alpha)}}{\mathcal{R}\sum_{i=1}^n {w_i}} = \sum_{i=1}^n {w_i^*(\alpha)}.  
\end{equation}
Note that in \eqref{eq_SI:F_w}, we utilize the fact that $w(\alpha)\times \mathcal{R} =v(\alpha)$ and $\sum_{i=1}^n v_i(\alpha) = \mathcal{R}(\alpha)$.  
Therefore, even without the knowledge of the estimated reproduction number, we can still compute the impact of the isolation rate by using the scaling factor 
$\mathcal{F}(\alpha)$, given the serial interval distribution of the spreading process. The proposed method in~\eqref{eq_SI:F_w} provides a foundation for reconstructing the spreading process under different isolation rates, where we need to compute the scaling factor by directly scaling the default serial interval distribution $w$.

Again, consider that the epidemic spreading processes \baike{without any testing-for-isolation strategies} over both campuses follow the initial serial interval distribution $w$ as shown in Figure~\ref{fig:COVID19_In_Pro_SI} (Right). Consider that UIUC tested the entire campus twice a week during Fall 2020 and these tests were evenly distributed throughout the week, we calculate the daily isolation rate for UIUC as $\alpha_I = 2/7=0.28$. Furthermore, using~\eqref{eq_SI:F_w} we compute $\mathcal{F}(\alpha_I) = 0.389$.  Based on our analyses, it is natural to believe that the  reproduction number without the implemented testing-for-isolation strategies would be the estimated one given in Figure~\ref{fig:Estimated_R_UIUC_SI} divided by $\mathcal{F}(\alpha_I) = 0.389$.  
However, by directly following this procedure, we might overlook a critical factor. In Section~SI-2-B, \baike{the population size is sufficiently large} when generating infected cases, thereby neglecting the impact of the susceptible population size on the reproduction number. For densely populated university campuses, it is more realistic to consider a fixed population size during a semester. Hence, although the infection profile (serial interval distribution) is only affected by the isolation rate, on the exact same date, the differences in the susceptible population size will also have an impact when reverse engineering the reproduction number.
For this reason, we further scale the reconstructed  reproduction number
\begin{equation}
\label{eq_SI:R_rec}
    \hat{\mathcal{R}}_t = \frac{\mathcal{R}_t S(t)}{\mathcal{F}(\alpha)N},
\end{equation}
where $N\in \mathbb{N}_{\geq 0}$ is the number of the total population, $S(t)\in[0,N]$ is the number of the susceptible population and
at time step $t$, and $\mathcal{R}_t$ is the estimated reproduction number at time step $t$.

\begin{figure}[p]
\centering
\includegraphics[width=\textwidth]{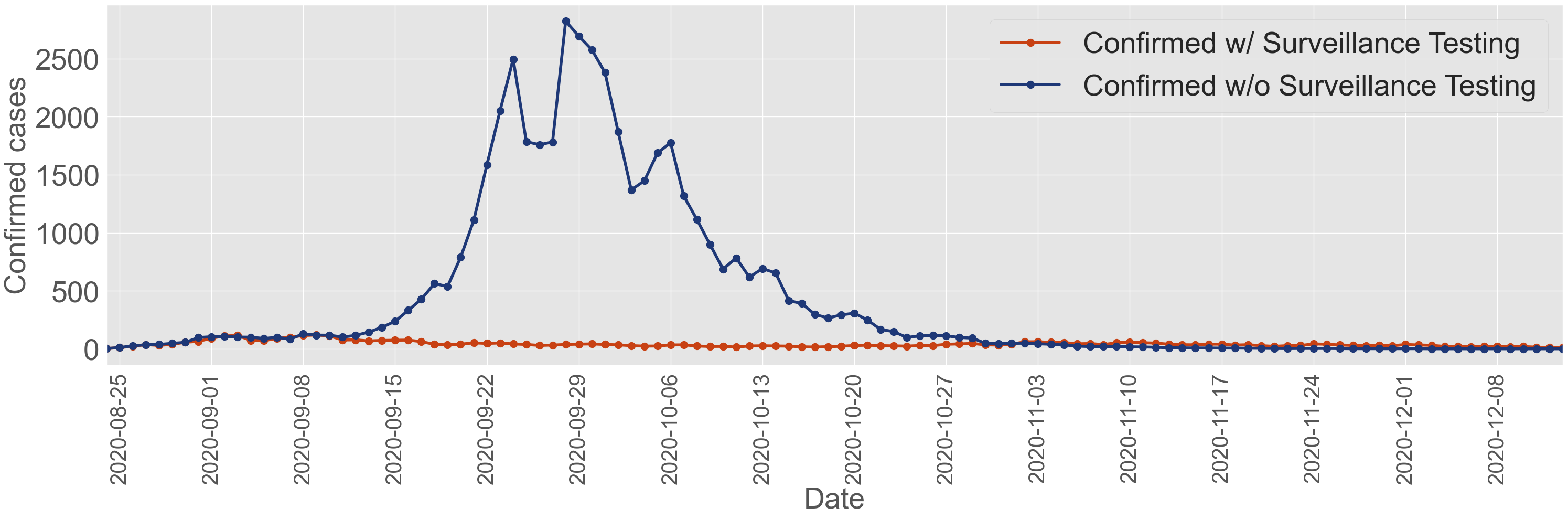}
\caption{Confirmed cases over the UIUC campus w/ and w/o testing-for-isolation strategies.}
\label{fig:UIUC_Rec_SI_w/o_SI}
\end{figure}

\begin{figure}[p]
\centering
\includegraphics[width=\textwidth]{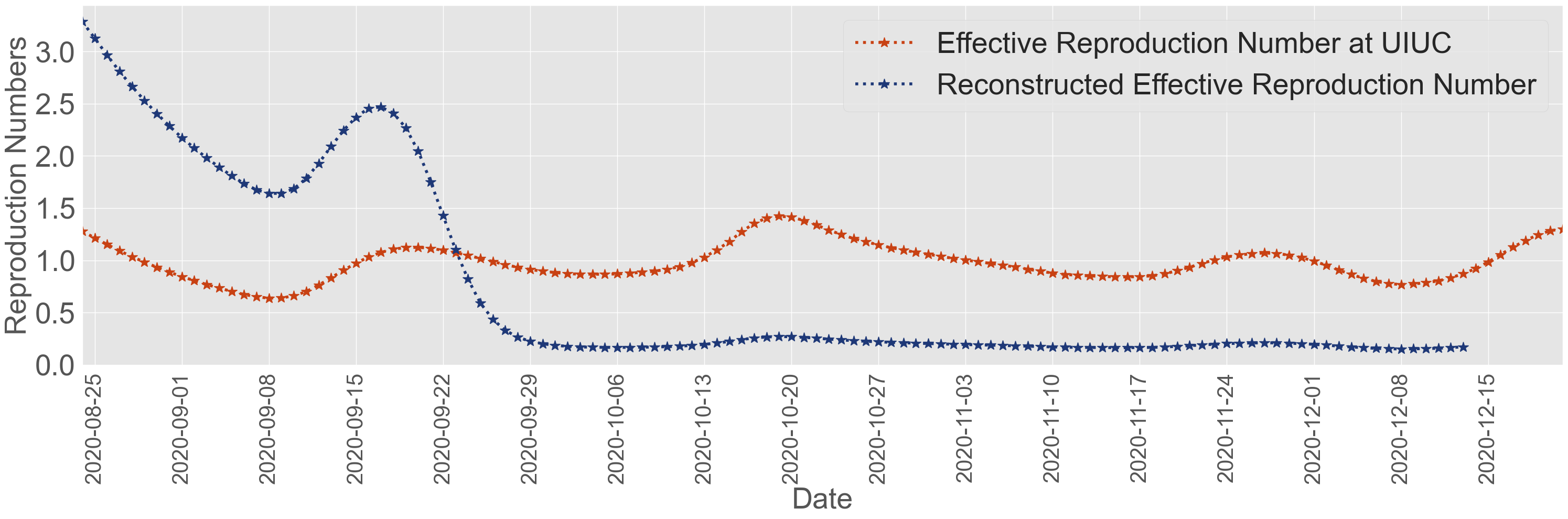}
\caption{The reproduction number at the UIUC campus w/ and w/o testing-for-isolation strategies.}
\label{fig:UIUC_R_100_SI}
\end{figure}

The total population on the UIUC campus during Fall 2020 is approximately $N=50,000$ and using the scaling factor $\mathcal{F}= \mathcal{F}(\alpha_I) = 0.389$, we reconstruct the spreading process by simulating the scenario where UIUC had not implemented any testing-for-isolation strategies. To capture the worst-case scenario, we explore a situation where the infected population behaves as if they were uninfected, either due to indifference or lack of awareness. This scenario is akin to assuming that all infected individuals are asymptomatic. In this scenario, we perform testing but no isolation, only recording the confirmed cases. Consequently, the infected population spreads the virus throughout the full infection period, adhering to the serial interval distribution represented by $w$ from Figure~\ref{fig:COVID19_In_Pro_SI} (Left). These parameters depict a potential worst-case scenario for COVID-19 spreading across the UIUC campus, as no infectious cases are isolated.
We generate the 'confirmed cases' without employing any testing-for-isolation strategies, as depicted in Figure~\ref{fig:UIUC_Rec_SI_w/o_SI} by the solid line in dark blue\footnote{These cases are confirmed positive through testing but no isolation actions are taken.}. Please note that although we label these instances as 'confirmed cases,' in the simulation, they are not actually isolated from the population. These cases are recorded by applying the same delay distributions to the infected cases, facilitating a comparison with the daily confirmed cases observed under the actual implemented testing-for-isolation strategy at UIUC, represented by the red solid line. In Figure~\ref{fig:UIUC_R_100_SI}, the corresponding reproduction number are presented. The marked line in dark 
blue illustrates the reproduction number of the reconstructed spreading process without the implementation of testing-for-isolation strategies. The red marked line represents the estimated reproduction number, as shown in Figure~\ref{fig:Estimated_R_UIUC_SI}, obtained from the data featuring the implemented testing-for-isolation strategy.

Note that Figure~\ref{fig:UIUC_Rec_SI_w/o_SI} illustrates that without any isolation strategies and \baike{under the condition} that the entire population on campus behaved as if they did not have COVID-19, the entire population on the UIUC campus would have been infected approximately two months after the start of the Fall 2020 semester. In comparison to the confirmed cases shown in Figure~\ref{fig:UIUC_P_SI}, where the peak number of confirmed cases was under 200, we observe a significant spike in daily confirmed cases, eventually exceeding 2,500 cases. In addition to studying the confirmed cases, based on the reproduction number in Figure~\ref{fig:UIUC_Rec_SI_w/o_SI}, we find that during the first month of the Fall 2020 semester, the reconstructed reproduction number without testing-for-isolation strategies is consistently higher than the estimated reproduction number obtained from real spreading data on campus. This phenomenon is caused by the fact that in the real-world scenario, the implementation of testing-for-isolation strategies, specifically testing everyone then isolating all confirmed cases twice a week, reshaped the infection profile of COVID-19 spreading at UIUC. Therefore, without the testing-for-isolation strategy, according to~\eqref{eq_SI:F_w}, the reconstructed reproduction number would have been scaled up. However, after mid-September 2020, due to the fact of a fixed total population on campus ($N=50,000$), the lack of a sufficient susceptible population outweighs the impact of the scaling factor from the testing-for-isolation strategies, as captured by the term $S(t)/N$ in~\eqref{eq_SI:F_w}. As a result, the reconstructed population of confirmed cases in Figure~\ref{fig:UIUC_Rec_SI_w/o_SI} starts to decrease, along with the reconstructed reproduction number becoming smaller than 1 in Figure~\ref{fig:UIUC_R_100_SI}. This situation can be explained by considering that the pandemic reaches the point of herd immunity, where the number of new confirmed cases naturally decreases due to a sufficiently small susceptible population.

 We apply the same reconstruction mechanism (\eqref{eq_SI:F_w}) to reconstruct the spread over the Purdue campus without the surveillance testing-for-isolation strategy, i.e., we implement surveillance testing but the tested positive cases take no actions. In reality, Purdue implemented a surveillance testing-for-isolation strategy by sampling approximately 50,000 individuals on campus each week. However, in this case, we consider that there was no isolation for asymptomatic cases under the surveillance testing, and symptomatic cases would be tested and isolated at a daily testing rate of $\alpha_s =1/7$ \baike{through voluntary testing}. Although the surveillance testing rate $\alpha_s =0.1/7$, 
 we acknowledge that testing may have been influenced by contact tracing strategies and population behaviors, as introduced in Section~SI-1-B. 
 Consequently, the testing rate will be biased and may not precisely reflect 10\% per week as intended. Hence, to align with our reproduction number estimation process, we consider 30\% isolation rate of the surveillance testing, indicating that the 10\%  surveillance testing caught at least 30\% asymptomatic infections at Purdue.

\begin{figure}[p]
\centering
\includegraphics[width=\textwidth]{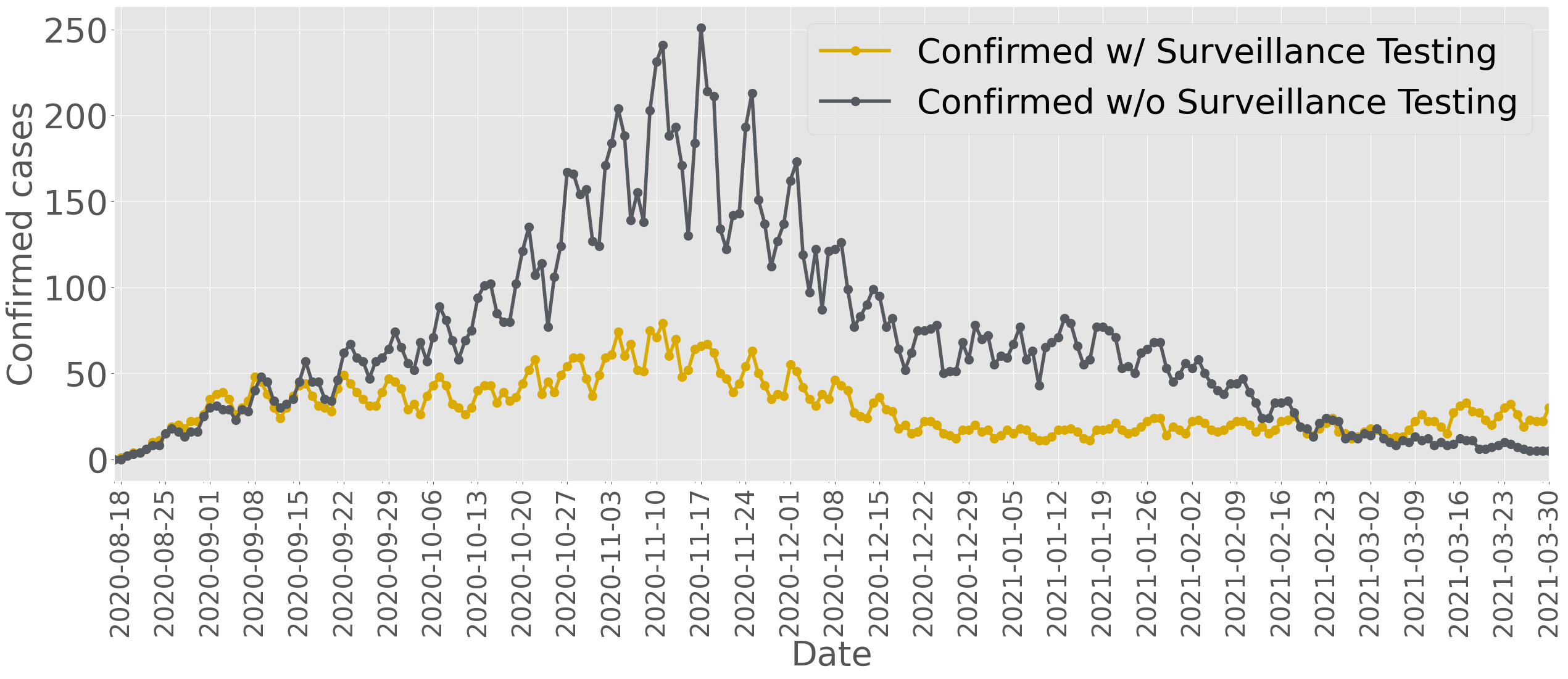}
\caption{Confirmed cases at the Purdue campus w/ and w/o  isolation under the surveillance testing.}
\label{fig:Purdue_0.3_SI}
\end{figure}

\begin{figure}[p]
\centering
\includegraphics[width=\textwidth]{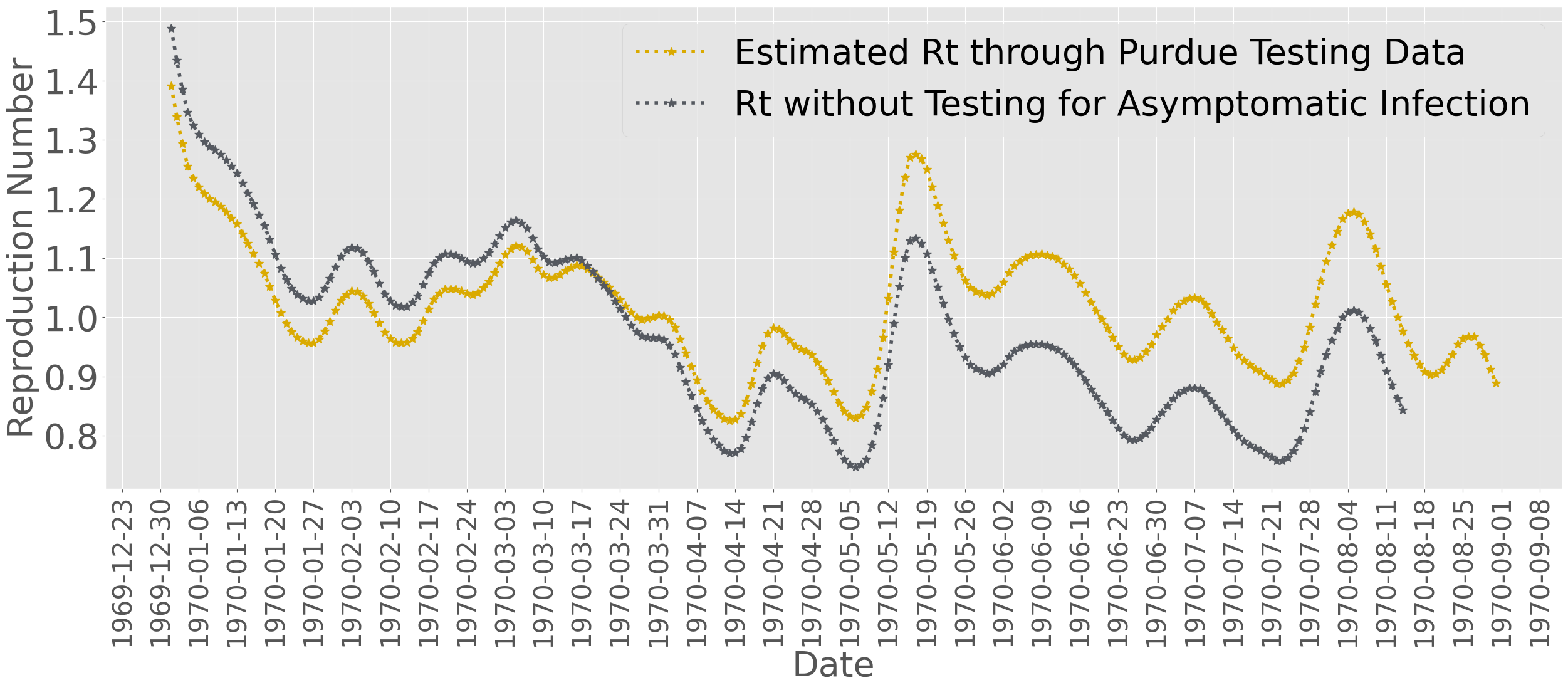}
\caption{The reproduction number at the Purdue campus w/ and w/o isolation under the surveillance testing.}
\label{fig:Purdue_0.3_SI_R}
\end{figure}

Figure~\ref{fig:Purdue_0.3_SI} illustrates that without any isolation under the surveillance testing and \baike{under the condition} that symptomatic infections are isolated through voluntary testing, the two 
outbreaks at Purdue during Fall 2020 and Spring 2021 will be higher. The peak infection values are nearly doubled. Compared to the reconstruction process at UIUC, the outbreak at Purdue is less severe since all symptomatic cases are caught and isolated through the voluntary testing strategy. 
In addition to studying the confirmed cases, based on the reproduction number in Figure~\ref{fig:Purdue_0.3_SI_R}, we find that during the first two months of the Fall 2020 semester, the reconstructed reproduction number without the isolation under surveillance testing are consistently higher than the estimated reproduction number obtained from real spreading data on campus.
Similar to UIUC, 
after October 2020, the fixed total population on campus ($N=50,000$) that results in the lack of a sufficient susceptible population outweighs the impact of the scaling factor from the testing-for-isolation strategies, as captured by the term $S(t)/N$ in~\eqref{eq_SI:F_w}. As a result, the reconstructed population of confirmed cases in Figure~\ref{fig:Purdue_0.3_SI}
starts to decrease, along with the reconstructed reproduction number becoming smaller than 1 in Figure~\ref{fig:Purdue_0.3_SI_R}.

The results illustrate that
the COVID-19 pandemic could have been far worse on both campuses, in terms of huge peak infection values, which could have collapsed the health-care system, and a larger infected population, which  would have forced the schools to close. Therefore, in order to safely operate smart communities such as large universities during a pandemic, it is critical to implement testing-for-isolation strategies with a sufficiently large testing/isolation rate. \baike{Note that we reconstruct the possible worst-case scenario at the UIUC campus. For the Purdue campus, we consider a $30\%$ isolation rate for asymptomatic cases. Further, all confirmed case via voluntary testing will isolate themselves. 
The reconstruction processes are heavily determined by these conditions. Since we have no further information on how confirmed cases behaved and the exact isolation rate of the testing strategy, we provide possible scenarios to primarily validate our counterfactual analysis in Figure~\ref{fig:Control_Framework_SI}. 
There could be other scenarios in the reconstruction process under different conditions. We further discuss
more reconstructions of the spreading process over UIUC and Purdue campuses under different scenarios.}

\subsubsection*{SI-2-E-3. Sensitivity analysis on reconstruction and evaluation}
The reconstruction and evaluation on the UIUC campus considers the worst-case scenario where 
all confirmed cases in Figure~\ref{fig:UIUC_P_SI}
isolated themselves after being caught by the testing, and
none of these confirmed cases will isolate themselves from the population without the testing-for-isolation strategy. Here, the sensitivity analysis on the reconstruction at the UIUC campus studies what would happen without the isolation
if $\alpha_I$ confirmed cases had isolated themselves from the total population, where $\alpha_I\in\{10\%, 20\%,\dots, 100\%, 120\%, \dots, 200\%\}$. Based on \eqref{eq_SI:R_rec}, we obtain the reconstructed reproduction number of UIUC through the scaling factor $\mathcal{F}(\alpha_I)$
\begin{equation}
\label{eq_SI:UIUC_RE_S}
\hat{\mathcal{R}}_t = \frac{\mathcal{R}_t S(t)}{\mathcal{F}(\alpha_I)N}
 = \frac{S(t)\sum^{n}_{i=1}v_i}{N\sum^{n}_{i=1}(v_i(\alpha_I))}.
\end{equation}

\begin{figure}[p]
\centering
\includegraphics[width=\textwidth]{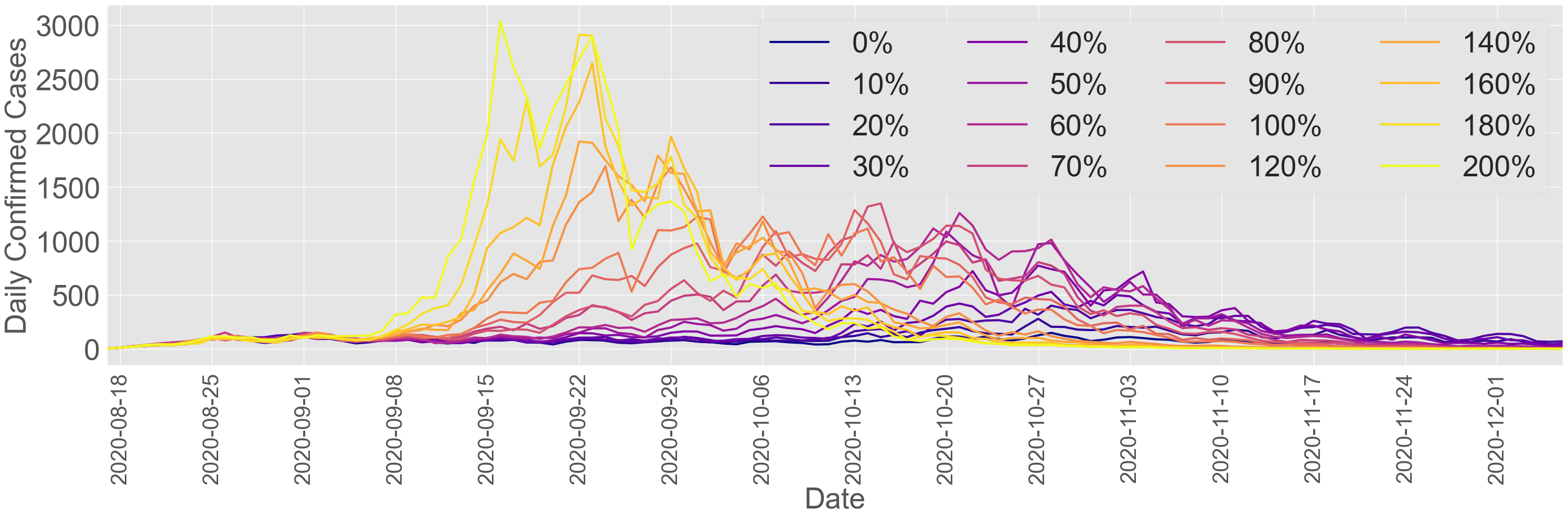}
\caption{Daily confirmed cases at the UIUC campus w/ different proportions of following the isolation rules.}
\label{fig:UIUC_I_S_SI}
\end{figure}

According to~\eqref{eq_SI:UIUC_RE_S},
a lower scaling factor $\mathcal{F}(\alpha_I)$ results in a higher reconstructed reproduction number. Furthermore, in real-world spreading processes, a higher isolation rate $\alpha_I$ leads to a higher $\hat{\mathcal{R}}_t$. 
For instance, with a $200\%$ testing rate at UIUC, only $\alpha_I = 100\%$ of the confirmed cases isolated themselves from the population, reflecting a $100\%$ isolation rate. Consequently, without the testing-for-isolation strategy, the remaining $100\%$ of confirmed cases not isolated from the population remain unaffected. The testing-for-isolation influences only half of the confirmed cases.
In an extreme scenario where $\alpha_I = 0\%$ of confirmed cases isolated themselves, Figure~\ref{fig:UIUC_P_SI} would record only the infected cases at UIUC under the testing strategy.
Hence, under this situation, even without the testing strategy, the infected cases would not change.
Based on this concept, 
Figure~\ref{fig:UIUC_I_S_SI} captures scenarios for $\alpha_I\in\{10\%, 20\%, \dots, 100\%, 120\%, \dots, 200\%\}$ illustrating what would occur if the testing-for-isolation strategy had not been implemented. 
Figure~\ref{fig:UIUC_I_S_SI}  illustrates that the more confirmed cases follow the isolation strategy upon testing positive during Fall 2020 and Spring 2021, the more severe the spread would be without the testing-for-isolation strategy.
When $\alpha_I = 200\%$ (where all confirmed cases follow the isolation rule), the reconstruction generates the worst-case scenario seen in Figure~\ref{fig:UIUC_Rec_SI_w/o_SI}.

\begin{figure}[p]
\centering
\includegraphics[width=\textwidth]{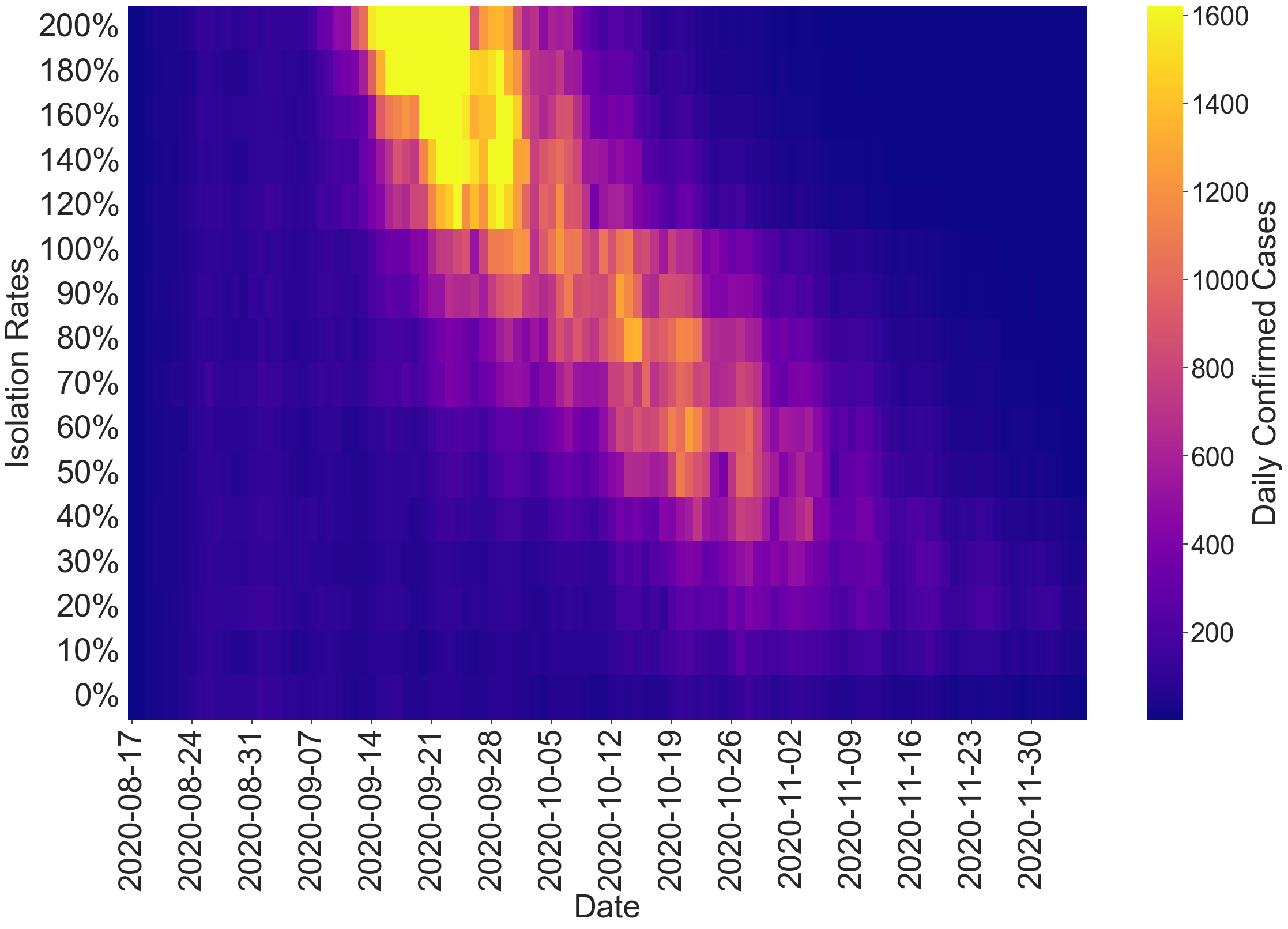}
\caption{Daily confirmed cases at the UIUC campus w/ different proportions of following the isolation rules.}
\label{fig:UIUC_C_Heatmap_SI}
\end{figure}

\begin{figure}[p]
\centering
\includegraphics[width=\textwidth]{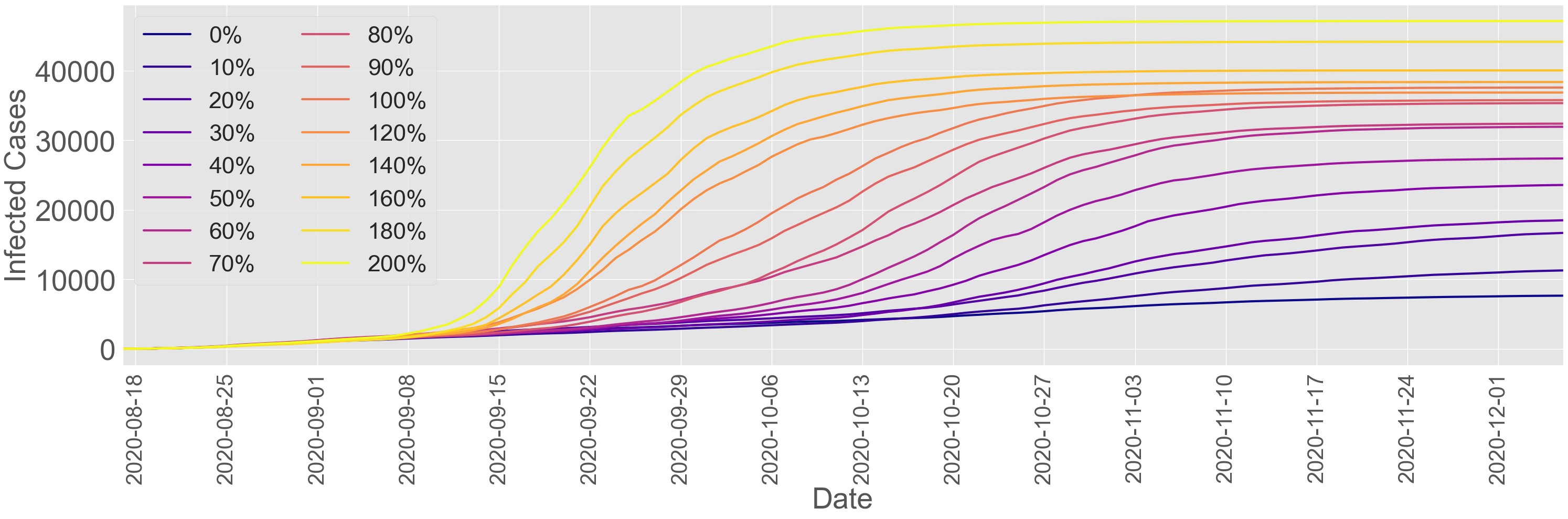}
\caption{Confirmed cumulative cases at the UIUC campus w/ different proportions of following the isolation rules.}
\label{fig:UIUC_C_total_SI}
\end{figure}

\begin{figure}[p]
\centering
\includegraphics[width=\textwidth]{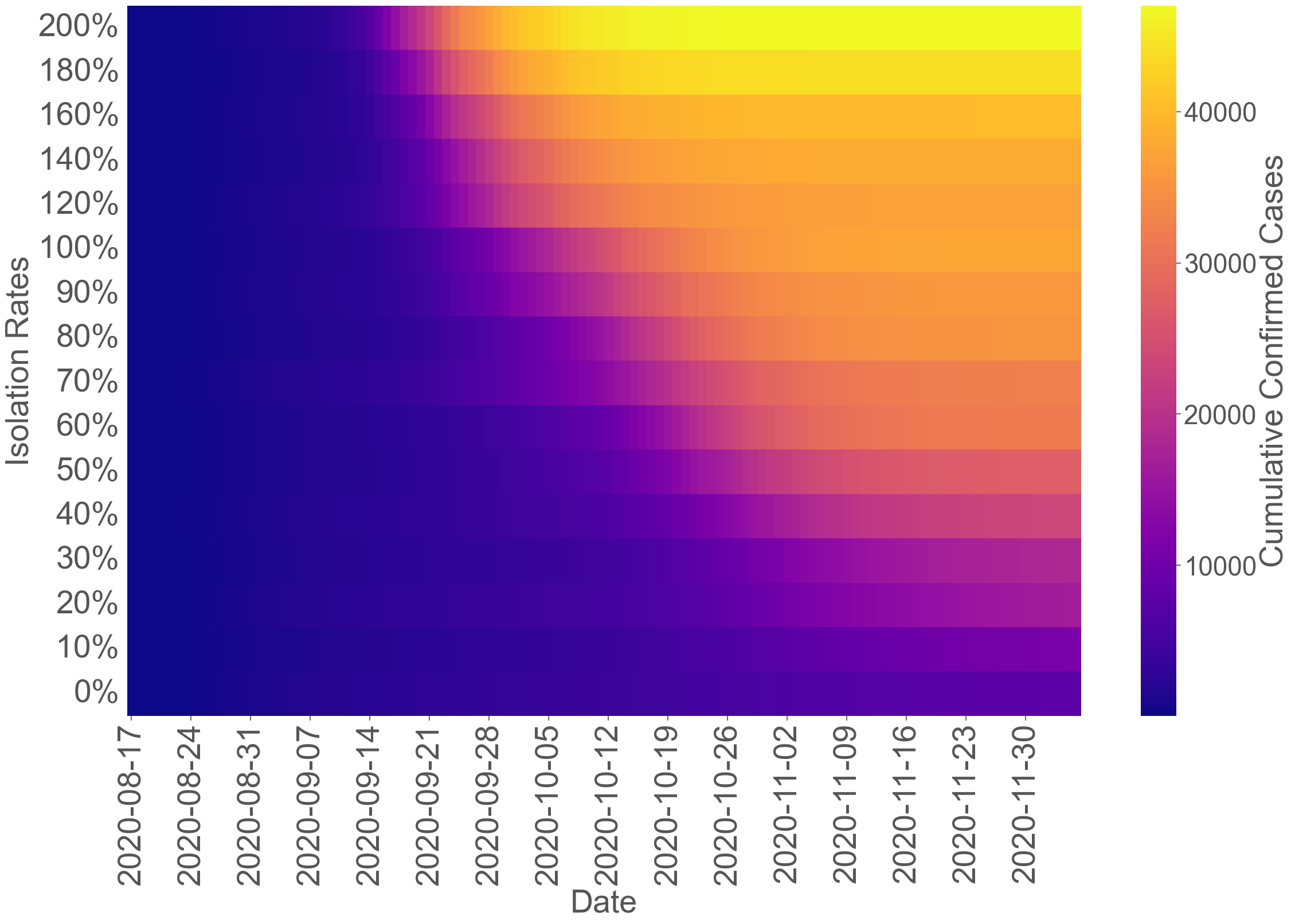}
\caption{Confirmed cumulative cases at the UIUC campus w/ different proportions of following the isolation rules.}
\label{fig:UIUC_T_SIotal_Heat_Map}
\end{figure}

We plot the heatmap of the daily confirmed cases in Figure~\ref{fig:UIUC_C_Heatmap_SI}, the cumulative confirmed cases in Figures~\ref{fig:UIUC_C_total_SI} and~\ref{fig:UIUC_T_SIotal_Heat_Map}. To summarize, our counterfactual analysis of the epidemic process at UIUC highlights the importance of understanding population behavior in modeling the epidemic spreading process. Analyzing the reconstructed spreading processes over the UIUC campus enables us to quantify the impact of the isolation rate on the spread. Note that not all individuals on campus adhere strictly to the testing-for-isolation strategy. It is common to observe confirmed positive cases not adhering to the isolation guidelines and continuing to spread the virus. Hence, it becomes crucial to include the study of population behavior in epidemic modeling and the reconstruction process. Considering the various aspects of population behavior enhances the accuracy and reliability of our models and reconstructions, leading to more effective strategies for controlling and managing epidemics.



\begin{figure}[p]
\centering
\includegraphics[width=\textwidth]{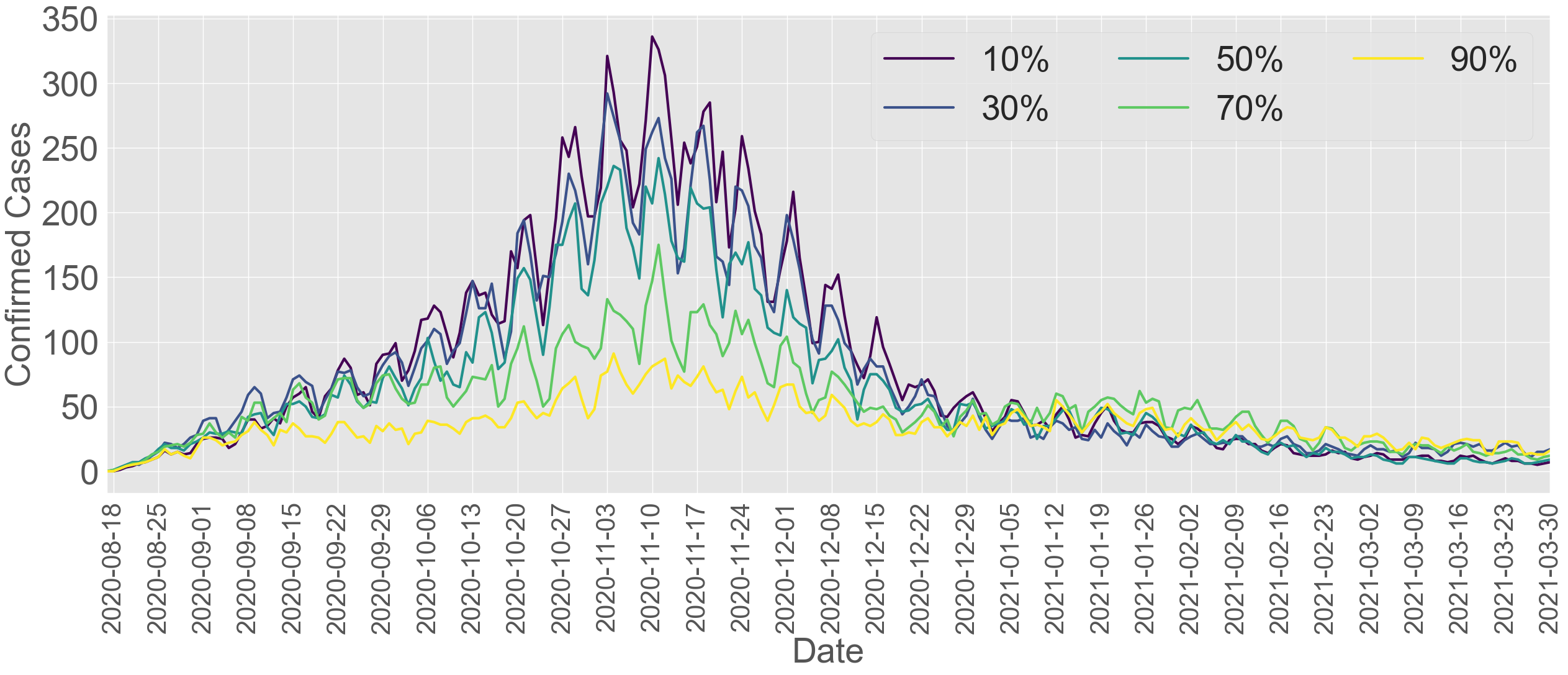}
\caption{Daily confirmed cases over the Purdue campus w/ different proportions of symptomatic cases}
\label{fig:Purdue_Asym_Daily_SI}
\end{figure}

\begin{figure}[p]
\centering
\includegraphics[width=\textwidth]{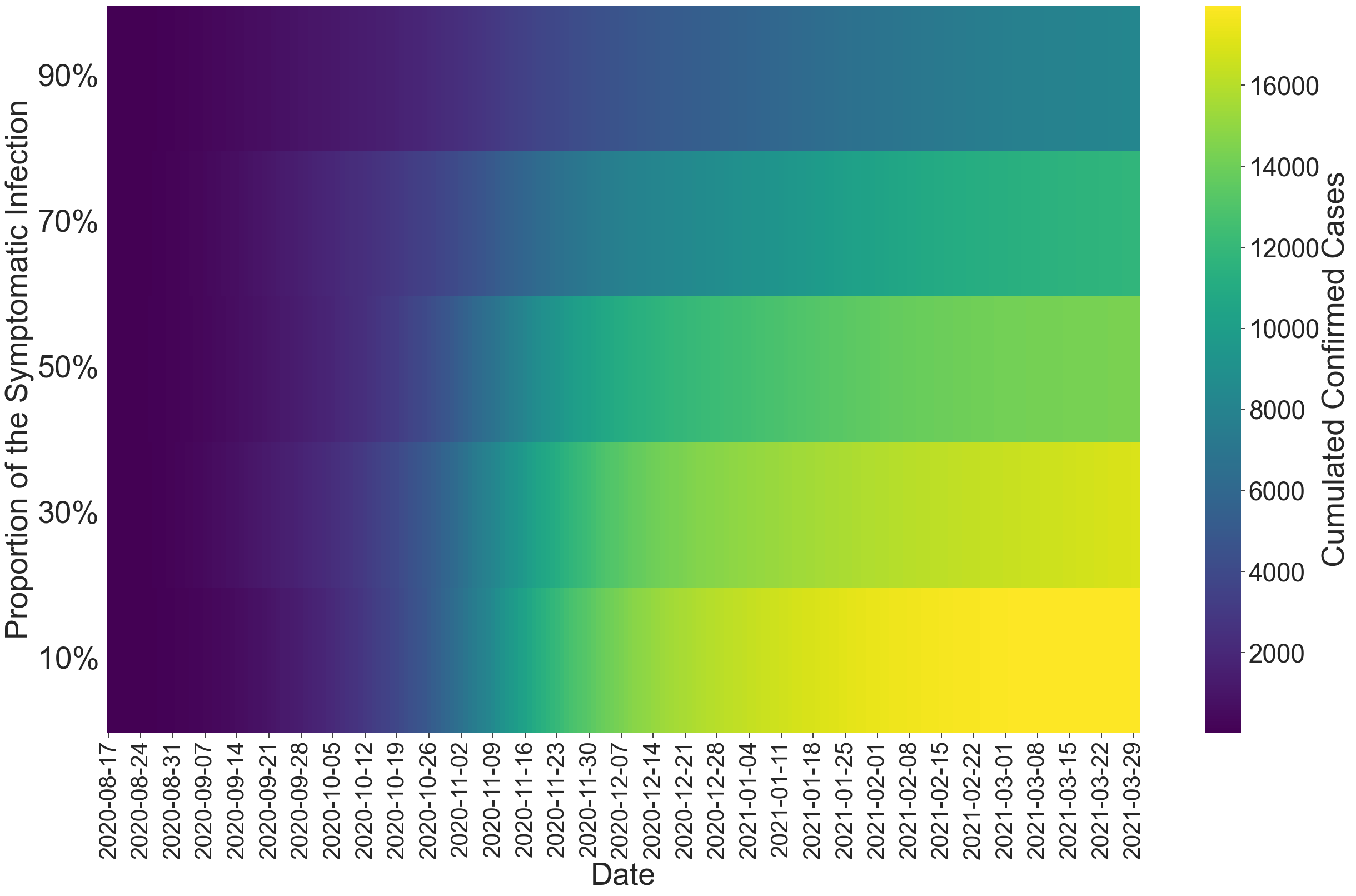}
\caption{Cumulative confirmed cases over the Purdue campus w/ different proportions of symptomatic cases: A heat map}
\label{fig:Purdue_Asym_Cumu_SI_Heat}
\end{figure}

When examining the impact of population behavior at the UIUC campus, we did not differentiate between symptomatic and asymptomatic infections. However, Purdue University employed distinct testing methods for symptomatic and asymptomatic cases, with voluntary testing for symptomatic infections and surveillance testing for asymptomatic ones. 
It becomes necessary to explore the effect of the symptomatic ratio $\theta$ on the counterfactual analysis of the reconstruction. Public health data indicates an average symptomatic ratio of $124.0M/144.6M = 0.85$ from February 2020 to September 2021 in the United States~\cite{CDC_sym_2021}.
Purdue categorizes cases identified by voluntary testing as symptomatic infections and those detected through surveillance testing as asymptomatic infections. Consequently, since not all symptomatic cases underwent voluntary testing, some entered the surveillance testing pool.
Hence, Purdue's recorded symptomatic ratio differs at 55\% compared to the public health record. To assess the impact of the symptomatic ratio on the epidemic reconstruction under Purdue's implemented testing-for-isolation strategy, we adjusted $\theta$ within the range of $\{10\%, 30\%,  50\%, 70\%, 90\% \}$.
In this scenario, we vary the ratio of the symptomatic infection while keeping the same isolation rate of 30\% for asymptomatic cases under surveillance testing, the same as in estimating the reproduction number and reconstructing the spreading process at the Purdue campus.
Hence, based on~\eqref{eq_SI:F_w},
the scaling factor is given by 
\begin{equation}
\label{eq_SI:F_P_Asym}
\mathcal{F}(\alpha_P) = \frac{S(t)(\theta \sum_{i=1}^n\underline{v}_i(\underline{\alpha}_P=1/7) +(1-\theta) \sum_{i=1}^n\overline{v}_i(\overline{\alpha}_P=0.3/7))}{N\sum_{i=1}^nv_i},
\end{equation}
where $\alpha_P$
is the overall isolation rate from the isolation rates under voluntary testing $(\underline{\alpha}_P=1/7)$ and surveillance testing $(\overline{\alpha}_P=0.3/7)$, respectively.
According to~\eqref{eq_SI:F_P_Asym}, a higher symptomatic ratio ($\theta$) results in a higher scaling factor $\mathcal{F}(\alpha_P)$, which leads to a lower reconstructed reproduction number. Therefore, we explore various ratios for symptomatic infections ($\theta$) from the set $\{10\%, 30\%, \dots, 90\% \}$. 
The reconstructed daily confirmed cases under different symptomatic infection ratios ($\theta$)
at the Purdue campus are illustrated in Figure~\ref{fig:Purdue_Asym_Daily_SI}, along with the corresponding heatmap in Figure~\ref{fig:Purdue_Asym_Cumu_SI_Heat}. Additionally, the cumulative confirmed cases under various symptomatic infection ratios at the Purdue campus are shown in Figure~\ref{fig:Purdue_Asym_Cumu_SI_Heat}, with the corresponding heatmap in Figure~\ref{fig:Purdue_Asym_Cumu_SI_Heat}.

\begin{figure}[p]
\centering
\includegraphics[width=\textwidth]{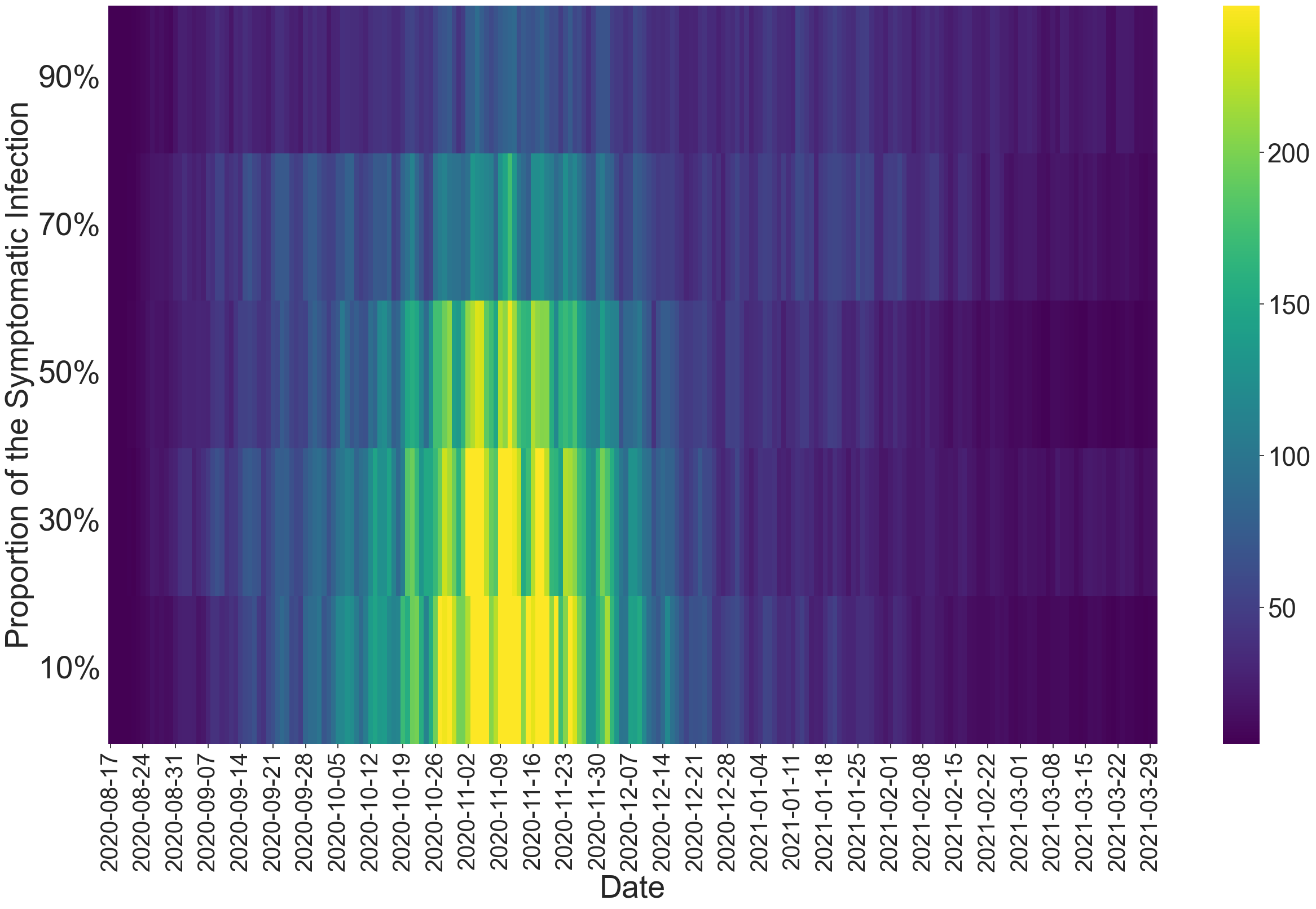}
\caption{Daily confirmed  cases over the Purdue campus w/ different proportions of symptomatic cases: A heat map}
\label{fig:Purdue_Asym_Daily_SI_Heat}
\end{figure}

Figures~\ref{fig:Purdue_Asym_Daily_SI}
and~\ref{fig:Purdue_Asym_Daily_SI_Heat}  demonstrate projections based on the scenario where Purdue's surveillance testing-for-isolation strategy isolated 30\% of asymptomatic cases.  The outbreaks appeared to be more severe when lower proportions of symptomatic infections were considered. This is reflected in the higher peak infection values and cumulative confirmed cases. 
 our analysis assumes that all symptomatic infections were subjected to the testing-for-isolation strategy, tested, and isolated through voluntary testing. Meanwhile, $\overline{\alpha}_P = 30\%$ of asymptomatic infections were tested and isolated through surveillance testing.
 Therefore, the severity of the outbreaks shown in Figure~\ref{fig:Purdue_Total_SI} 
 depended on the proportion of confirmed cases that were symptomatic or asymptomatic. 
 For instance, a higher proportion of confirmed symptomatic cases (captured by the $\theta= 90\%$ line in Figure~\ref{fig:Purdue_Asym_Daily_SI}) resulted in milder outbreaks compared to scenarios with a higher proportion of asymptomatic cases (as depicted by the $\theta= 10\%$ line in Figure~\ref{fig:Purdue_Asym_Daily_SI}). This outcome correlates with the condition that a higher symptomatic infection ratio generates a higher scaling factor in~\eqref{eq_SI:F_P_Asym}, leading to less severe outbreaks when the surveillance testing-for-isolation strategy is not in place during our reconstruction process. 

In summary, considering the testing-for-isolation strategies implemented by UIUC and Purdue, which aim to manage infected cases and prevent significant infection spikes (as shown in
\baike{Figures}~\ref{fig:UIUC_C_total_SI} and~\ref{fig:Purdue_Total_SI} ), we conclude that without these strategies, the epidemic would have been worse.
It is critical to emphasize again that the impact of the testing-for-isolation strategy is dependent on the isolation rate.  Solely conducting testing without encouraging isolation would make little difference, as testing can only detect infectious cases. Fortunately, the isolation rate is typically proportional to the testing rate due to other policies and shifts in behavior resulting from heightened awareness.
Furthermore, the reconstruction process is affected by population behavior, the symptomatic ratio, and various other factors influencing the counterfactual analysis. Therefore, obtaining precise and accurate information about the spread is critical to leverage the reverse engineering of the reproduction number methodology for analyzing the spreading process. While the analytical results are specific to particular conditions and their corresponding reconstructions, the methodology proposed for reconstruction and intervention strategy evaluation is innovative. It provides a new perspective for researchers to assess the impact of implemented pandemic intervention strategies through the reproduction number in a model-free way.

\begin{figure}[p]
\centering
\includegraphics[width=\textwidth]{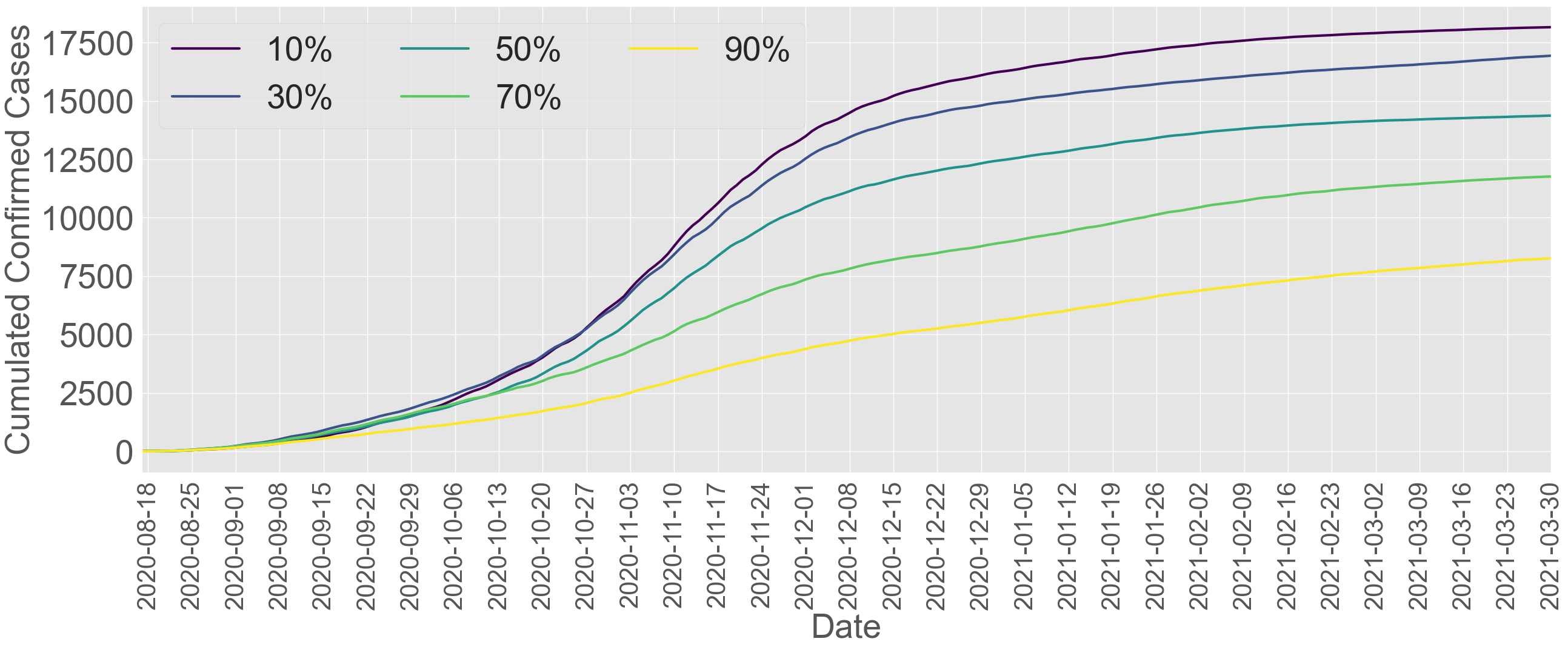}
\caption{Cumulative confirmed cases over the Purdue campus w/ different proportions of symptomatic cases}
\label{fig:Purdue_Asym_Cumu_SI}
\end{figure}


\subsection*{SI-2-F. The Impact of The Isolation Rate On Spreading Processes}
\label{secSI: Fix_Control}
In this section, we introduce new methodologies to explore the impact of isolation rates on epidemic spreading. We first focus on a fixed isolation rate, often referred to as the open-loop control strategy. Our analysis involves the reconstructed spreading processes over both campuses, creating an experimental environment to examine potential outcomes if universities had implemented varying isolation rates during Fall 2020.
As shown in~\ref{fig:Control_Framework_SI},
we propose a mechanism for generating confirmed cases using real-world data from \baike{Figures}~\ref{fig:UIUC_P_SI} and~\ref{fig:Purdue_Total_SI}. The core methodology incorporates utilizing the scaling factor to reverse engineer the reproduction number, as defined in~\eqref{eq_SI:F_w}. Generally, accounting for the implemented daily isolation rate in reality as $\alpha$, we denote the scaling factor as $\mathcal{F}(\alpha)$ according to~\eqref{eq_SI:F}. Given the estimated reproduction number under the implemented isolation rate $\alpha$, denoted as $\mathcal{R}_t(\alpha)$ from~\eqref{eq_SI:F_w},the reproduction number without the isolation rate is
$\frac{\mathcal{R}_t(\alpha)S(t)}{\mathcal{F}(\alpha)N}$. 
When introducing a different isolation rate $\alpha^*$ at the moment $t$, we denote the new scaling factor as $\mathcal{F}(\alpha^*)$. We compute the new reproduction number with the update isolation rate from $\alpha$ to $\alpha^*$ by
\begin{equation}
\label{eq_SI:New_R}
 \mathcal{R}_t(\alpha^*) = \mathcal{R}_t(\alpha)\frac{S^*(t)}{S(t)}\frac{\mathcal{F}(\alpha^*)}{\mathcal{F}(\alpha)}.
\end{equation}
Using~\eqref{eq_SI:New_R},
we \baike{are able to} calculate the reproduction number $\mathcal{R}_t(\alpha^*)$ under a different fixed isolation rate  $\alpha^*$ by utilizing the estimated reproduction number $\mathcal{R}_t(\alpha)$ under the implemented isolation rate $\alpha$ (e.g., in \baike{Figure}~\ref{fig:Estimated_R_UIUC_SI}). The computed scaling factors  are obtained by scaling the default serial interval distribution $w$ with the isolation rates $\alpha$ and $\alpha^*$ as described in~\eqref{eq_SI:SI_Und_Test} and~\eqref{eq_SI:F_w}, and the ratio between the susceptible population at any given time $t$.
Further, with~\eqref{eq_SI:New_R} we generate the reproduction number $\mathcal{R}_t(\alpha^*)$ under a different fixed isolation rate $\alpha^*$. \baike{
The isolation rate is not affected by the severity of the epidemic spread, making it an open-loop control strategy.}
By applying~\eqref{eq_SI:SI_Und_Test}
we use $w(\alpha^*)$ to calculate the infection profile $v(\alpha^*)$  under the fixed isolation rate $\alpha^*$. 
Therefore, utilizing the synthetic data generation mechanism outlined in Section~SI-2-B, we generate cases under different fixed isolation rates.

\baike{In the previous section, we conclude that the reconstructed spreading environment is sensitive to various factors.}
In this section, we utilize the worst-case \baike{reconstructed scenario} at the UIUC campus (\baike{Figure}~\ref{fig:UIUC_Rec_SI_w/o_SI}) as our testing environment. The worst-case scenario \baike{considers} that a case caught by the testing will not \baike{ be isolated}, and the case will behave as uninfected until recovery. We \baike{compare} the outcomes under different fixed isolation rates (open-loop control strategy) on the UIUC campus during Fall 2020, considering rates that are less than $200\%$ weekly. Under the fact that the testing rate at UIUC during Fall 2020 is around $200\%$ weekly, 
the isolation rates are drawn from the set $\{0\%, 10\%, 20\%, \dots, 90\%, 100\%, 120\%,\dots 190\%, 200\%\}$. Additionally, the testing process does not differentiate between symptomatic and asymptomatic cases, \baike{since we leverage the same infection profile and population behavior for all infected cases.}

\begin{figure}[p]
\centering
\includegraphics[width=\textwidth]{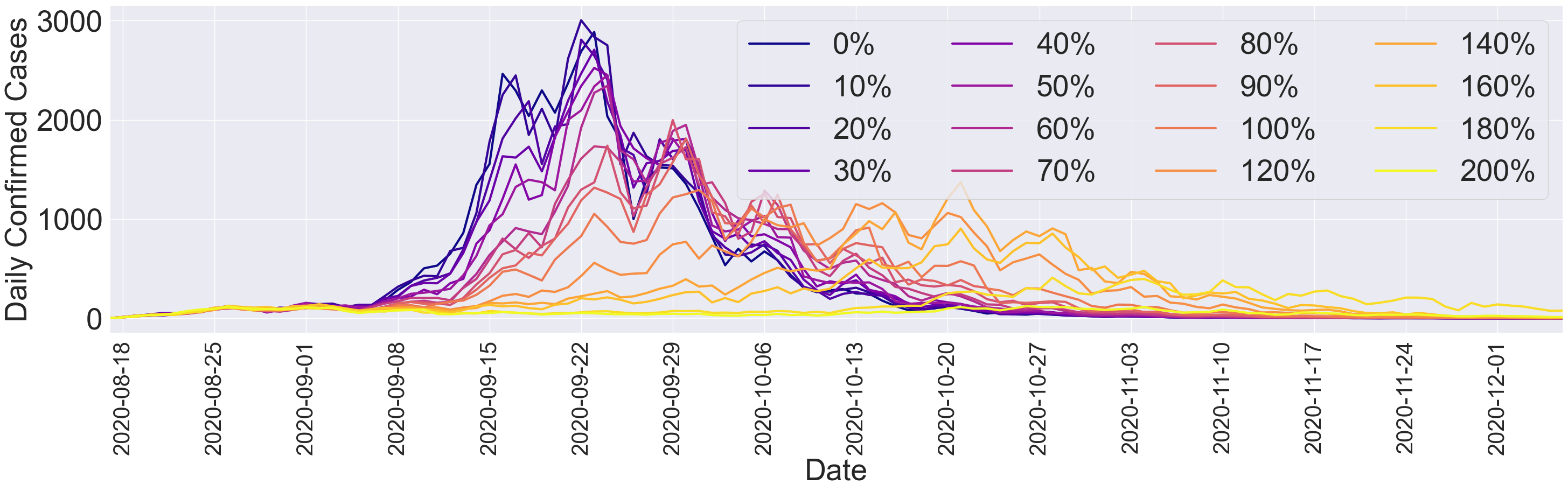}
\caption{Daily confirmed cases over the UIUC campus with different isolation rates}
\label{fig:UIUC_Control_Fixed_SI}
\end{figure}

\begin{figure}[p]
\centering
\includegraphics[width=\textwidth]{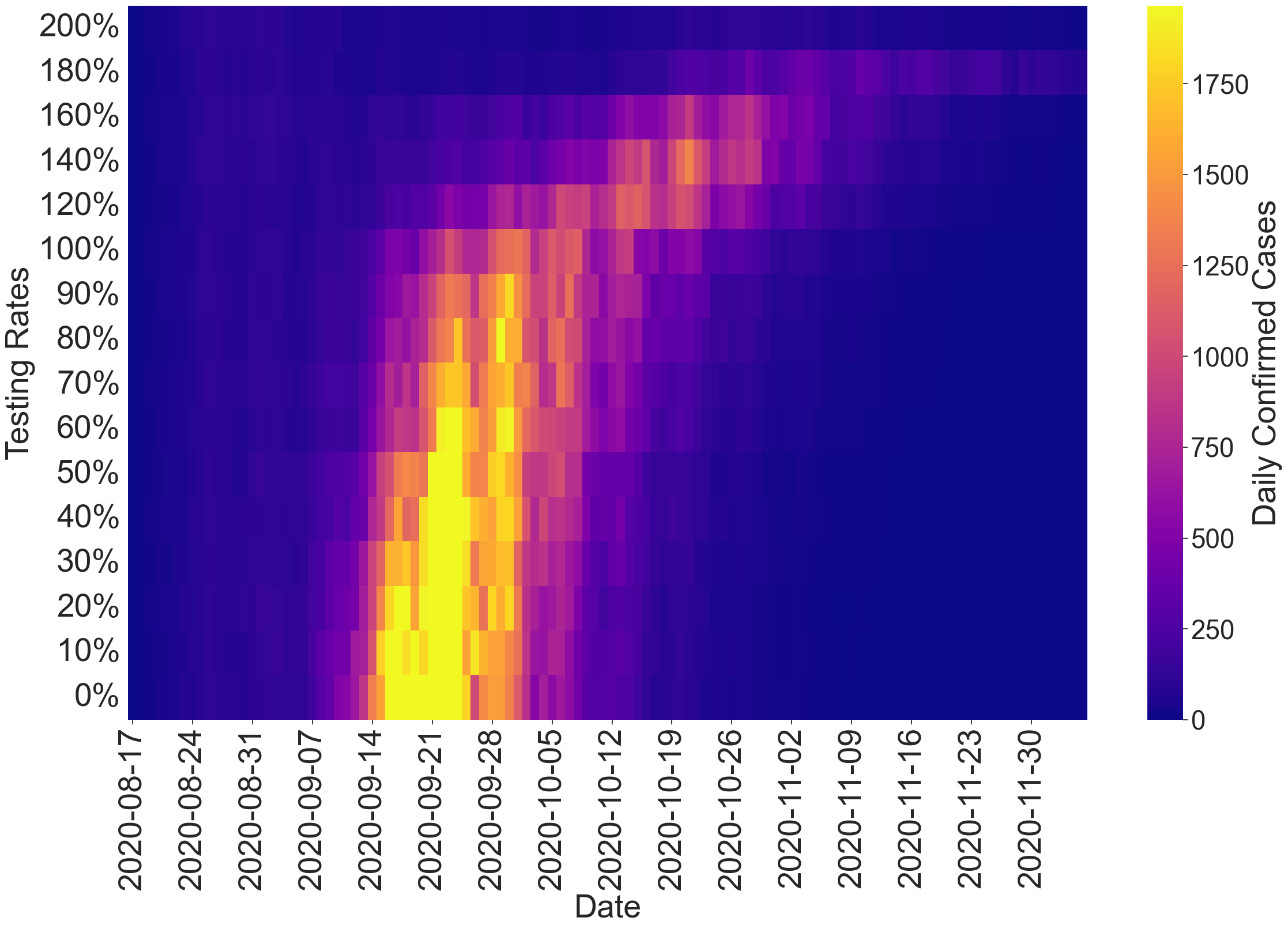}
\caption{Daily confirmed cases over the UIUC campus with different isolation rates: A heat map}
\label{fig:UIUC_Control_Fixed_SI_Heat}
\end{figure}

\begin{figure}[p]
\centering
\includegraphics[width=\textwidth]{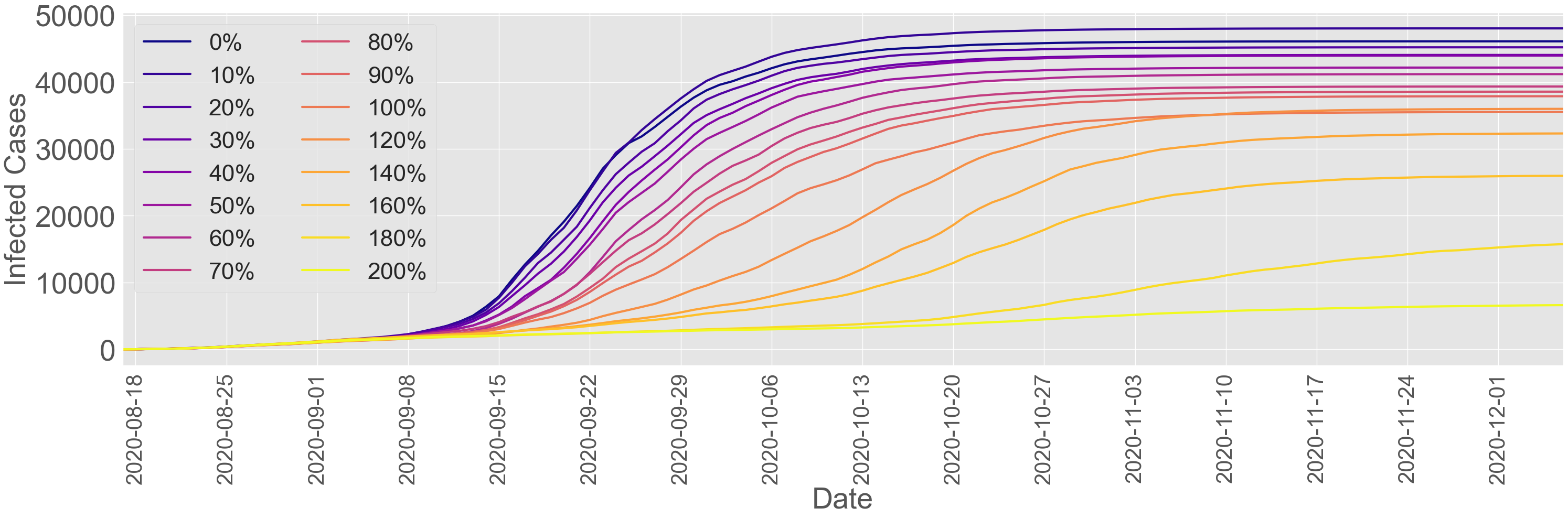}
\caption{Cumulative confirmed cases over the UIUC campus with different isolation rates}
\label{fig:UIUC_Control_Fixed_SI_Cumu}
\end{figure}

\begin{figure}[p]
\centering
\includegraphics[width=\textwidth]{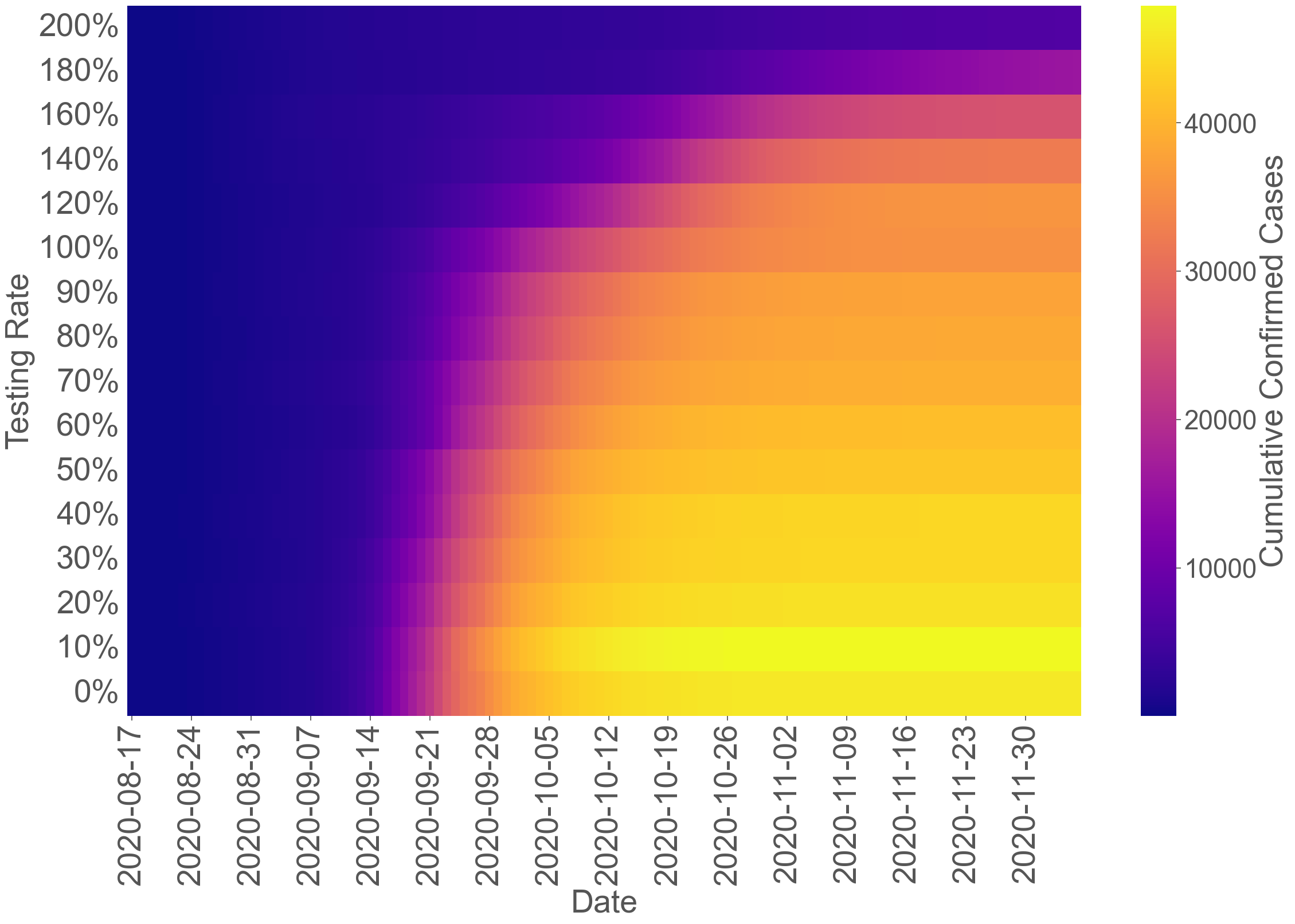}
\caption{Cumulative confirmed cases over the UIUC campus with different isolation rates: A heat map}
\label{fig:UIUC_Control_Fixed_SI_Cumu_Heat}
\end{figure}

Based on the daily confirmed cases presented in \baike{Figure}~\ref{fig:UIUC_Control_Fixed_SI}, we observe that higher isolation rates generate relatively smoother and flatter curves in terms of confirmed cases. The bright area in \baike{Figure}~\ref{fig:UIUC_Control_Fixed_SI_Heat} indicates that higher isolation rates result in lower and delayed spikes. Examining the cumulative confirmed cases in \baike{Figure}~\ref{fig:UIUC_Control_Fixed_SI_Cumu}, we conclude that higher isolation rates generally  \baike{lead to a lower number of} total cumulative cases.
An important distinction was observed between isolation rates below $100\%$ per week and isolation rates above $100\%$ per week. When the isolation rate was $100\%$ or lower, the cumulative confirmed cases amounted to approximately 35,000 out of a total population of 50,000. The primary difference between isolating at $100\%$ and isolating at $0\%$ per week was that isolating at $100\%$ per week flattened the curve more and resulted in lower peak infection values. Figure~\ref{fig:UIUC_Control_Fixed_SI_Cumu_Heat} \baike{shows} that when the isolation rate exceeded $100\%$ per week, increasing the rate by $20\%$ per week notably decreased the total cumulative cases.
This phenomenon can be explained by considering the threshold conditions of an epidemic spreading process. Epidemics often exhibit threshold conditions that determine their spreading behavior. Thus, for the spreading processes on the UIUC campus, when the isolation rate was below $100\%$ per week, significant outbreaks were inevitable, as they approached an equilibrium where nearly everyone would eventually be infected. We conclude that, in order to effectively mitigate the spreading process over the UIUC campus under the worst-case scenario situation, it is necessary to test everyone at least once a week. 


Note that when the weekly isolation rate reached  $200\%$, the daily confirmed cases and cumulative confirmed cases matched the UIUC confirmed cases introduced in \baike{Figure}~\ref{fig:UIUC_Rec_SI}.
Similarly, when the isolation rate was zero, the daily confirmed cases and cumulative confirmed cases aligned with the reconstruction results in \baike{Figure}~\ref{fig:UIUC_Rec_SI_w/o_SI}. This phenomenon demonstrates the idea of reverse engineering the reproduction number to study the spread under different intensity of the isolation.
Therefore, \baike{under the conditions that we leverage for reconstructing the spread at the UIUC campus,} we conclude that in order to avoid major outbreaks in the worst-case scenario at the UIUC campus, it is necessary to test the entire campus once a week, then ensure that all confirmed cases are isolated. Increasing the isolation rate beyond that further flattened the curve and reduced the total number of confirmed cases. Note that all these analyses were based on the worst-case scenario assumption at UIUC, shown in Figure~\ref{fig:UIUC_Rec_SI_w/o_SI}. If we were to test these fixed isolation rates in a different testing environment over UIUC, such as the \baike{other reconstructed spreading environment} provided by \baike{Figure}~\ref{fig:UIUC_I_S_SI}, we would obtain significantly different results. Hence, when evaluating epidemic mitigation strategies like isolation rates, it is critical to consider the factors on population behavior and \baike{the} virus spreading behavior. 

It is important to note the differences between Figure~\ref{fig:UIUC_C_Heatmap_SI} and Figure~\ref{fig:UIUC_Control_Fixed_SI}. Both figures conduct a counterfactual analysis on the spread at the UIUC campus with varying isolation rates. Figure~\ref{fig:UIUC_C_Heatmap_SI} reconstructs the spread on the UIUC campus with different implemented isolation rates. Despite UIUC testing everyone twice a week, the isolation rate under this high testing rate remains unknown in reality. Therefore, when reconstructing what would have happened under the $200\%$ testing rate, different isolation rates under this $200\%$ testing rate need to be considered to capture varying population behavior.

Figure~\ref{fig:UIUC_C_Heatmap_SI} studies how the fixed isolation rate we implemented affects an environment where the original isolation rate at UIUC is $200\%$ per week. It simulates the scenario at UIUC without the implemented isolation strategy as the experimental environment. Thus, Figure~\ref{fig:UIUC_C_Heatmap_SI} provides multiple scenarios to simulate the spread at UIUC without their implemented testing-for-isolation strategy. Then, Figure~\ref{fig:UIUC_Control_Fixed_SI} tests the impact of different isolation strategies on one scenario, which is the worst-case scenario captured by~\ref{fig:UIUC_Rec_SI_w/o_SI}.


\begin{figure}[p]
\centering
\includegraphics[width=\textwidth]{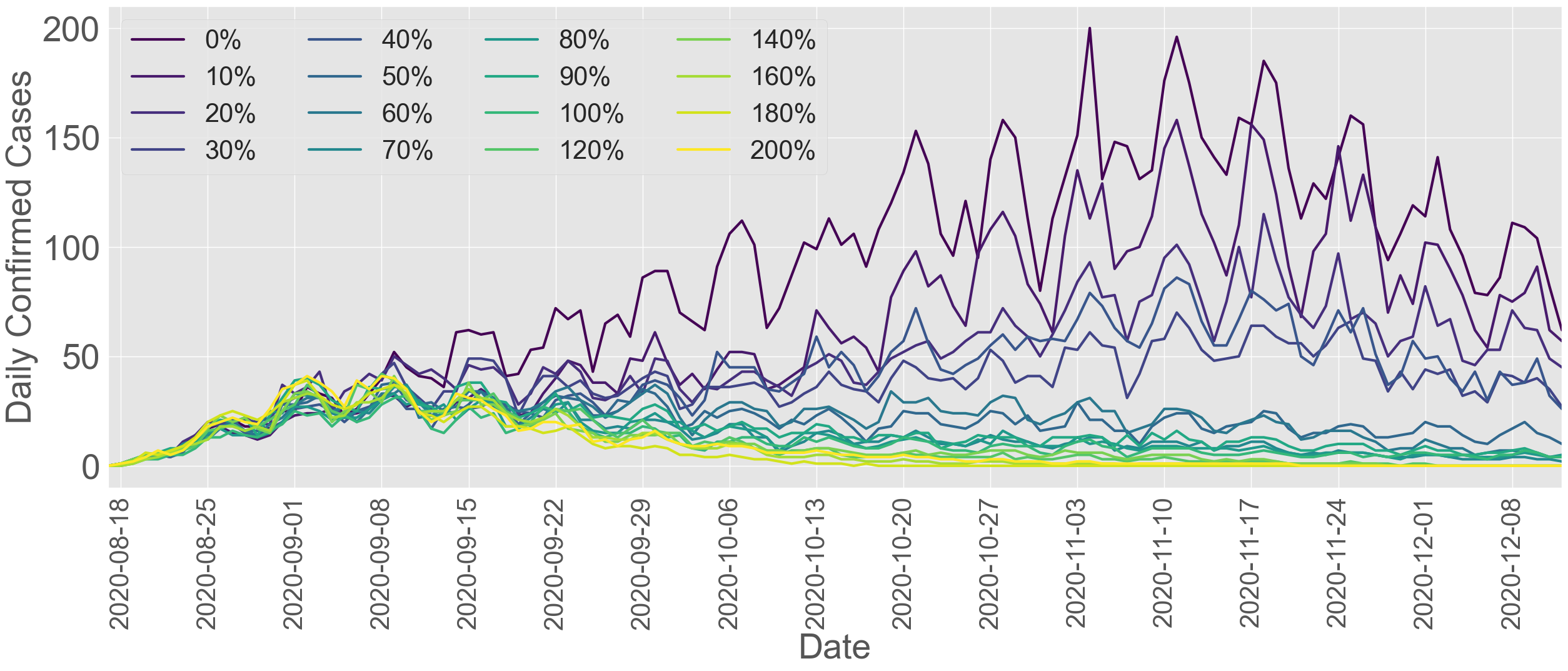}
\caption{Daily confirmed cases over the Purdue campus with different isolation rates under the surveillance testing}
\label{fig:Purdue_Control_Daily_SI}
\end{figure}

\begin{figure}[p]
\centering
\includegraphics[width=\textwidth]{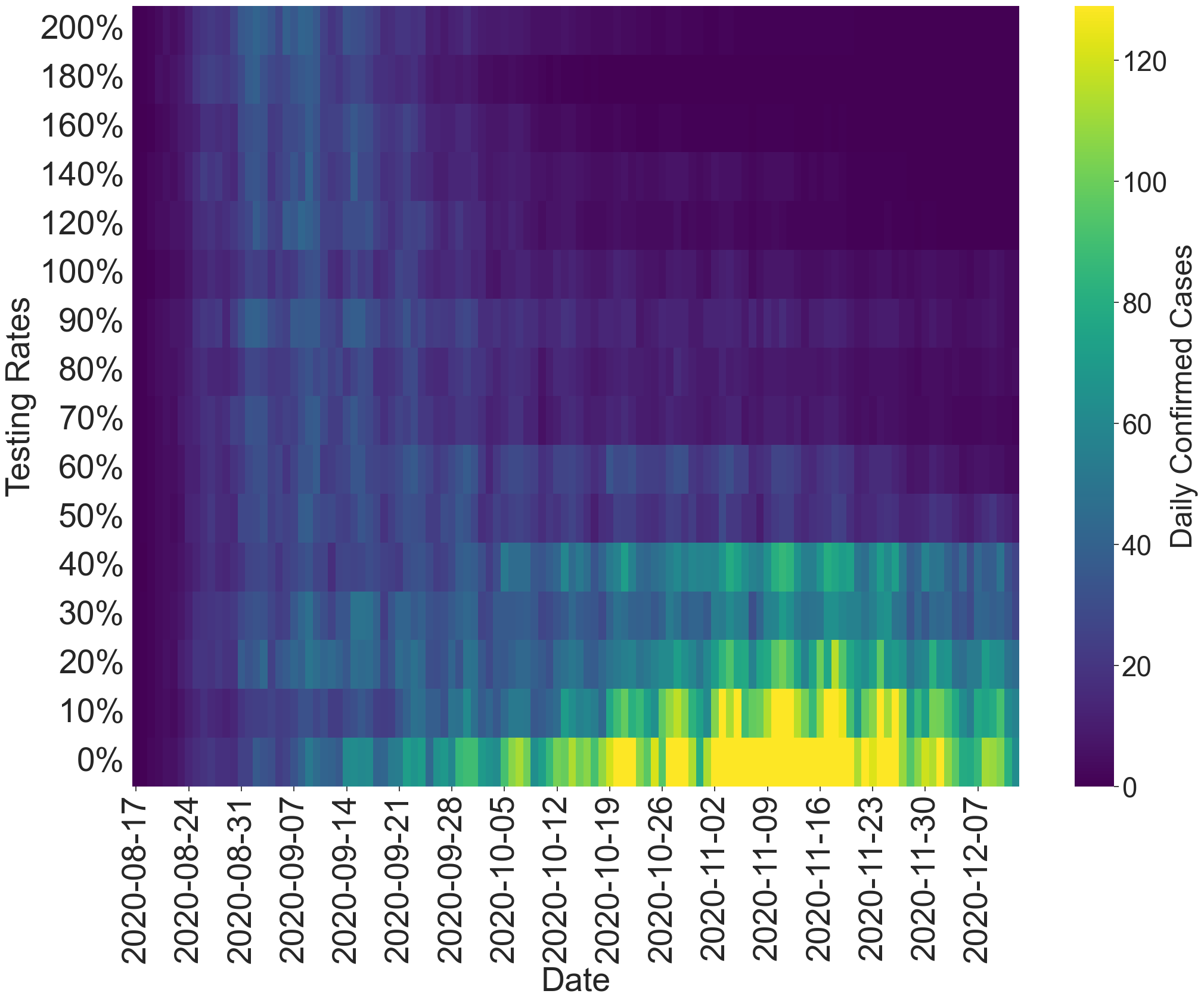}
\caption{Daily confirmed cases over the Purdue campus with different isolation rates under the surveillance testing: A heat map}
\label{fig:Purdue_Control_Daily_SI_Heat}
\end{figure}

Compared to UIUC, we focus on evaluating the impact of  different isolation rates on the spreading process over the Purdue campus, considering a different setting:
\begin{itemize}
    \item Symptomatic and asymptomatic cases have the same spreading behavior.
    \item Symptomatic cases on the Purdue campus seek testing-for-isolation within a week after becoming infectious, equivalent to a $100\%$ weekly testing and isolation rate for symptomatic cases.
    \item The isolation rate for asymptomatic cases is $30\%$ per week.
    \item From the testing data, $55\%$ of the population on the Purdue campus is symptomatic.
\end{itemize}
To capture the reconstructed environment without the implemented isolation under surveillance testing, we referred to Figure~\ref{fig:Purdue_0.3_SI} with the $30\%$ isolation rate scenario. We consider a milder spread over Purdue in order to evaluate the impact of the surveillance testing under a different spread behavior. We expect  that the spreading process with different isolation rates under surveillance testing will be less severe compared to UIUC.
Similarly to UIUC, we consider isolation rates under surveillance testing for asymptomatic cases from $\{0\%, 10\%, 20\%, \dots, 90\%, 100\%, 120\%,\dots 190\%, 200\%\}$. Utilizing the reconstruction mechanism proposed in~\eqref{eq_SI:New_R} and Section~SI-2-B, we generate the daily confirmed cases in \baike{Figure}~\ref{fig:Purdue_Control_Daily_SI} and the corresponding heatmap in \baike{Figure}~\ref{fig:Purdue_Control_Daily_SI_Heat}. As expected, the daily confirmed cases at the Purdue campus are notably lower than those at UIUC. Higher isolation rates under the surveillance testing result in smoother and flatter curves in terms of confirmed cases, as observed in Figure~\ref{fig:Purdue_Control_Daily_SI}. Moreover, Figure~\ref{fig:Purdue_Control_Daily_SI_Heat} demonstrates that higher isolation rates also lead to lower spikes.

\begin{figure}[p]
\centering
\includegraphics[width=\textwidth]{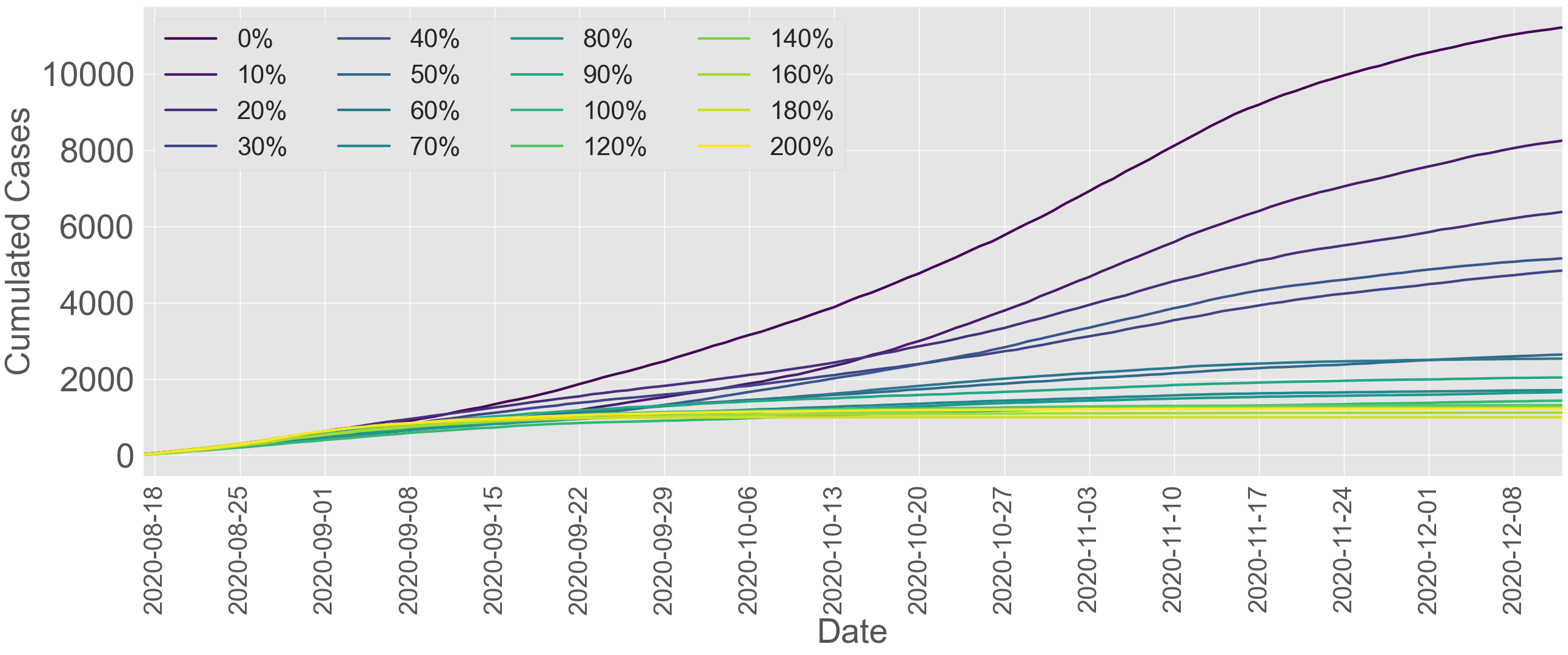}
\caption{Cumulative confirmed cases over the Purdue campus with different isolation rates under the surveillance testing}
\label{fig:Purdue_Control_Cumu}
\end{figure}

\begin{figure}[p]
\centering
\includegraphics[width=\textwidth]{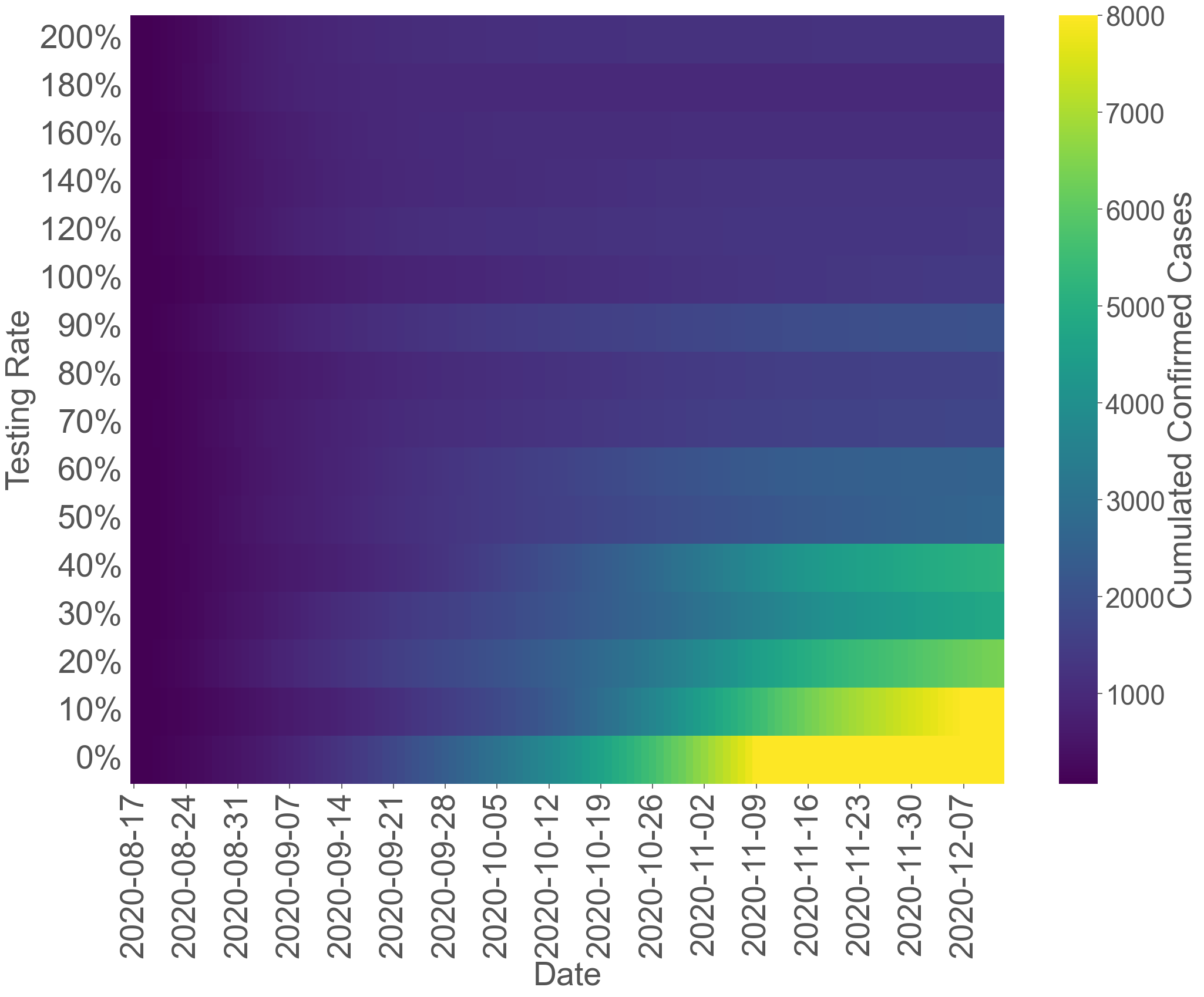}
\caption{Cumulative confirmed cases over the Purdue campus with different isolation rates under the surveillance testing: A heatmap}
\label{fig:Purdue_Control_Cumu_Heat}
\end{figure}

We further plot the cumulative confirmed cases in \baike{Figure}~\ref{fig:Purdue_Control_Cumu} with the corresponding heatmap in \baike{Figure}~\ref{fig:Purdue_Control_Cumu_Heat}. 
The cumulative confirmed cases in Figure~\ref{fig:Purdue_Control_Cumu} suggest that higher isolation rates generally lead to fewer cumulative confirmed cases. Analyzing the heatmaps in \baike{Figures}~\ref{fig:Purdue_Control_Daily_SI_Heat} and ~\ref{fig:Purdue_Control_Cumu_Heat}, a marked difference is observed between implementing isolation rates for asymptomatic cases below $50\%$ per week and rates above $50\%$ per week.
When the isolation rate under surveillance testing for asymptomatic cases is set at $50\%$  or lower per week, a notable outbreak occurs with large spikes towards the end of the Fall 2020 semester at Purdue. During this period, the cumulative confirmed cases exceed 5,000. However, when the isolation rate for asymptomatic cases surpasses $50\%$ per week, no significant outbreak is observed throughout the Fall 2020 semester at Purdue. 
Furthermore, increasing the isolation rate per week does not substantially decrease the total number of cumulative cases. In the reconstructed spreading process over the Purdue campus, if the isolation rate for asymptomatic cases remains below 50\% per week, the cumulative confirmed cases range from 5,000 to 10,000 during Fall 2020. Yet, when the rate exceeds 50\%, an increase in the isolation rate leads to a larger decrease in the cumulative confirmed cases. In this situation, the cumulative confirmed cases range from 500 to 2,500 during Fall 2020.
This pattern could be explained by considering the potential threshold conditions of the epidemic spreading process over the Purdue campus. It is hypothesized that an isolation rate under the surveillance testing rate of 50\% serves as the threshold condition, determining whether there will be an outbreak at the end of the Fall 2020 semester. Notably, the implemented surveillance testing-for-isolation by Purdue, which caught approximately 30\% of asymptomatic cases (as estimated in \baike{Figure}~\ref{fig:Estimated_R_Purdue_SI} and reconstructed in \baike{Figure}~\ref{fig:Purdue_Rec_SI}), is lower than the 50\% isolation rate threshold condition.
Hence, based on the real-world confirmed cases at Purdue in Figure~\ref{fig:Purdue_Total_SI}, an outbreak is observed in the middle and at the end of the Fall 2020 semester. From the analysis in this example, it can be inferred that testing and then isolating more than 50\% of asymptomatic cases during Fall 2020 at Purdue could help avoid the significant spikes observed on campus.

This section employs the spread data from the UIUC and Purdue campuses to validate the reverse engineering of the reproduction number for evaluating mitigation strategies. The goal is to conduct counterfactual analysis to refine outcomes by adjusting the strength of these strategies. Examining the potential outcomes at both UIUC and Purdue under varying fixed isolation rates reveals threshold conditions associated with isolation rates, which might have prevented possible outbreaks. It is essential to note that computational results are contingent on the environmental conditions during the reconstruction of the testing setup. Nonetheless, with precise information about the spread, this model-free framework offers valuable insights into isolation rates and their threshold conditions.
The next section delves into the prospect of altering isolation rates during the semester based on the epidemic's severity, introducing a model-free feedback control mechanism. This approach aims to effectively manage the spread of the virus.

\subsection*{SI-2-G. Model-Free Feedback Control Mechanism}
\label{secSI:feedback_control}
In Section~SI-2-F, we study the impact of implementing different fixed testing rates, i.e., open-loop control strategies, on the spread of infections at UIUC and Purdue campuses. It is demonstrated that when the isolation rate exceeds a certain threshold, significant reductions in peak infections and cumulative cases can be achieved. Therefore, it is reasonable to implement a relatively higher isolation rate on the campus to ensure safe operations. The success of implementing higher testing and isolation rates to mitigate the spread of the virus has been demonstrated by both UIUC and Purdue University. While Purdue conducted 5000 tests for surveillance testing (Figure~\ref{fig:Purdue_SVL_T_SI}), they also performed a significant number of voluntary tests on symptomatic and suspected infection cases (Figure~\ref{fig:Purdue_SR_T_SI}). However, for other smart communities like industrial parks, implementing high testing and isolation rates poses challenges due to resource constraints. Maintaining a high testing rate over an extended period, such as several months, can be costly in terms of testing kits, human resources, testing sites, and other resources.

Therefore, it becomes crucial to design epidemic mitigation strategies that can potentially save  resources while still keeping the infection level below a safe threshold. To achieve this goal, the idea of feedback control design methodologies can be leveraged. Feedback control systems are prevalent in both natural and engineered systems. In natural systems, such as the human body, complex feedback control mechanisms work to maintain various parameters within specific ranges, such as body temperature, blood pressure, and blood sugar levels. Similarly, well-designed feedback control engineering systems, like air conditioning systems, monitor room temperature and adjust system parameters to maintain the temperature at a desired level. Utilizing feedback control policies enables robustness against inaccurate evaluations of the spreading environment. Thus, even in the presence of uncertainties or inaccuracies in assessing the current situation~\cite{van2023effective}, the feedback control system can dynamically adjust epidemic mitigation strategies to effectively respond to changing conditions and maintain the spread of the virus within manageable levels.

Effective mitigation strategies
are crucial during epidemics like the recent COVID-19 pandemic. Researchers have applied systems and control theory to address various epidemic mitigation challenges~\cite{di2020covid,sharomi2017optimal, watkins2019robust, morris2021optimal, hewing2020learning, allgower2012nonlinear, tsay2020modeling,kohler2020robust,morato2020optimal,peni2020nonlinear,carli2020model,grundel2020much,sereno2021model, tsay2020modeling,kohler2020robust,morato2020optimal,peni2020nonlinear,carli2020model,grundel2020much,sereno2021model, acemoglu2021optimal,casella2020can,nowzari2016epidemics}, including modeling, estimating, and controlling the spread of epidemics, particularly in the context of the COVID-19 pandemic~\cite{stewart2020control,richard2021age,casella2020can, zino2021analysis, kohler2020robust,acemoglu2021optimal,morris2021optimal,she2022learning}.
Some studies have investigated policies, such as the "on-off" policy~\cite{tsay2020modeling}, which assesses the social and economic costs of strict social distancing measures, and calibrated epidemic models to examine the impact of social distancing restrictions~\cite{perkins2020optimal}. Other strategies have optimized epidemic mitigation by combining molecular and serology testing~\cite{acemoglu2021optimal} and addressing the vulnerabilities of optimal approaches~\cite{morris2021optimal}.
In addition to optimal control methods, control scientists and engineers have explored model predictive control frameworks~\cite{kohler2020robust,carli2020model,zino2021analysis,she2022learning,tsay2020modeling,kohler2020robust,morato2020optimal,peni2020nonlinear,carli2020model,grundel2020much,sereno2021model} and other strategies~\cite{khadilkar2020optimising,scarabaggio2021nonpharmaceutical,mubarak2022individual} to develop optimal or suboptimal epidemic mitigation policies. For instance, one study identified vaccination targets using transmission network structures~\cite{scarabaggio2021nonpharmaceutical}.  Furthermore, there are other studies that consider epidemic mitigation and resource allocation~\cite{bloem2009optimal, nowzari2016epidemics,di2017optimal,sharomi2017optimal, di2019optimal,liu2019analysis,dangerfield2019resource,preciado2014epidemic_optimal,han2015data}.
Regarding epidemic state observation problems, a comprehensive closed-loop framework was proposed in one study. This framework includes a nonlinear observer for estimating system states, forecasting the spread, and controlling outbreaks~\cite{hota2020closed}. Furthermore, recognizing uncertainties in implementing theoretical control designs in real-world scenarios is crucial~\cite{casella2020can, van2023effective}. However, the current control analyses and designs are reliant on specific epidemic spreading models. Furthermore, none of the control designs utilize the reproduction number as feedback information to adjust the intensity of the intervention strategy.

In order to introduce and validate our proposed model-free feedback control framework, we first present the optimal resource allocation problem for testing-for-isolation as the following control design challenge. This issue served as inspiration for our model-free feedback control design. Our goal is to mitigate the epidemic by minimizing the total number of tests (usually proportional to the isolated cases) during each period of the epidemic, as demonstrated in the following cost function
\begin{equation} 
\label{eq_SI:cost_function}
J(u(t))=\int_{t_0}^{t_0+T}u(\tau)d\tau,
\end{equation}
\baike{where $[t_0, t_0+T]$ is the period of consideration and $u(t)$ denotes the testing rate at time step $t$.
Further, to simplify the formulation, we assume the testing rate is equal to the isolation rate. For instance, conducting random daily tests on 20\% of the total population allows the identification and isolation of 20\% of the infectious cases each day. As we  are discussing resources, it is more practical to consider tests based on testing rates rather than the isolation rate. Furthermore, $u(t)\in [\underline{u}, \overline{u}]$}, where $\underline{u}$ and $\overline{u}$ represent the lower and upper bounds on testing/isolation rates, respectively. To derive the testing-for-isolation strategy that minimizes the total number of tests needed during the period and simultaneously ensures that the number of infected individuals remains below a desired threshold $\bar{I}$, we formulate the following optimization problem, 
\begin{subequations}\label{eq_SI:prob}
\begin{align}
&\min_{u(t), t_0\leq t \leq t_0+T}  \, \, J(u(t)) \\
&\text{s.t.}  \, \, 
\dot{\boldsymbol{x}}(t)=f(\boldsymbol{x}(t),u(t)),  \\ 
\label{eq_SI:constraints}
&0 \leq I(t) \leq \bar{I}, 
\underline{u} \leq {u}(t) \leq \bar{u},
\end{align}
\end{subequations}
where $\dot{\boldsymbol{x}}(t)=f(\boldsymbol{x}(t),u(t))$ denotes the unknown spreading dynamics.
We use $\boldsymbol{x}(t)$ to represent all possible spreading states and $I(t)$ to represent infected cases at any given time $t$.
The state constraint $\bar{I}$ outlines the infection threshold concerning factors like available public resources, such as hospital capacity.
As we lack direct access to the model  $\dot{\boldsymbol{x}}(t)=f(\boldsymbol{x}(t),u(t))$, we cannot solve the optimization problem directly. Nevertheless, we will establish a connection between the solution for the optimization problem and the reproduction number.

Based on previous analyses~\cite{she2022optimal,acemoglu2021optimal}, when $T$  is sufficiently large to ensure the epidemic fades away by $T$ in problem~\eqref{eq_SI:prob}, the optimal solution is to 
adjust the testing rates to maintain the reproduction number at 1 when the infected population hits the infection threshold $\bar{I}$ (assuming the ability to adjust the testing rate arbitrarily). Theoretically, when the reproduction number is sustained at 1 and the infected population equals the infection threshold $\bar{I}$, the number of new infections will stabilize at the infection threshold $\bar{I}$.
Under these conditions, based on~\eqref{eq_SI: Reproduction_Num}, conducting more tests will consume additional testing resources but will result in a decrease in the number of daily new infections. However, since the primary objective is to prevent the number of new infections from surpassing the infection threshold $\bar{I}$, it becomes unnecessary to allocate additional testing resources if we can maintain the reproduction number at 1 through a lower testing rate. In practice, due to uncertainties arising from modeling, estimation, and computation, maintaining the reproduction number at 1 is not robust against these disturbances, which may lead to the reproduction number easily exceeding 1.

As indicated by~\eqref{eq_SI: Reproduction_Num}, conducting fewer tests will result in an increase in the infected population, surpassing the infection threshold, as the reproduction number will be higher than 1. Furthermore, observations from the successful implementation of testing-for-isolation strategies at both UIUC and Purdue revealed that the estimated reproduction number fluctuated around 1, as shown in \baike{Figures}~\ref{fig:Estimated_R_UIUC_SI} and~\ref{fig:Estimated_R_Purdue_SI}. Therefore, Drawing inspiration from both theoretical analyses and real-world testing-for-isolation results, we aim to simplify the control problem formulated in~\eqref{eq_SI:prob} to the goal of maintaining the reproduction number at a certain value $\mathcal{R}^*$ ($\mathcal{R}^*\in (0,1))$ during the spread. We intend to validate this idea in the testing environments constructed for UIUC (Figure~\ref{fig:UIUC_Rec_SI_w/o_SI}) and Purdue (Figure~\ref{fig:Purdue_0.3_SI}). Additionally, we aim to compare the effectiveness of our proposed method with the implemented fixed testing rates at UIUC and Purdue 

\begin{figure}[p]
\centering
\includegraphics[width=\textwidth]{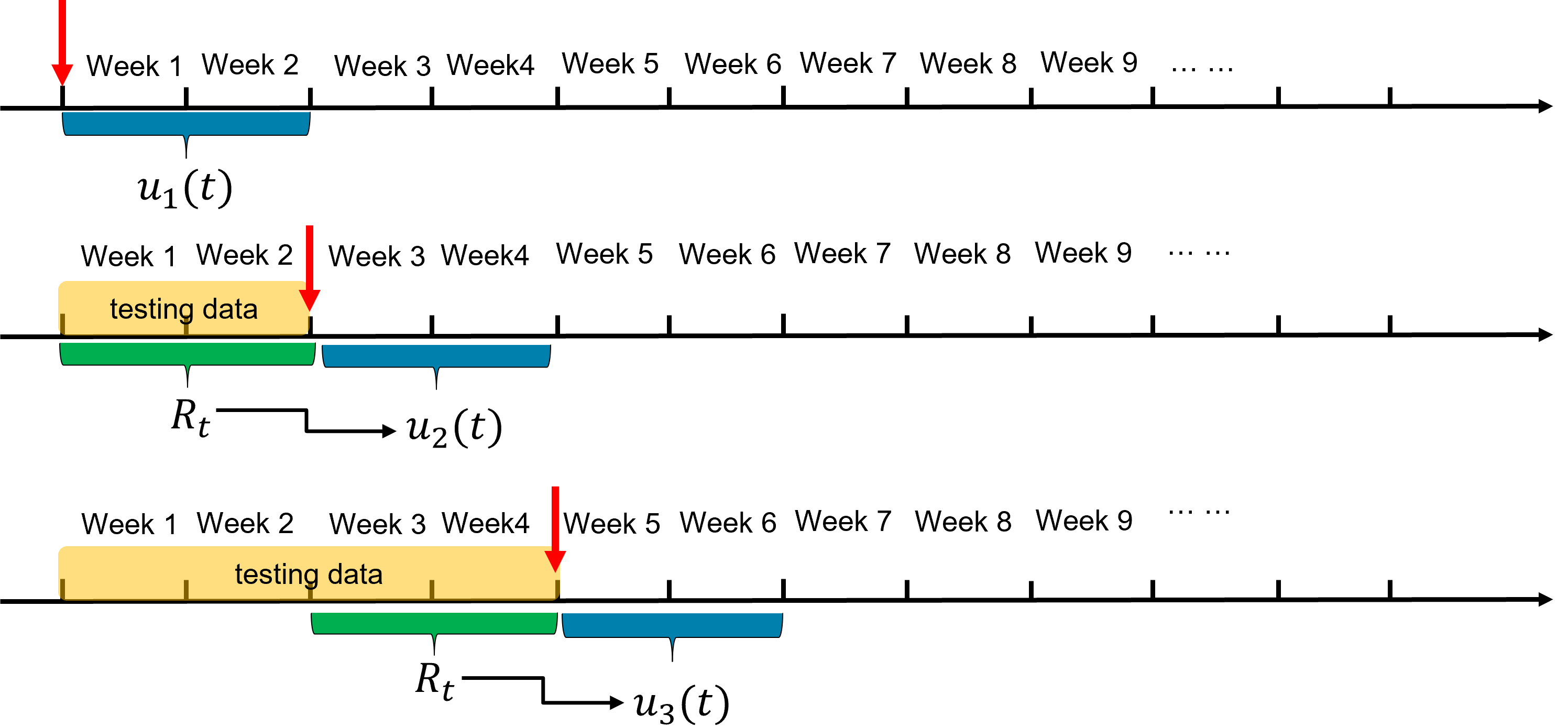}
\caption{Illustration of the model-free feedback control framework.}
\label{fig:Control_Illustration}
\end{figure}

In order to implement the concept of controlling the reproduction number at $\mathcal{R}^*$, we need to 1) propose the model-free feedback control framework, and 2) design a method to update the control input, i.e., the testing rate, based on the feedback information (the estimated reproduction number). 
We will illustrate the model-free feedback control framework and how it is applied in practice with an example. Consider a scenario where the testing rate can be adjusted bi-weekly. We implement a reasonable daily testing rate $u_1(t)$ for the first two weeks, as shown in the top \baike{plot in} Figure~\ref{fig:Control_Illustration}. After the two-week period, at the beginning of the third week, we employ the testing rate  $u_1(t)$  and the confirmed cases from the first and second weeks to estimate the reproduction number $\mathcal{R}_t$ during week 1 and week 2. Using the default serial interval distribution $w$, the scaling factor computed through~\eqref{eq_SI:F_w}, and and the estimation method outlined in Section~SI-2-D, we determine the average estimated reproduction number for weeks 1 and 2.
If the average estimated reproduction number matches our target value, $\mathcal{R}^*$, we maintain the same testing rate, setting $u_2(t) = u_1(t)$ for week 3 and week 4. However, if the estimated reproduction number does not align with~$\mathcal{R}^*$, we compute a new testing rate $u_2(t)$ to regulate the reproduction number at $\mathcal{R}^*$ for week 3 and week 4. This process is repeated iteratively, adapting the testing rate based on the estimated reproduction number, thereby establishing a model-free closed-loop feedback control framework.  
After introducing the general feedback control framework, we will now explain how to update the testing rate at each iteration without utilizing spreading models.  
For simplification purposes, we do not differentiate between symptomatic and asymptomatic cases. \baike{Therefore, we use $v$ to represent the initial infection profile and $w$ to represent the corresponding serial interval distribution of the spreading process.} However, the developed techniques can be applied to compute testing rates for infections that involve a mix of symptomatic and asymptomatic cases using the mechanisms proposed in~\eqref{eq_SI: Reproduction_Num} and~\eqref{eq_SI:F}.
Consider an epidemic spreading process with a known time-invariant serial interval distribution as given in~\eqref{eq_SI:SI_Dist}. During a specific time period $k$, such as the first two weeks in the previous example, we can derive the average estimated reproduction number $R_k(\alpha_k)$ under the implemented testing rate $\alpha_k$, using the mechanism proposed in Section~SI-2-D. Based on this average estimated reproduction number, we define the following novel mechanism to update the testing rate.
We first compute the reconstructed infection profile through the average estimated reproduction number $\mathcal{R}_k (\alpha_k)$ during the time period $k$, 
given by 
\begin{equation*}
 v(\alpha_k) = \sum_{i=1}^n w(\alpha_k)\mathcal{R}_k(\alpha_k),
\end{equation*}
where $\alpha_k$ is the fixed testing rate during the period $k$.
If $\mathcal{R}_k(\alpha_k)\neq \mathcal{R}^*$, we compute a new testing rate $\alpha_{k+1}$ during the next period $k+1$ (e.g., the following next two weeks in the previous example). We propose the following methodology to update the testing rate $\alpha_{k+1}$:
\begin{equation}
\label{eq_SI:Testing_Rate}
\sum_{i=1}^n v_i(\alpha_k)(1-\alpha_{k+1})^i =\mathcal{R}_k(\alpha_k)\sum_{i=1}^n w_i(\alpha_k)(1-\alpha_{k+1})^i =\mathcal{R}^*.
\end{equation}
Note that the only unknown variable in~\eqref{eq_SI:Testing_Rate} is the testing rate for the subsequent period, $\alpha_{k+1}$. Therefore, by solving~\eqref{eq_SI:Testing_Rate}, we can directly compute the testing rate for the next time step using the feedback information derived from the estimated reproduction number $R_k(\alpha_k)$ and and the pre-defined serial interval distribution $w$, under the previously implemented testing rate, $\alpha_k$. The methodology proposed in~\eqref{eq_SI:Testing_Rate} necessity to construct explicit spreading models to generate the new testing rate, making it a model-free feedback control approach to epidemic mitigation.
As a feedback control mechanism,~\eqref{eq_SI:Testing_Rate} is more robust to uncertainties compared to generating the testing rate through an open-loop mechanism.

When updating the testing rate for a future period $k+1$, we employ the average estimated reproduction number during period $k$ to compute the future testing rate. This approach considers that if we maintain the same testing rate from period~$k$ to~$k+1$, i.e., $\alpha_{k+1}=\alpha_{k}$, Consequently, this method evaluates the spreading process as time-invariant over a short period, or at least predicts no significant changes in the reproduction number under the same implemented testing rate. Note that real-world spreading processes are time-varying, as evident from the observed variations in the estimated reproduction number over the UIUC campus during Fall 2020 under the fixed 200\% per week testing rate, as shown in Figure~\ref{fig:Estimated_R_UIUC_SI}. Therefore, the proposed methodology, without predicting the trend of the reproduction number in the near future, may present a drawback in the current control framework.

We use the reconstructed spreading environment at UIUC and Purdue to validate the model-free feedback control mechanism.
As a result, Our control design goal is to maintain the reproduction number at  $\mathcal{R}^*$, which is slightly less than 1~\cite{van2023effective}. We initially implement the feedback control framework in the testing environment constructed for the UIUC campus during Fall 2020. This setting allows us to compare the feedback control framework with the implemented fixed testing rate (isolation rate) at UIUC, involving testing the entire campus twice a week.

\begin{figure}[p]
\centering
\includegraphics[width=\textwidth]{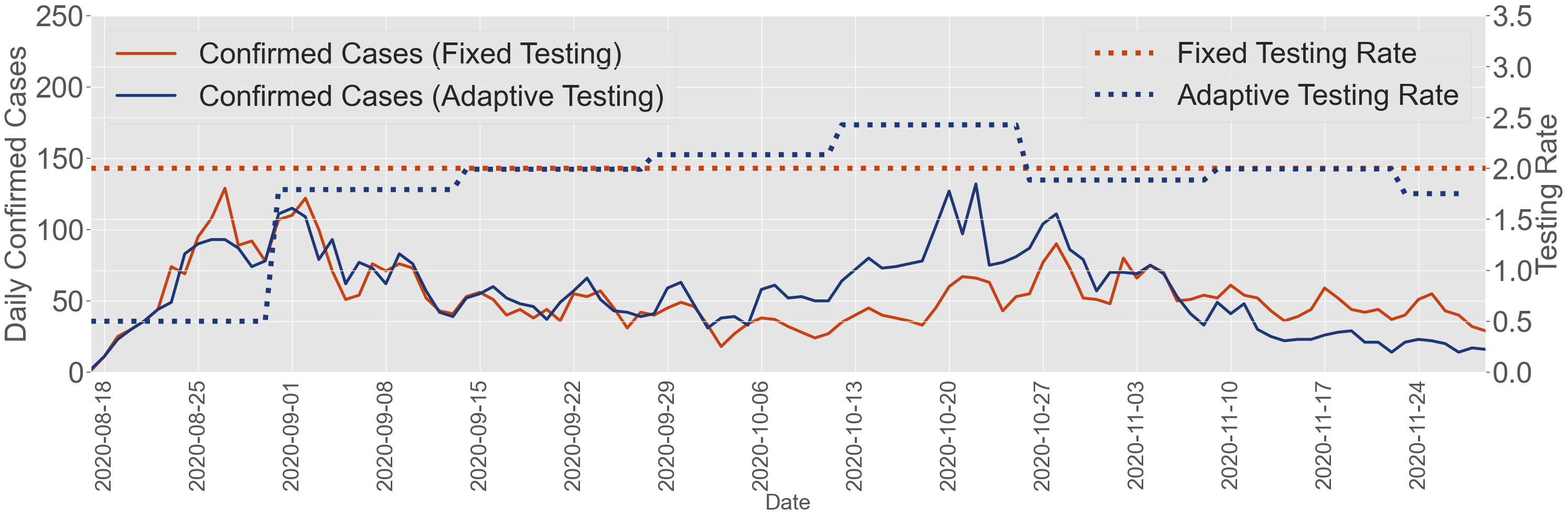}
\caption{Comparison between the fixed and feedback testing-for-isolation strategies at UIUC}
\label{fig:UIUC_Control_0.9_SI}
\end{figure}

Considering the same worst-case scenario testing environment and assumptions for the UIUC campus as discussed in Sections~SI-2-C and Section~SI-2-F. We implement the feedback control framework proposed in this section in the testing environment and compared it to UIUC's testing-for-isolation strategy, which involves testing/isolating the entire campus twice a week. Our objective is to control the target reproduction number at $\mathcal{R}^*=0.95$. 
The results are shown in Figure~\ref{fig:UIUC_Control_0.9_SI}.
Using the worst-case scenario reconstructed testing environment at UIUC captured by Figure~\ref{fig:UIUC_Rec_SI_w/o_SI}, tests are conducted on the entire campus twice a week, and our proposed feedback control methodology from~\eqref{eq_SI:Testing_Rate} is separately implemented.
Figure~\ref{fig:UIUC_Control_0.9_SI} demonstrates that the feedback control framework we proposed aligns with the testing policy implemented by UIUC in terms of daily and total confirmed cases, which are around 6300.
Furthermore, through simulation, we find that the implemented testing-for-isolation strategy by UIUC will result in a total of 16 tests per individual, whereas our proposed feedback control strategy will only require 14 tests per individual. Throughout most of the Fall 2020 period, the feedback control strategy utilizes a lower testing rate compared to UIUC’s 200\% testing rate. However, in October, the feedback control framework applies higher testing rates than the rates implemented by UIUC. Recall that real-world confirmed data (\baike{Figure}~\ref{fig:UIUC_P_SI}) and the estimated reproduction number (\baike{Figure}~\ref{fig:Estimated_R_UIUC_SI}) at UIUC during Fall 2020 reveal a significant spike in cases due to the return of the football season. Consequently, the feedback control framework raises the testing rate to mitigate the potential outbreak. As observed, the feedback control framework implements fewer tests when there is a lower risk of outbreaks and increases the testing rate when there is a potential spike. This example effectively illustrates the core concept of the model-free feedback control mechanism proposed for the 
pandemic mitigation framework, where we can leverage the connection between the intervention strategy and the reproduction number to design feedback control.

\begin{figure}[p]
\centering
\includegraphics[width=\textwidth]{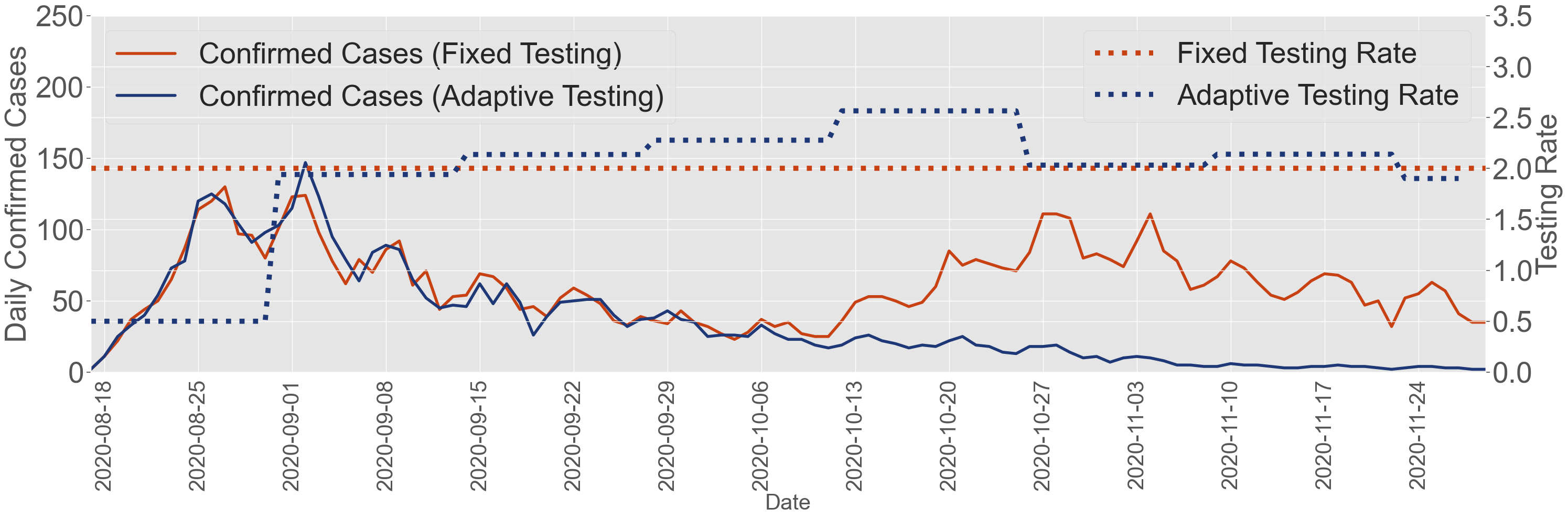}
\caption{Comparison between the fixed and feedback testing-for-isolation strategies at UIUC}
\label{fig:UIUC_Control_0.8_SI}
\end{figure}

In addition to aiming to match the confirmed cases under the testing-for-isolation strategy implemented by UIUC (200\% testing rate), it is also interesting to investigate if the feedback control framework can generate more conservative testing-for-isolation strategies. In this case, we adjust the goal of controlling the target reproduction number at  $\mathcal{R}^*=0.90$. The resulting confirmed cases and corresponding testing rates are presented in \baike{Figure}~\ref{fig:UIUC_Control_0.8_SI}.
When aiming to control the reproduction number at $\mathcal{R}^*=0.90$, we observe fewer daily and total confirmed cases during Fall 2020 compared to the 200\% weekly testing rate. While the 200\% testing rate results in around 6300 total confirmed cases, the feedback control framework generates approximately 3500 confirmed cases during the same period. Although the feedback control framework requires around the same number of tests, specifically 15.6 tests per person during Fall 2020 compared to the 16 tests under the 200\% testing rate, the total number of confirmed cases is reduced by approximately 45\%. The significant reduction in total confirmed cases, with a similar amount of testing resources, is due to the feedback control framework saving tests when there is no immediate risk of potential outbreaks, allowing for increased testing during potential spikes, as shown in Figure~\ref{fig:UIUC_Control_0.8_SI}. Specifically, the feedback control framework increases the testing rate during a potential outbreak in October. By utilizing the estimated reproduction number to identify potential outbreaks, the feedback control framework adjusts the testing rate accordingly. If we distribute the tests uniformly instead of implementing the feedback control framework to save tests for targeting potential outbreaks, simply adding 1 test per person to the implemented testing-for-isolation strategy during Fall 2020 might not have mitigated the spread to the same extent achieved by the feedback control framework. The feedback control framework effectively optimizes the allocation of tests and enables a reduction in total confirmed cases by approximately 45\% through strategic testing adjustments. 


Both \baike{Figure}~\ref{fig:UIUC_Control_0.9_SI} and \baike{Figure}~\ref{fig:UIUC_Control_0.8_SI} demonstrate the potential value of the proposed model-free feedback control framework in the context of the testing-for-isolation strategy. Leveraging the estimated reproduction number at each iteration to update the testing rate helps correct uncertainties associated with the design framework in revere engineering the reproduction number.
Moreover, the core idea behind this framework shares similarities with model predictive control techniques, which employ feedback design to learn the model while designing control strategies~\cite{hewing2020learning}. 

After implementing the control mechanism on the reconstructed environment at the UIUC campus, we proceed to explore the framework at the Purdue campus.
As discussed earlier, we can generalize the feedback control methodology in~\eqref{eq_SI:Testing_Rate}to encompass spreading with both symptomatic and asymptomatic infections. Therefore, we investigate the model-free feedback control framework over the Purdue campus, where the reconstructed testing environment is based on a specific reconstruction presented in Figure~\ref{fig:Purdue_0.3_SI}. Recall that the reconstruction allows for a $30\%$ isolation rate and involves a symptomatic infection ratio of $\theta = 55\%$.


Unlike UIUC, where we do not distinguish between symptomatic and asymptomatic infections while implementing the testing-for-isolation strategy, at Purdue, we only adjust the surveillance testing rate for asymptomatic infections. Similar to the implementation of different fixed testing rates for the Purdue campus in Section~SI-2-F, all symptomatic cases will test and isolate themselves within a week once they become infectious and symptomatic.
Hence, based on the estimated $\mathcal{R}_k$ under the testing/isolation rate for asymptomatic cases $\overline{\alpha}_k$ and the testing/isolation rate for symptomatic cases $\underline{\alpha}_k$ at period $k$, based on the defined initial serial interval distributions $\overline{w}$ and $\underline{w}$ for asymptomatic and symptomatic infections, respectively,
the reconstructed infection profile is as follows
\begin{equation*}
 \sum_{i=1}^n v_i(\overline{\alpha}_k) = \theta\sum_{i=1}^n\underline{w}_i(\underline{\alpha}_k)\mathcal{R}_k +
 (1-\theta)\sum_{i=1}^n \overline{w}_i(\overline{\alpha}_k)\mathcal{R}_k, 
\end{equation*}
where $\mathcal{R}_k$ denotes the estimated reproduction number under 
the testing rate for
symptomatic ($\underline{\alpha}_k$) and asymptomatic ($\overline{\alpha}_k$) infections.
If $\mathcal{R}\neq \mathcal{R}^*$, we compute \baike{a} new testing rate for asymptomatic cases $\overline{\alpha}_{k+1}$ during the next period $k+1$. 
Based on the proposed feedback control mechanism in~\eqref{eq_SI:Testing_Rate},
we can further compute the updated testing rate for asymptomatic cases $\overline{\alpha}_{k+1}$ by solving the equation:
\begin{equation}
\label{eq_SI:Testing_Rate_Asym}
\mathcal{R}_k\sum_{i=1}^n \theta\underline{w}_i(\underline{\alpha}_{k})(1-\underline{\alpha}_{k+1})^i
+\mathcal{R}_k
 \sum_{i=1}^n(1-\theta)\overline{w}_i(\overline{\alpha}_{k})(1-\overline{\alpha}_{k+1})^i =\mathcal{R}^*.
\end{equation}
Unlike~\eqref{eq_SI:Testing_Rate} where we have only one unknown $\alpha_{k+1}$,~\eqref{eq_SI:Testing_Rate_Asym} involves two unknowns: the updated testing rates for asymptomatic ($\overline{\alpha}_k$) and symptomatic ($\underline{\alpha}_k$) infections. Therefore, it is necessary to determine one testing rate in advance in practice to obtain the other testing rate. In this particular instance, along with the reconstruction process at Purdue, we consider that all symptomatic infections will be detected through voluntary testing, resulting $\underline{\alpha}_k = \underline{\alpha}_{k+1}=1$. Thus, the only testing rate we will update in~\eqref{eq_SI:Testing_Rate_Asym} is for the asymptomatic infection, i.e., $\overline{\alpha}_{k+1}$. 

Recall that we use the same serial interval distribution for symptomatic and asymptomatic infections, i.e., $\underline{w} = \overline{w} = w$.  In the scenario we study, the only unknown in~\eqref{eq_SI:Testing_Rate_Asym}
is the asymptomatic testing rate for the next time period, $\overline\alpha_{k+1}$. By solving~\eqref{eq_SI:Testing_Rate_Asym}, we can compute the updated asymptomatic testing rate for the next period. Furthermore,~\eqref{eq_SI:Testing_Rate_Asym} can be considered as a generalized formulation of~\eqref{eq_SI:Testing_Rate}, where we can substitute different serial interval distributions for symptomatic and asymptomatic infections  ($\underline{w}$ and $\overline{w}$), the ratio of the symptomatic infection ($\theta$), and the expected reproduction number $\mathcal{R}^*$.
Moreover, if we want to update both symptomatic ($\underline{\alpha}$) and asymptomatic testing rates ($\overline{\alpha}$), we can utilize~\eqref{eq_SI:Testing_Rate_Asym} to establish a relationship between $\underline{\alpha}_{k+1}$ and $\overline{\alpha}_{k+1}$.Based on this relationship and the available testing resources, we can allocate resources to symptomatic (voluntary) and asymptomatic (surveillance) testing rates accordingly.

\begin{figure}[p]
\centering
\includegraphics[width=\textwidth]{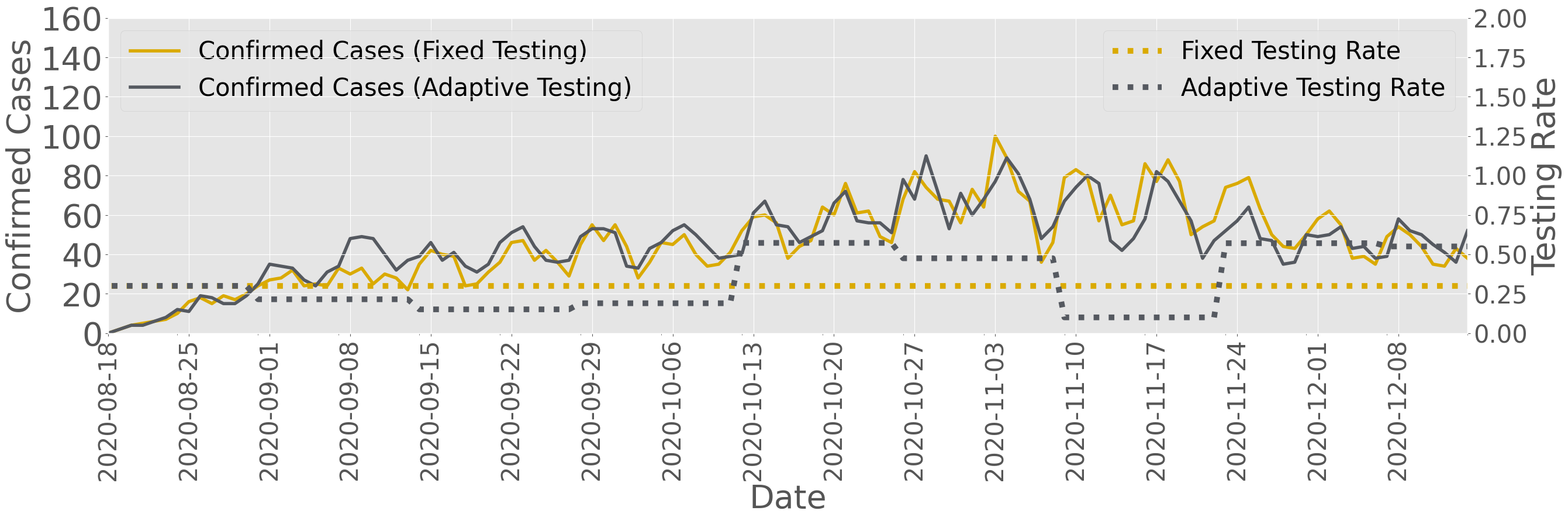}
\caption{Comparison between the fixed and feedback testing-for-isolation strategies at Purdue}
\label{fig:Purdue_Control_1.05}
\end{figure}

Through the implementation of the proposed model-free feedback control methodology in~\eqref{eq_SI:Testing_Rate_Asym} into pandemic mitigation framework, our goal is to match the confirmed cases at Purdue University during Fall 2020, as illustrated in Figure~\ref{fig:Purdue_Total_SI}. To achieve this goal, we conduct simulations in the reconstructed environment at Purdue during Fall 2020, considering both the strategies implemented by Purdue and our proposed feedback control methodology based on~\eqref{eq_SI:Testing_Rate_Asym}.
In order to generate a total number of confirmed cases similar to those observed over the Purdue campus during Fall 2020, we set the expected reproduction number slightly higher than 1, specifically at $\mathcal{R}^*=1.05$\footnote{Ideally, $\mathcal{R}^*$ needs to be smaller than 1. Here, we attempt to generate the similar spreading process as Purdue through the feedback mechanism. Hence, we set $\mathcal{R}^*=1.05$ to be slightly higher than 1.}. Using these settings, we plot the daily confirmed cases with solid lines and the weekly testing rates with dotted lines. Illustrated by the solid lines in Figure~\ref{fig:Purdue_Control_1.05}, we observe around 5200 total confirmed cases under the testing-for-isolation strategy implemented by Purdue and the feedback control strategy, which can adjust the testing rates for asymptomatic cases. 

Comparing the fixed surveillance testing rate implemented by Purdue (30\% per week, represented by the dark grey dotted line)  and the testing rate generated by our feedback control framework (average 28\% per week, represented by the yellow dotted line) in \baike{Figure}~\ref{fig:Purdue_Control_1.05}, we notice that the feedback control framework significantly increases the testing rate for asymptomatic cases, particularly during October 2020. This outcome is in line with the observation made in  
Section~SI-1, where both UIUC and Purdue experienced substantial spikes in the infected population on campus during that month, primarily due to various gathering events associated with the college football season. Thus, similar to the simulation for UIUC, the feedback control framework adjusts the testing rates for asymptomatic cases, as indicated by the dark grey dotted line in  \baike{Figure}~\ref{fig:UIUC_Control_0.9_SI}, surpassing the fixed testing rates implemented by Purdue during October 2020. The simulated result suggests that we can allocate testing resources more efficiently to mitigate potential outbreaks.

\begin{figure}[p]
\centering
\includegraphics[width=\textwidth]{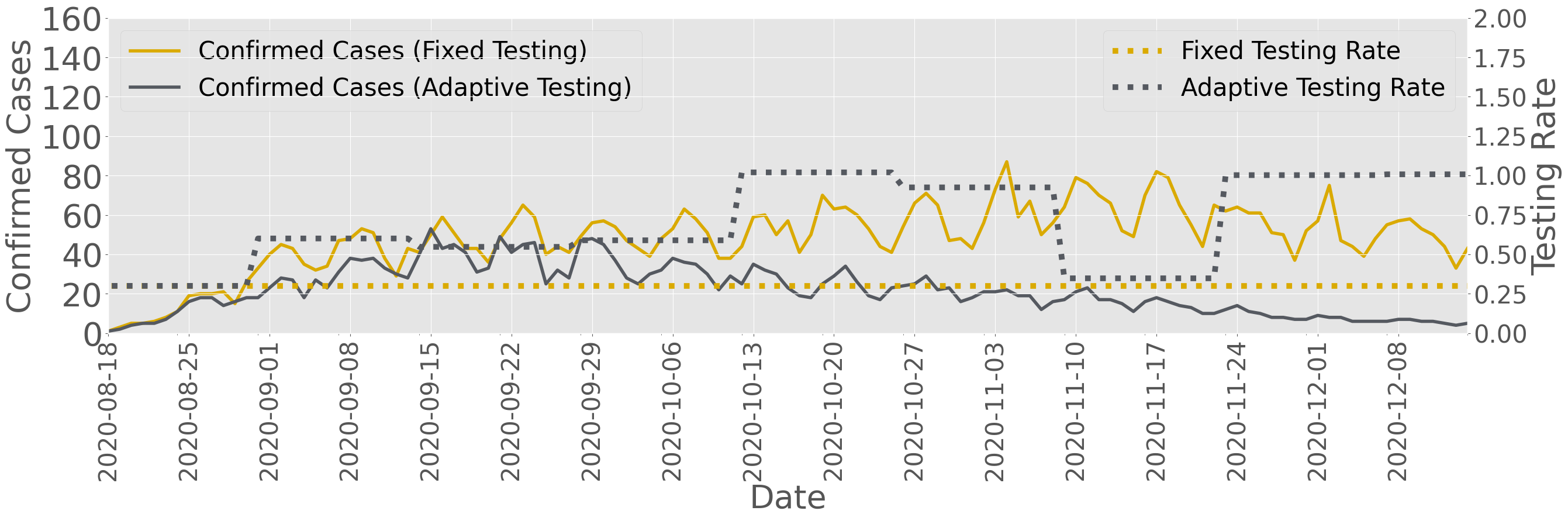}
\caption{Comparison between the fixed and feedback testing-for-isolation strategies at Purdue}
\label{fig:Purdue_Control_1.0}
\end{figure}

In addition to evaluating the testing needs for asymptomatic cases to achieve the same number of confirmed cases as at Purdue, we establish the mitigation problem's objective to control the reproduction number at $\mathcal{R}^*=0.95$. Using the same reconstructed testing environment as in the prior example to capture the spread across the Purdue campus, we implement both the testing-for-isolation strategies from Purdue and our proposed feedback control mechanism based on~\eqref{eq_SI:Testing_Rate_Asym}. We illustrate the confirmed cases with solid lines and the weekly testing rates with dotted lines in \baike{Figure}~\ref{fig:Purdue_Control_1.0}.

Comparing the confirmed cases resulting from Purdue's fixed testing rate strategy (yellow solid lines) and our control framework (dark solid lines), we observe that our control framework effectively prevents potential outbreaks, especially during October, by implementing a higher testing rate for surveillance testing. Additionally, while the fixed testing rate would result in around 5200 confirmed cases during Fall 2020, our feedback testing rates would yield approximately 2800 confirmed cases. To achieve a decrease of about 30\% in the total number of confirmed cases, the feedback control framework would implement an average 59\% testing rate (isolation rate) for asymptomatic infections. These testing resources are predominantly utilized during October 2020. Despite the feedback control framework requiring a higher consumption of testing rates for asymptomatic cases, it effectively helps prevent potential outbreaks during the middle of the Fall 2020 semester at the Purdue campus.


In this section, we introduce a mechanism that utilizes the estimated reproduction number as feedback to automatically update the testing rate. This mechanism is based on the proposed model-free feedback control methodologies in methodologies in~\eqref{eq_SI:Testing_Rate} and~\eqref{eq_SI:Testing_Rate_Asym}. We validate the application of these methodologies by comparing the implemented testing-for-isolation strategies and the strategies generated by the feedback control mechanism within the framework for counterfactual analysis, strategy evaluation, and feedback control of epidemics, at both the UIUC and Purdue campuses, illustrated in Figure  \baike{Figure}~\ref{fig:Control_Framework_SI}. Note that in this work, we utilize the reconstructed environment to execute the feedback control framework. This choice is due to our access solely to historical data, allowing us to illustrate and validate our framework. In real-time scenarios where we have access to current testing data, we can obtain all the necessary information to directly update the testing rate in the near future. However, we still recommend using a reconstructed environment to evaluate the potential impact of different testing/isolation rates. As demonstrated in Section~SI-2-E, the results generated through the framework depend on conditions related to population behaviors and the characteristics of the spreading process.
\section*{SI-3. Discussion}
As indicated in \baike{Figure}~\ref{fig:Control_Framework_SI} and the previous sections of this document, the \baike{tool that enables} of our framework for model-free counterfactual analysis, strategy evaluation, and feedback control of epidemics, is the reverse engineering of the reproduction number. Estimating and scaling the reproduction number is crucial for the assessment of intervention strategies and the design of feedback controls. However, solely relying on the reproduction number for a comprehensive evaluation of a pandemic may pose several limitations and challenges in real-world application. Hence, in this section, we delve into potential limitations and challenges within our proposed framework by examining the restrictions associated with relying solely on the reproduction number.

The reproduction number can indicate the trend in the newly infected population. However, relying solely on the reproduction number may result in incorrect assessments of the spread if the precise number of infected or confirmed cases is not considered. For example, consider the estimated reproduction number from the UIUC in \baike{Figure}~\ref{fig:Estimated_R_UIUC_SI} and Purdue in \baike{Figure}~\ref{fig:Estimated_R_Purdue_SI} during Fall 2020. The highest estimated reproduction number at UIUC is approximately 1.3, while at Purdue, it is around 1.1. Furthermore, the estimated reproduction number for UIUC shows higher variance compared to Purdue. If we are to base our judgment solely on the reproduction number, one may infer that Purdue performed better in mitigating the spread on campus. However, when comparing the confirmed cases at UIUC in \baike{Figure}~\ref{fig:UIUC_P_SI} and Purdue in \baike{Figure}~\ref{fig:Purdue_SVL_P_SI} during Fall 2020, it is apparent that UIUC reported significantly fewer daily confirmed cases and lower spikes in terms of peak infections. Additionally, UIUC conducted more tests, indicating a potential detection of more infected cases compared to Purdue.

So, what could be amiss with our analysis? Consider a scenario where the daily infected cases increase from 100 to 200, resulting in a reproduction number of 2. In another scenario, the daily infected cases rise from 1000 to 1200, which yields a reproduction number of 1.2. Despite both scenarios displaying an increase of 200 infected cases, the difference in the reproduction number is significant. 
This disparity occurs because the reproduction number only captures relative changes in infected cases while disregarding absolute changes. Hence, since UIUC had fewer confirmed cases than Purdue, even a small increase in confirmed cases at UIUC would lead to a larger rise in the reproduction number compared to Purdue.

To address this issue, it is critical to integrate the analysis of the reproduction number with the assessment of infected or confirmed cases. In our proposed work, our objective is to regulate the reproduction number to an anticipated value $\mathcal{R}^*$, as shown in~\eqref{eq_SI:Testing_Rate} and~\eqref{eq_SI:Testing_Rate_Asym}. Nonetheless, as discussed, if the daily infected cases are notably high, managing the reproduction number slightly below 1 might still yield a high daily infected population, potentially resulting in an outbreak or leading to a slow reduction in the daily infected cases. To prevent such circumstances, we can enhance our  framework, particularly the model-free feedback control methodology, by considering both infected or confirmed cases and the reproduction number as feedback information. If the daily confirmed cases substantially exceed our anticipated threshold $\bar{I}$, upon updating the testing rate, we can select an anticipated reproduction number $\mathcal{R}^*$ considerably lower than 1.
Conversely, if the daily confirmed cases fall far below our expected threshold $\bar{I}$, as per the analyses in Section~SI-2-G,the framework may allocate more tests than necessary. In this scenario, when adjusting the testing rate, the updated rate could scale the reproduction number $\mathcal{R}^*$ slightly below than 1. 
In summary, when deploying the  model-free framework for strategy evaluation and feedback control, it becomes essential to consider both the reproduction number and the daily infected cases (or confirmed cases) to overcome the limitations of solely relying on the reproduction number as the exclusive feedback information for control design. By adapting the expected reproduction number based on the daily infected cases, we can enhance the precision and effectiveness of the framework.

\section*{SI-4. Future Works}
We acknowledge several limitations of the proposed framework for model-free counterfactual
analysis, strategy evaluation, and feedback control of epidemics,
and provide potential avenues for improvement through future work. First, when estimating the  reproduction number, we rely on existing infection profiles from the literature. However, different communities may experience different spreading behavior.
Therefore, it is crucial to incorporate contact tracing data from the testing-for-isolation strategies to update the infection profile and the serial interval distribution during the pandemic.

Furthermore, in our control strategy design, \baike{we leverage the reproduction number estimated in the past to refer the reproduction number in the near future when the testing strategy is not changed,} without considering predictions of its trend. In reality, the reproduction number can vary due to factors such as seasonal changes, breaks, or events on campus. Thus, incorporating a predictive control mechanism 
 with machine learning techniques
 to account for these fluctuations is a challenging yet important extension.

Additionally, the goal of maintaining the reproduction number at $\mathcal{R}^*$ that is slightly smaller than 1 aligns with the objective of keeping the infected population at an acceptable level for both universities. However, if the initial infected population is substantial, maintaining the reproduction number at a fixed reproduction number $\mathcal{R}^*$ that is slightly smaller than 1 can still lead to a large number of infections. Hence, adjusting the control goal of the framework becomes highly significant~\cite{vegvari2022commentary, parag2022epidemic}. 

While the framework has the potential to save testing resources, implementing frequent changes in testing policy as designed in the control framework (e.g., weekly or bi-weekly adjustments) may be unrealistic in practice. Even in more flexible environments such as universities, UIUC and Purdue have maintained their testing policies constant for an entire semester. \baike{In addition, the testing rate generated by the feedback control design may exceed the testing capacity of a university. Hence, exploring a constrained optimization problem on the developed mechanism, where implementing constraints on control input, the testing rate, is also necessary in the future. }


We leverage aggregated data to validate the framework and perform all analyses. We can enhance the framework by incorporating spatial and heterogeneous spreading data. For instance, both UIUC and Purdue have recorded infection data that can facilitate the construction of contact-tracing networks. We can estimate the reproduction number to characterize spreading over sub-regions and adjust the testing strategies for these regions separately.
By leveraging data at high resolution, future work can focus on studying a distributed model-free strategy evaluation and feedback control framework that allocates varying degrees of intervention strategies based on the severity of the spread in different connected regions. In addition, the connections between different regions can be inferred through machine learning techniques such as graph learning and causal inference.

Another potential limitation in the feedback control framework is the absence of predictions for the reproduction number in the near future. We have discussed the drawback of lacking predictions when updating the testing/isolation rates in Section~SI-2-G. While the feedback control mechanism~\cite{van2023effective} helps mitigate the uncertainties introduced by the current framework, it is crucial to have accurate predictions of the reproduction number to generate and implement more precise testing and isolation rates and to anticipate potential changes in the \baike{epidemic} spreading processes. 
Researchers have demonstrated the effectiveness of implementing reproduction number prediction method based on historical estimated reproduction number, as showcased in the Epinow2 package~\cite{abbott2020estimating, gostic2020practical, bracher2021pre}. In our future work, we explore the relationship between data sizes and prediction accuracy to further enhance the feedback control framework, by accounting the uncertainty from the variance. By incorporating predictive capabilities, we aim to improve the framework's ability to adapt to evolving epidemic dynamics and optimize epidemic mitigation strategies accordingly.

Nevertheless, we hope that this work sparks discussions on the role and limitations of reverse engineering the reproduction number for model-free 
counterfactual
analysis, strategy evaluation, and feedback control of epidemics,
from both analytical and computational perspectives. We believe that leveraging statistical information and inference in the field of epidemic spreading processes can inspire and benefit the design of model-free strategy evaluation and feedback control for the control society. Furthermore,
our work not only advances the field of mathematical epidemiology but also strengthens systems engineering in controlling complex systems under uncertainty, civil and environmental engineering in investigating disease spread on urban life, the development of smart and sustainable communities, the construction of modern cyber-physical-human systems, and urban ecology in stabilizing our ecosystem.

\end{document}